\documentclass[12pt, draftclsnofoot, onecolumn]{IEEEtran}
\usepackage{amsfonts}
\usepackage{slashbox}
\usepackage{amsmath}
\usepackage{multirow}
\usepackage{graphicx}
\usepackage{amsfonts}
\usepackage{amsmath,epsfig}
\usepackage{mathbbold}
\usepackage{stfloats}
\usepackage{amssymb}
\usepackage{url}
\usepackage{bm}
\usepackage{cite}
\usepackage{amsmath}
\usepackage{amssymb}
\usepackage{latexsym}
\usepackage{multirow}
\usepackage{epsfig}
\usepackage{graphics}
\usepackage{mathrsfs}
\usepackage{algorithmic}
\usepackage{algorithm}
\usepackage{subfigure}
\usepackage{slashbox}
\usepackage[all]{xy}

\usepackage{pifont}
\usepackage{bbding}
\usepackage{stmaryrd}
\usepackage{amssymb}
\usepackage{amsfonts}
\usepackage{epic}
\usepackage{stfloats}
\usepackage{latexsym}
\usepackage{epstopdf}
\usepackage{epic}
\usepackage{bm}
\usepackage{xcolor}
\usepackage{mathrsfs}
\usepackage{pifont}
\usepackage{bbding}
\usepackage{amsmath,epsfig}
\usepackage{mathbbold}
\usepackage{stmaryrd}
\usepackage{amssymb}
\usepackage{amsfonts}
\usepackage{epic}
\usepackage{graphicx}
\usepackage{subfigure}

\usepackage{enumerate}

\usepackage{stfloats}
\usepackage{latexsym}
\usepackage{epstopdf}
\usepackage{epic}
\usepackage{multirow}
\usepackage{stfloats}
\usepackage{bm}
\usepackage{amsmath}
\usepackage{amssymb}
\usepackage{amsthm}

\usepackage{booktabs}
\usepackage{color}
\usepackage{cases}
\usepackage{array}
\usepackage{makecell}
\usepackage{times}

\newtheorem{remark}{Remark}

\newcommand{\mv}[1]{\mbox{\boldmath{$ #1 $}}}

\newcommand{\A}{\bm A}

\newcommand{\OO}{\mathcal O}

\newcommand{\Ss}{\bm S}
\newcommand{\G}{\bm G}

\newcommand{\0}{\bm 0}

\newcommand{\ttheta}{\mathbf \Theta}

\newcommand{\g}{\bm g}
\newcommand{\h}{\bm h}

\graphicspath{{./figs/}}

\begin{document}
\title{Intelligent Reflecting Surface Aided  Wireless Communications: A Tutorial}
\author{\IEEEauthorblockN{Qingqing~Wu, Shuowen Zhang,  Beixiong Zheng, Changsheng You, and~Rui~Zhang,~\IEEEmembership{Fellow,~IEEE}}
\thanks{Q. Wu is  with the State Key Laboratory of Internet of Things for Smart City and Department of Electrical and Computer Engineering, University of Macau, Macau,  China 999078 (email: qingqingwu@um.edu.mo). He was with the Department of Electrical and Computer Engineering, National University of Singapore, Singapore 117583.}
\thanks{S. Zhang, B. Zheng, C. You, and R. Zhang are with the Department of Electrical and Computer Engineering, National University of Singapore, Singapore 117583 (e-mail:\{elezhsh, elezbe, eleyouc, elezhang\}@nus.edu.sg). }
\thanks{Corresponding author: Rui Zhang. }
}
\IEEEspecialpapernotice{(Invited Paper)}
\maketitle
%

\begin{abstract}
Intelligent reflecting surface (IRS) is an enabling   technology to engineer the radio signal prorogation in  wireless networks. By smartly tuning the signal reflection via a large number of low-cost passive  reflecting elements, IRS is capable of dynamically altering wireless channels to enhance the communication performance. It is thus expected that the new IRS-aided hybrid wireless network comprising both active and passive components will be  highly promising  to achieve a sustainable  capacity growth cost-effectively in the future.  Despite its great  potential, IRS faces new challenges to be efficiently integrated into wireless networks, such as reflection optimization, channel estimation, and deployment from communication design perspectives.  In this paper, we provide a tutorial overview of IRS-aided wireless communication to address the above issues, and elaborate  its reflection and channel models, hardware architecture and practical constraints, as well as various appealing  applications in wireless networks.  Moreover, we highlight important  directions worthy of further investigation in future work.
\end{abstract}





\section{Introduction}
\subsection{Motivation}
Although the  fifth-generation (5G) wireless network  is still under deployment worldwide, both academia and industry have been enthusiastically  looking into future beyond 5G (B5G) such as the sixth-generation (6G) wireless network that targets at meeting more stringent requirements than 5G, such as ultra high data rate and energy efficiency, global coverage and connectivity, as well as extremely high reliability and low latency. These requirements, however, may not  be fully achieved with the existing technology trends  for accommodating 5G services (e.g., enhanced mobile broadband (eMBB), ultra-reliable and low latency communication (URLLC), and massive machine-type communication (mMTC)), which mainly include \cite{boccardi2014five,shafi20175g,zhang2016fundamental,wu2016overview}
\begin{itemize}
\item deploying increasingly more active nodes such as base stations (BSs),  access points (APs), relays, and distributed antennas/remote radio heads (RRHs) to shorten the communication distance for achieving enhanced network coverage and capacity, which, however, incurs higher energy consumption and deployment/backhaul/maintenance cost, as well as the more severe and complicated  network interference issue;
\item packing substantially  more antennas at the BSs/APs/relays  to harness the enormous massive multiple-input-multiple-output (M-MIMO) gains, which requires  increased hardware and energy cost as well as  signal processing complexity;
\item  migrating to higher frequency bands such as millimeter wave (mmWave) and even terahertz (THz)  frequencies to utilize their large and available bandwidth, which inevitably results in deploying even more active nodes and mounting them even more antennas (i.e., super MIMO) so as to compensate for their higher propagation loss over distance.
\end{itemize}
In view of the above issues and limitations, it is imperative to develop disruptively new and innovative technologies to achieve a sustainable capacity growth of future wireless networks with low and affordable  cost, complexity, and energy consumption.

On the other hand, the fundamental challenge for achieving ultra-reliable wireless communications  arises from the time-varying wireless  channels due to user mobility. Traditional approaches  for tackling this challenge either compensate for the channel fading by exploiting various modulation, coding and  diversity techniques, or adapt to it  via adaptive power/rate control and beamforming techniques  \cite{tse2005fundamentals,goldsmith2005wireless}. However, they not only need additional overhead but also  have limited  control over the largely random wireless channels, thus leaving  the ultimate barrier to achieving high-capacity and ultra-reliable wireless communications  unconquered.

\subsection{What is IRS?}
Motivated by the above,  intelligent reflecting surface (IRS) has  recently emerged  as a promising new paradigm  to achieve smart and reconfigurable wireless channels/radio propagation environment for B5G/6G wireless communication systems   \cite{wu2018IRS,JR:wu2018IRS,JR:wu2019IRSmaga}.
Generally speaking, IRS is a planar surface comprising a large number of  passive reflecting elements,  each of which is able to induce a controllable amplitude and/or  phase change to  the incident signal independently.
By  densely deploying IRSs in wireless network and smartly coordinating their reflections, the signal propagation/wireless channels  between transmitters and receivers can be flexibly reconfigured to achieve
desired realizations and/or distributions, which thus provides a new means  to fundamentally tackle the wireless  channel  fading impairment and interference issue, and potentially achieves a quantum leap improvement for wireless communication capacity and reliability.

\begin{figure*}[t]
\centering
\subfigure[M-MIMO system  versus IRS-aided small/moderate  MIMO system.]{\includegraphics[width=0.9\textwidth]{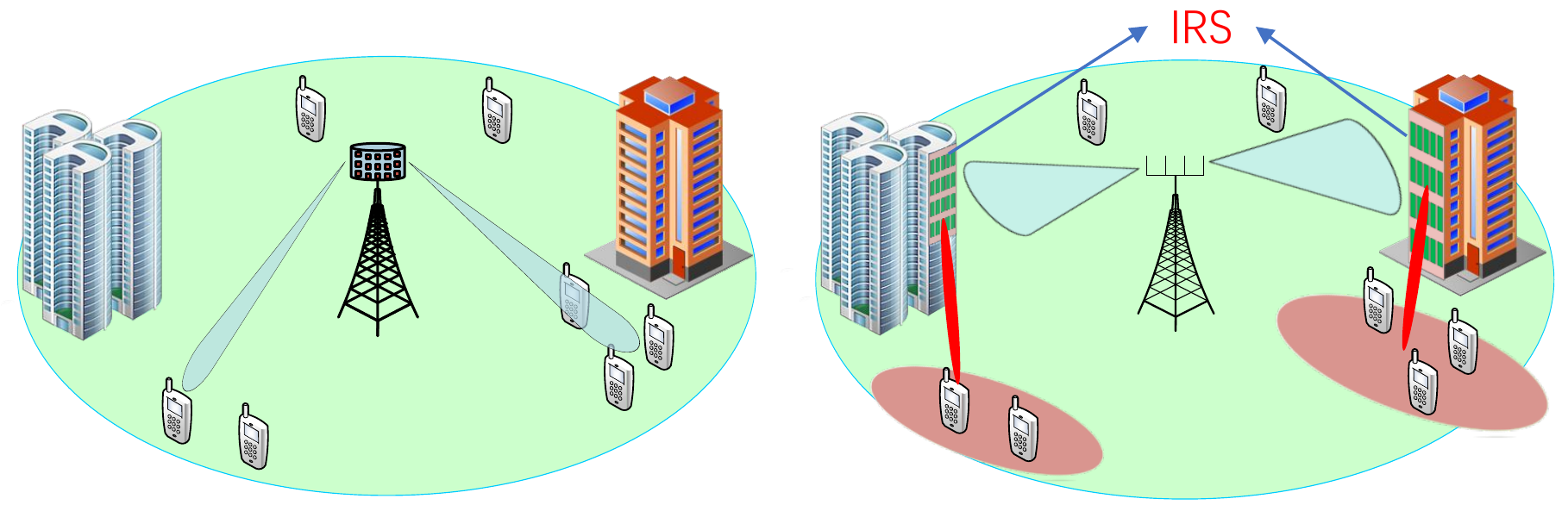}} ~~
\subfigure[Heterogenous wireless network versus IRS-aided hybrid wireless network.]{\includegraphics[width=0.9\textwidth]{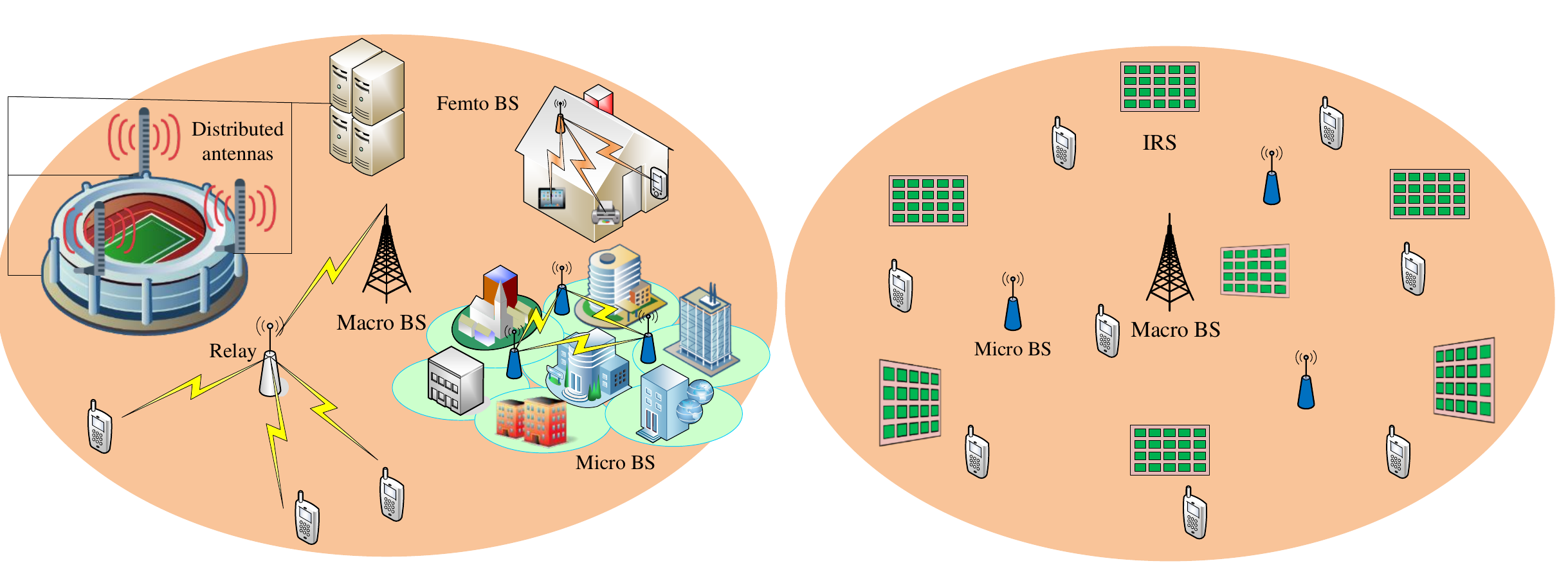}} 
\caption{Potential paradigm shifts of wireless system/network designs with IRS.} \label{simulation:pow2}
\end{figure*}

Not only conceptually appealing,  IRS also  possesses various practical advantages for implementation. First, its  reflecting elements (e.g., low-cost printed dipoles) only passively reflect the impinging signals without requiring any  transmit radio-frequency (RF) chains, thus can be implemented/operated with  orders-of-magnitude lower hardware/energy cost as compared to traditional active antenna arrays or the recently proposed active surfaces  \cite{hu2017beyond}. Besides,  IRS operates in full-duplex (FD) mode  and is free of any antenna noise amplification as well as self-interference, which thus offers competitive advantages over traditional active relays, e.g.,  half-duplex (HD) relay that suffers from low spectral efficiency as well as   FD relay that needs sophisticated techniques for  self-interference cancellation.  Furthermore, since IRS is generally of low  profile,  light weight,  and  conformal geometry, it can be easily mounted on/removed from environment objects for deployment/replacement. Finally,  IRS serves as an auxiliary  device in  wireless networks and can be integrated into them transparently, thus providing great  flexibility and compatibility with existing wireless systems (e.g., cellular or WiFi).

Due to  the above promising advantages, IRS is suitable to be  massively deployed in wireless networks to significantly enhance its spectral and energy efficiency cost-effectively. As such, it is envisioned that IRS will lead to fundamental paradigm shifts of wireless system/network designs, namely, from the existing M-MIMO system without IRS  to the new IRS-aided small/moderate MIMO system, as well as  from the existing  heterogenous wireless network to the new IRS-aided hybrid network in the future, as shown in Figs. \ref{simulation:pow2} (a) and (b), respectively.  On one hand,  different from M-MIMO  that leverages  tens and even hundreds of active antennas to generate sharp beams directly, an IRS-aided MIMO system  allows the BS to be equipped with substantially less  antennas without compromising the users' quality-of-service (QoS), by exploiting the large aperture of  IRS to create fine-grained reflect beams via smart passive reflection \cite{JR:wu2018IRS}. As such, the system hardware cost and energy consumption can be significantly reduced, especially for wireless systems migrating to higher frequency bands in the future. On the other hand, although existing wireless networks rely on a  heterogenous multi-tier architecture consisting of macro and small BSs/APs, relays, distributed antennas, etc., they are all active nodes that generate new signals in the network, thus requiring sophisticated coordination and interference management among them in order to achieve the premise of enhanced network spatial capacity with more active nodes deployed. However, this approach inevitably aggravates the network operation  overhead and thus may not be able to sustain the wireless network capacity growth cost-effectively in the future.  In contrast, integrating IRSs into wireless network will shift the existing heterogeneous network with active components only  to a new hybrid architecture comprising both active and passive components co-working in an intelligent way. Since IRSs are of much lower cost as compared to their active counterparts, they can be more densely deployed in wireless network at even lower cost, yet without the need of sophisticated interference management between IRSs thanks to their passive reflection and resultant local coverage. By optimally setting the ratios between active BSs and passive IRSs deployed in the hybrid network given their total cost, a sustainable network capacity scaling with  cost can be achieved \cite{lyu2020hybrid}.



\begin{figure}[!t]
\centering
\includegraphics[width=1\textwidth]{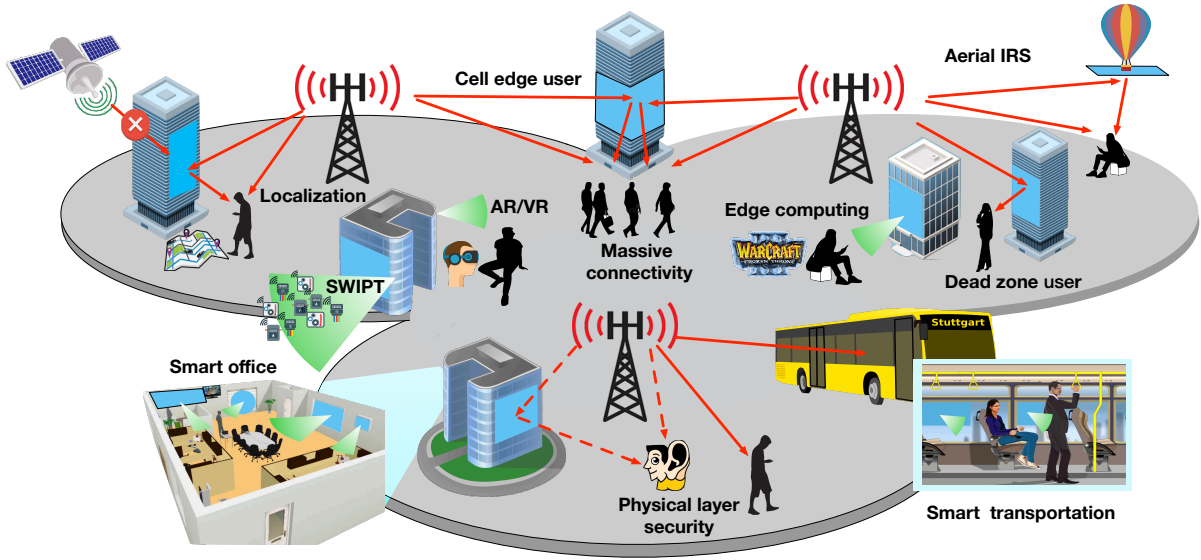}
\caption{Illustration of IRS applications in future wireless network.} \label{system:model}
\end{figure}
In Fig. \ref{system:model}, we show an envisioned future wireless network aided by IRSs with a variety of promising applications. 
For instance, for users located in a service dead zone, IRS can be deployed to create a virtual  line-of-sight (LoS) link between the users and their serving BS/AP that bypasses the obstacle between them. This is particularly useful for the coverage extension in mmWave and THz  communications that are highly vulnerable to blockage. Besides, deploying IRSs at the cell edge not only helps improve the desired signal power at cell-edge users but also facilitates the suppression of co-channel interference from neighboring cells to them.  Moreover, to improve the efficiency of simultaneous wireless information and power transfer (SWIPT) from the AP to wireless devices in e.g., smart office/home,  the large aperture of IRS can  be leveraged to compensate for the significant power loss over long distance via reflect beamforming to its  nearby devices.
In indoor environment,  IRS can also  be attached to the ceilings, walls, furniture, and even  behind the paintings/decorations, to help achieve enhanced coverage and high-capacity hot-spot, which is particularly appealing for eMBB and mMTC applications in factories, stadiums,  shopping centers, airports, etc.
While in outdoor environment, IRS can be coated on the building facade, lamppost, advertising board, and even the surface of high-speed moving vehicles, to support various applications, e.g., URLLC for remote control and smart transportation by effectively compensating the Doppler effects \cite{basar2019reconfigurable}. Therefore, IRS is a disruptive technology for making our current ``dumb''  environment intelligent, which can potentially benefit a wide range of vertical  industries in 5G/6G such as transportation, manufacturing, smart city, etc. Recently, IRS has been  recognized as one promising technology for  the future 6G ecosystem  \cite{latva2019key2,rajatheva2020white}. In addition, there has been an upsurge of interest in industry on implementing and commercializing  IRS-like technologies to create new value chains, and in the meanwhile, several  pilot projects have been launched to advance the research in this new field, with their more information given in Table \ref{Table:industry}.   It is worth noting that there have been other terminologies similar to IRS named   in the literature, such as  intelligent wall \cite{subrt2012intelligent},  smart reflectarray \cite{Infocom_Sun}, and reconfigurable metasurface/intelligent surface (RIS) \cite{di2019smart,basar2019wireless}, large intelligent surface/antennas (LISA) \cite{liang2019large}, RFocus \cite{arun2019rfocus}, among others, which, despite different numerologies,   are essentially based on the same principle of passive and tunable reflecting/refracting surfaces.\footnote{Also note that such surfaces can be practically manufactured as a mirror or lens for signal reflection and refraction, respectively, depending on the application scenarios where wireless transmitter and receiver are located on the same side of the surface, or on its opposite sides.}


\begin{table*}[t!]
\centering
{
\caption{List of main industry progress, prototypes, and projects related to IRS.}\label{Table:industry}
\resizebox{\textwidth}{!}{
\begin{tabular}{| c | c |l|}
\hline
\textbf{Company}& \textbf{Year}& \textbf{\qquad\qquad\qquad\qquad\qquad\qquad Main activity and achievement} \\
\hline
NTT DOCOMO and Metawave & 2018& Demonstrate 28 GHz-band 5G using the first meta-structure reflectarray \cite{DOCOMO2018}. \\
\hline
Lumotive and TowerJazz & 2019 &  Demonstrate the first true solid-state beam steering using liquid crystal metasurface\cite{Lumotive2019}.\\
\hline
Pivotal Commware & 2019 & Demonstrate holographic beamforming technology using software-defined antennas\cite{PivotalCommware2019}.\\
\hline
NTT DOCOMO and AGC Inc. & 2020 & Demonstrate the first prototype transparent dynamic metasurface for 5G\cite{DOCOMO2020}.\\
\hline
Greenerwave & -- &  Develop physics-inspired algorithms for reconfigurable metasurfaces\cite{greenerwave}.\\
\hline
\textbf{Research project}& \textbf{Start year}& \textbf{\qquad\qquad\qquad\qquad\qquad\qquad\qquad\quad  Main objective}\\
\hline
VisorSurf&2017&Develop a hardware platform  for software-driven functional metasurface\cite{visorsurf}.\\
\hline
ARIADE&2019&Design metasurface integrated with new radio and artificial intelligence (AI) techniques\cite{ariadne}.\\
\hline
\multirow{2}{*}{PathFinder}
&\multirow{2}{*}{\begin{tabular}[c]{@{}c@{}}2021\end{tabular}}&
\multirow{1}{*}{Establish the theoretical and algorithmic foundations for  intelligent metasurface enabled}\\
& &  wireless $2.0$ networks\cite{PathFinder}.\\
\hline
\end{tabular}}}
\end{table*}

\subsection{What's New?}

Although IRS can be regarded as a reconfigurable metasurface, it extends the traditional applications of metasurface via controlling electromagnetic (EM) waves such as  invisibility cloaking, imaging, radar sensing, and hologram \cite{he2019tunable}, to the new frontier of  wireless communication as an innovative enabler for smart and reconfigurable propagation environment.  Moreover, compared with the traditional  reflectarray  \cite{hum2013reconfigurable} where a passive mirror/lens with fixed/reconfigurable beam patterns is placed in the near field of the wireless transceiver for saving the active antennas/RF chains,  IRS is flexibly located in the network to  help alter  the wireless communication channel via smart reflection. As such, the design of IRS-aided wireless systems/networks faces new and unique challenges that arise from a communication standpoint, which are elaborated as follows.

Firstly, the passive reflections of all reflecting elements at each  IRS need to be properly designed to achieve  cooperative signal focusing and/or interference cancellation in its local proximity. Meanwhile, to serve all the users  in the network regardless of whether there is any associated IRS nearby each user, the IRS passive reflections also need to be jointly designed with the BSs'/users' transmissions so as to optimize their end-to-end communications over the reconfigured wireless channels by IRSs.
Secondly, as IRS  does not possess  RF chains in general, how to  acquire the channel state information (CSI) between IRS and its serving BSs/users that is essential to the aforementioned IRS reflection optimization becomes a practically difficult task, especially considering that IRS typically has a large number of reflecting elements and thus associated channel coefficients to be estimated.
Thirdly, the optimal deployment strategy of IRSs in wireless network to maximize the network capacity   is expected  to be significantly different from that for the conventional wireless networks with active BSs/APs and  relays only  due to their different array architectures (passive versus active) and operating mechanisms (reflect versus transmit/receive), which thus needs to be re-investigated thoroughly.  In a nutshell, the efficient integration of IRSs into wireless network brings both new opportunities as well as challenges, which deserve a new and dedicated study.


   \begin{table*}[!htbp]
\centering
{\caption{List of representative survey/overview papers related to IRS.}\label{Table:liteature:survey}
\resizebox{\textwidth}{!}{
\begin{tabular}{| c | l|}
\hline
\textbf{Reference}& \textbf{\qquad\qquad\qquad\qquad\qquad\qquad\qquad\qquad\qquad     Main contributions} \\
\hline
\multirow{2}{*}{\cite{JR:wu2019IRSmaga}}&\multirow{1}{*}{Provide an overview of IRS technology for wireless communication, and discuss its  main applications and   key }\tabularnewline
& technical challenges.\\
\hline
\multirow{2}{*}{\cite{di2019smart}}&\multirow{1}{*}{Discuss the applications of reconfigurable metasurface in improving communication, sensing and  computing }\\
& performance.\\
\hline
\multirow{2}{*}{\cite{basar2019wireless} }&\multirow{1}{*}{Summarize the state-of-the-art solutions for RIS-empowered wireless networks with an emphasis on applying}\\
& RIS  as multipath controller and energy-efficient transmitter.\\
\hline
\multirow{2}{*}{\cite{liang2019large} }&\multirow{1}{*}{Discuss the implementations, applications, and open research problems  of LISA as well as its differences and }\\
&  connections with backscatter communication.\\
\hline
\multirow{2}{*}{\cite{Liaskos2018} }&\multirow{1}{*}{Introduce the functional and physical architecture of software-controlled metasurface and discuss its network-layer }\\
&   integration.\\
\hline
\multirow{1}{*}{\cite{ntontin2019reconfigurable2} }&\multirow{1}{*}{Discuss key differences and similarities between RIS and relay, and  compare their communication performance.}\\
\hline
\multirow{2}{*}{\cite{elmossallamy2020reconfigurable} }&\multirow{1}{*}{Overview different implementations of RIS using metasurface and reflectarray, and  discuss suitable RIS channel }\\
&  modelling as well as challenges and opportunities in RIS-aided wireless networks.\\
\hline
\multirow{2}{*}{\cite{huang2019holographic} }&\multirow{1}{*}{Introduce the holographic MIMO surface (HMIMOS) and summarize its hardware architectures, classifications,}\\
& as well as main characteristics.\\
\hline
\multirow{2}{*}{\cite{di2020smart_JSAC} }&\multirow{1}{*}{Provide a comprehensive overview of RIS applications, technological advantages,  state-of-the-art research and}\\
&  future  research directions. \\
\hline
\multirow{2}{*}{\cite{yuan2020reconfigurable}}&\multirow{1}{*}{Briefly discuss three design issues on RIS, including IRS channel estimation, passive information transfer, and }\\
& resource allocation.\\
\hline
\multirow{2}{*}{\cite{gong2019towards}}&\multirow{1}{*}{Provide literature survey of IRS-aided wireless network and overview different performance metrics and  }\\
& analytical approaches.\\
\hline
\multirow{1}{*}{\cite{bjornson2020reconfigurable}}&\multirow{1}{*}{Review the fundamentals of RIS/IRS and debunk three specific myths about their  functionalities and performance gains. }\\
\hline
\end{tabular}}}
\end{table*}

 \subsection{Objective, Contribution, and Organization}

The promising prospects of IRS for future wireless  networks have spurred extensive research recently.
A handful of articles have appeared in the literature providing  overview or survey of the research work on IRS and its variants  from different perspectives, such as IRS implementation, channel modelling, applications, etc. \cite{JR:wu2019IRSmaga,di2019smart,basar2019wireless,ntontin2019reconfigurable2,liang2019large,Liaskos2018,elmossallamy2020reconfigurable,huang2019holographic,di2020smart_JSAC,yuan2020reconfigurable,gong2019towards,bjornson2020reconfigurable}, which are summarized  in Table~\ref{Table:liteature:survey} for ease of reference.

Compared with the above works, this paper is the first tutorial on IRS-aided wireless communications, with an emphasis on addressing its main technical challenges from a communication standpoint. In addition to reviewing the state-of-the-art results on IRS, this paper aims to provide in-depth technical discussion  to facilitate and inspire future research in  modelling, analysis,  design, optimization, and implementation of IRS-aided wireless networks. To this end, this tutorial paper offers a new and  systematic treatment on how to address three key design issues in them, namely, IRS passive reflection optimization, IRS channel estimation, and IRS deployment from various communication perspectives.


The rest of this paper is organized as follows. Section II introduces the fundamentals of IRS-aided wireless communication, including its signal and channel models, hardware architecture as well as practical constraints. Section III addresses  the passive reflection optimization for IRS-aided wireless communication under various system setups, namely, from single-user to multi-user, single-antenna to multi-antenna, narrow-band to broadband, as well as from single-cell to multi-cell. In Section IV, we present  promising methods  for  IRS channel estimation under different IRS configurations and communication setups.  In Section V, we address the optimal deployment of IRSs at both the link  and network levels.  In Section VI, we discuss other relevant topics on IRS for broadening its scope.  Finally, we conclude this paper in Section VII.

\emph{Notations:} In this paper, scalars are denoted by italic letters, vectors and matrices are denoted by bold-face lower-case and upper-case letters, respectively. $\mathbb{C}^{x\times y}$ denotes the space of $x\times y$ complex-valued matrices. For a complex-valued vector $\bm{x}$, $\|\bm{x}\|$ denotes its Euclidean norm and $\text{diag}(\bm{x})$ denotes a diagonal matrix with the elements in $\bm{x}$ on its main diagonal. The distribution of a circularly symmetric complex Gaussian (CSCG) random vector with mean vector  $\bm{x}$ and covariance matrix ${\bm \Sigma}$ is denoted by  $\mathcal{CN}(\bm{x},{\bm \Sigma})$; and $\sim$ stands for ``distributed as''. For a square matrix $\Ss$, ${\rm{tr}}(\Ss)$ and $\Ss^{-1}$ denote its trace and inverse, respectively, while $\Ss\succeq \bm{0}$ means that $\Ss$ is positive semi-definite, where $\bm{0}$ is a zero matrix of proper size.  For a general matrix $\A$, $\A^*$, $\A^H$,  ${\rm{rank}}(\A)$, and  $\left[{\A}\right]_{i,j}$ denote its conjugate,  conjugate transpose, rank, and the $(i,j)$th entry, respectively. $ \jmath $ denotes the imaginary unit, i.e., $\jmath ^2 = -1 $. $\mathbb{E}(\cdot)$ denotes the statistical expectation. $ \mathrm{Re}\{\cdot\}$ denotes the real part of a complex number.  The operator  mod($a,b$) returns the remainder after division of $a$ by $b$, where $a$ is the dividend and $b$ is the divisor; $\circledast$ denotes the convolution operation; $\lfloor x \rfloor$ and $\lceil x \rceil$ return the nearest integer less than or equal to, and  greater than or equal to    the real number  $x$, respectively.


%


\section{ IRS Fundamentals}\label{Sec:IRSBasics}
In this section, we present the  basics pertinent to IRS-aided wireless communication, where the fundamental  IRS signal and channel models are first introduced, followed by its hardware architecture and practical constraints as well as their induced issues worthy of future investigation. 
\subsection{IRS Signal and Channel Model}  
For the purpose of exposition, we consider the  basic point-to-point communication system where  an IRS comprising  $N$ passive reflecting elements on a planar surface is deployed to assist in the communication from a transmitter  to its intended receiver. For ease of illustration, we assume a single antenna at both the transmitter and receiver and the communication system is of narrow band, while the more general multi-antenna and/or broadband systems will be considered later in this paper.  The carrier frequency and the system bandwidth are denoted by $f_c$ and $B$ in hertz (Hz), respectively, with $B\ll f_c$.

 Let $x(t)$ denote the equivalent complex-valued baseband transmit signal.
Without loss of generality, we first focus on the signal propagation from the transmitter to the receiver via one particular reflecting element of the IRS, denoted by $n$, with $n\in\{1,\ldots,N\}$.
Denote by ${\alpha}_{1,n}   e^{-\jmath \xi_{1,n}}$ the equivalent baseband complex channel coefficient  from the transmitter to IRS element $n$  in which ${\alpha}_{1,n}$ and $\xi_{1,n}$ represent   the amplitude attenuation and phase shift of the frequency-flat channel of the narrow-band system, respectively. Then the passband signal impinging on IRS element $n$  is given by
\begin{align}   
y_{in, n}(t)  
  = \mathrm{Re}\left\{ {\alpha}_{1,n}   e^{-\jmath \xi_{1,n}}  x(t)     e^{\jmath 2\pi f_c t}   \right\}.  
\end{align}
Denote the amplitude attenuation and time  delay induced by IRS element $n$ by $\beta_n\in [0, 1]$\footnote{This is because IRS reflecting elements  are passive without signal amplification. } and $t_n \in [0, 1/f_c]$, respectively. By ignoring the hardware imperfections such as circuit non-linearity and phase noise,  the  reflected signal by IRS element $n$ is expressed as
\begin{align}\label{reflection}
y_{out, n}(t)  =   \beta_n y_{in,n }(t-t_n)  &= \mathrm{Re}\left\{ \beta_n   {\alpha}_{1,n} e^{-\jmath \xi_{1,n}}      x(t- t_n)   e^{\jmath 2\pi f_c (t-t_n)}   \right\} \nonumber \\
 & \approx  \mathrm{Re}\left\{  \left[  \beta_n e^{-\jmath \theta'_n}  \alpha_{1,n}  e^{-\jmath \xi_{1,n}}   x(t )  \right] e^{\jmath 2\pi f_c t}  \right\},      
\end{align}
where  we have assumed $ x(t-t_n) \approx  x(t ) $ due to the fact that $t_n\leq 1/f_c \ll 1/B$, and $-\theta'_n \triangleq -2\pi f_ct_n \in [ -2\pi,  0]$ is the phase shift induced by element $n$.
Let $s_{in,n}(t) \triangleq {\alpha}_{1,n}     e^{-\jmath \xi_{1,n}}  x(t) $ and $s_{out,n}(t)\triangleq  \beta_n e^{-\jmath \theta'_n}  \alpha_{1,n}  e^{-\jmath \xi_{1,n}}    x(t )  $, which denote the equivalent baseband signals of $y_{in, n}(t)$ and  $y_{out, n}(t)$, respectively. The IRS signal reflection model in the baseband is thus given by
\begin{align}\label{element:reflection}
s_{out, n}(t) = \beta_n e^{-\jmath \theta'_n}s_{in,n}(t)  \overset{(a)}{ =} \beta_n e^{ \jmath \theta_n}s_{in,n}(t),
\end{align}
where $\theta_n \in  [0,  2\pi]$ and ${(a)}$ is due to the fact that the phase shift is periodic with respect to $2\pi$. Thus, we consider the IRS phase shift in $[0,2\pi)$ for convenience in the sequel of this paper. From \eqref{element:reflection}, it is observed that in the baseband signal model,  the output/reflected signal of IRS element $n$ is given by multiplying the corresponding input/impinging signal  by a complex reflection coefficient, $\beta_n e^{ \jmath \theta_n}$.

From IRS element $n$ to the receiver, the reflected signal undergoes a similar equivalent narrow-band frequency-flat channel given by ${\alpha}_{2,n}   e^{-\jmath \xi_{2,n}}$. Then the passband signal arriving  at the receiver via IRS element $n$'s reflection is expressed as
\begin{align}\label{baseband:model}
y_{r,n}(t) & =   \mathrm{Re}\left\{    \left[  \alpha_{1,n}  e^{-\jmath \xi_{1,n}}  \beta_n e^{\jmath \theta_n}   \alpha_{2,n}  e^{-\jmath \xi_{2,n}}     x(t  ) \right] e^{\jmath 2\pi f_c t}   \right\}.  
\end{align}
Thus, the cascaded  channel from the transmitter to the receiver  via IRS element $n$ has been modeled.
Let  $h^*_{r,n} \triangleq  {\alpha}_{1,n}   e^{-\jmath \xi_{1,n}}$  and $  g_n \triangleq  {\alpha}_{2,n}   e^{-\jmath \xi_{2,n}}$. The corresponding baseband signal model of \eqref{baseband:model} is given by
\begin{align}\label{eq:7}
y_n(t)  =  \beta_n e^{\jmath \theta_n}   h^*_{r,n}g_n   x(t).
\end{align}
From \eqref{eq:7}, it is observed that the IRS reflected channel is a multiplication of three terms, namely, transmitter-to-element $n$ channel, IRS reflection, and element $n$-to-receiver channel.

For simplicity, we assume that there is no signal  coupling in the reflection by neighbouring IRS elements, i.e., all IRS elements reflect the incident signals independently. Due to the substantial path loss, we only consider signals reflected by the IRS for the first time and ignore those reflected by it two or more times. As such,  the  received signal from all IRS elements can be modeled as a superposition of their respective reflected signals; thus, the  baseband signal model accounting for  all the $N$ IRS elements is given by
\begin{align}\label{eq:8}
y(t)  =  \left(\sum_{n=1}^N \beta_n e^{\jmath \theta_n}   h_{r,n}^*g_n \right)  x(t) =  \h^H_r \ttheta \g x(t),
\end{align}
where $\h^H_r = [h^*_{r,1}, \cdots, h^*_{r,N}]$,  $\g  = [g_1, \cdots, g_N]^T$, and $\ttheta = \text{diag} (\beta_1 e^{\jmath\theta_1}, \cdots,  \beta_N e^{\jmath\theta_N})$.
Note that the IRS with $N$ elements essentially performs a linear mapping from the incident (input) signal vector to a reflected (output) signal vector by an $N\times N$  diagonal complex reflecting matrix $\ttheta$, which is diagonal because  each IRS element reflects the signal  independently and there is no signal coupling or joint processing over the IRS elements.

\begin{figure*}[!t]
\centering
\includegraphics[width=0.6\textwidth]{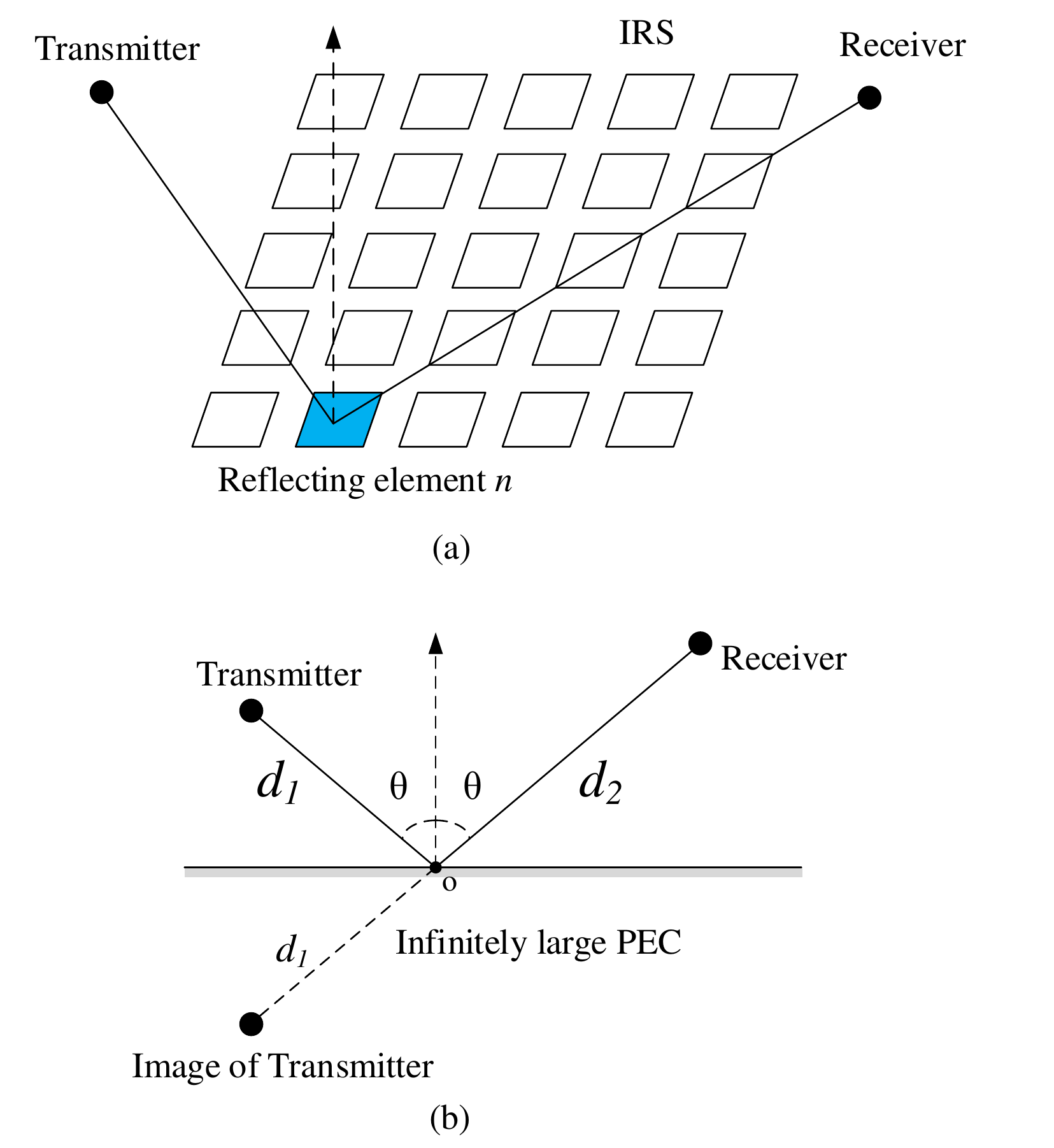}
\caption{Illustration for the path loss models of IRS-reflected channel. } \vspace{-0.0cm}\label{system:pathloss}
\end{figure*}

Note that the channel coefficients in $\h_r^H $ and $\g$ generally depend on  distance-related path loss, large-scale shadowing, and small-scale multipath fading.  In particular, the path loss of IRS-reflected channel captures its average power and is thus essential to the link budget analysis and performance evaluation of IRS-aided communications. Without loss of generality, consider IRS element $n$, which is assumed to be located sufficiently far from both the transmitter and receiver, with the distances from them given by $d_{1,n}$ and $d_{2, n}$, respectively, as shown in Fig.~\ref{system:pathloss}~(a).
 Under the far-field prorogation condition, we can assume that $d_{1,n}=d_1$ and $d_{2,n}=d_2$, $\forall n$.   Then, it follows that
$\mathbb{E}(|h_{r,n}|^2) \varpropto c_1(\frac{d_1}{d_0})^{-a_1}$ and $\mathbb{E}(|g_n|^2) \varpropto c_2 (\frac{d_2}{d_0})^{-a_2}$, $\forall n$, where $c_1$($c_2$) denotes the corresponding  path loss at the reference distance $d_0$, while $a_1$($a_2$) denotes the corresponding path loss exponent with typical values from 2 (in free-space propagation) to 6 \cite{goldsmith2005wireless}. From \eqref{eq:7}, it then follows that the average received signal power via the reflection by IRS element $n$, denoted by $P_{r,n}$,  is inversely proportional to $d_1^{a_1}d_2^{a_2}$, i.e.,
\begin{align} \label{eq:product:dis}
P_{r,n} \varpropto \frac{1}{d^{a_1}_1d^{a_2}_2}.
\end{align}
In other words, the IRS reflected channel via element $n$ suffers from double path loss, which is thus referred to as the  \emph{product-distance path loss model}.
As such, a large number of IRS reflecting elements are needed in practice to compensate for the severe  power loss due to double attenuation, by jointly designing their reflection amplitudes and/or phases to achieve high passive beamforming gains, as will be detailed later in Section III.

\begin{remark}
\emph{ In Fig. \ref{system:pathloss} (b), the IRS is replaced by an infinitely large perfect electric conductor (PEC) (or metallic plate). Assuming free-space propagation and applying the image theory \cite{baker2003mathematical},  it can be shown  that the {signal power} received at the receiver  via the PEC's reflection, denoted by $P_r$, is inversely proportional to the square of the sum distance of the two-hop links, i.e.,
\begin{align}\label{eq:sum:dis}
P_r \varpropto \frac{1}{ (d_1+d_2)^2}.
\end{align}
This model is usually referred to as the \emph{sum-distance path loss model}. Intuitively, due to the reflection of the infinitely large PEC, the received signal at the receiver were as if from an equivalent transmitter located at the image point of the original transmitter as shown in Fig.~1~(b), with the same link distance $d_1+d_2$, which is also known as specular reflection.
Note that this model is valid for the free-space propagation with an  infinitely large PEC, but in general inapplicable to the IRS-reflected channel modeled from the element level as given in \eqref{eq:7}--\eqref{eq:product:dis}. In particular, it is inappropriate to apply the sum-distance model to the scenario with one or more finite-size tunable PECs and conclude that the received  signal power scales with the number of PECs by exploiting their multiplicative passive beamforming gains and at the same time following   the more favorable (as compared to the product-distance model)  sum-distance based path loss, even under the free-space propagation.}
\emph{More theoretical and/or experimental studies on this issue can be found in  \cite{garcia2019reconfigurable,tang2019wireless,ozdogan2019intelligent,ellingson2019path,di2020analytical}.}
\end{remark}

\subsection{IRS Architecture, Hardware, and Practical Constraints}
The highly controllable reflection of IRS can be practically achieved by leveraging the existing digitally reconfigurable/programmable metasurface \cite{cui2014coding}. Specifically, metasurface is a planar array composed of massive properly designed reflecting elements/meta-atoms whose electrical thickness is typically in the order of subwavelength of the signal of interest.
By designing their  geometry shape (e.g., square or split-ring), size/dimension, orientation, arrangement, and so on, desired signal response (e.g., reflection amplitude and/or phase shift) of each element/atom can be realized. However, in wireless communication, the channel is generally time-varying   due to the mobility of the transmitter/receiver as well as the surrounding objects, thus calling for real-time tunable response of IRS based on the channel variation.  To this end, IRS elements need to be manufactured with dynamically adjustable reflection coefficients and IRS is required to connect to the wireless network to learn the exterior communication environment to enable its real-time adaptive reflection.

\begin{figure*}[!t]
\centering
~~~~~~~~~\includegraphics[width=0.90\textwidth]{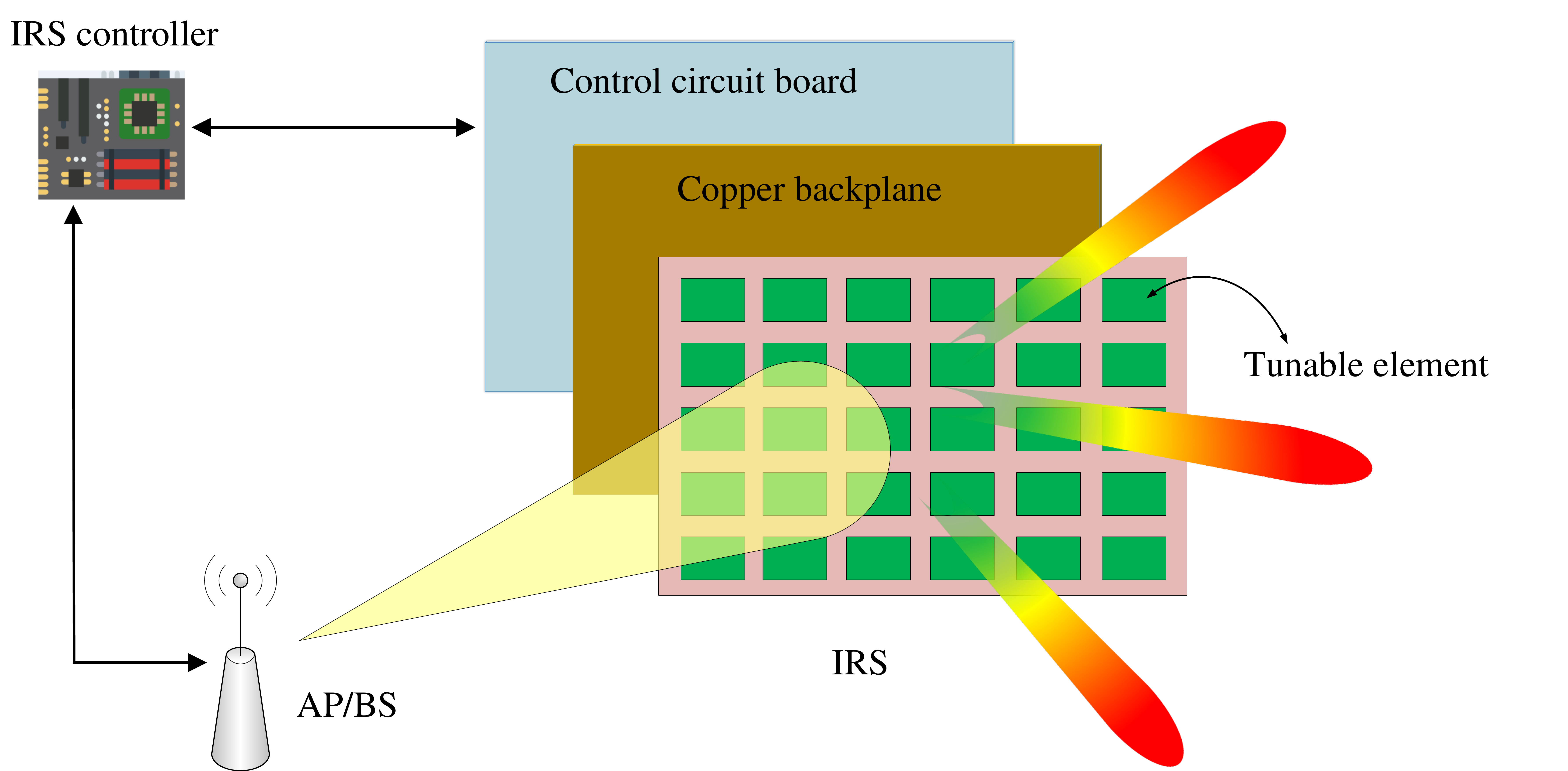}  
\caption{Architecture of IRS. } \vspace{-0.0cm}\label{system:hardware}
\end{figure*}

In Fig. \ref{system:hardware}, we illustrate one typical  architecture of  IRS, which consists of three layers and a smart controller. The first/outside  layer is composed of a large number of tunable/reconfigurable metallic patches printed on a dielectric substrate to directly manipulate incident signals. In the second/intermediate  layer, a copper plate is usually employed to minimize the signal  energy leakage during IRS's reflection. It is followed by the third/inside layer that is a control  circuit board responsible for exciting  the reflecting  elements as well as tuning their reflection amplitudes and/or phase-shifts in real time.
Moreover, the reflection adaptation is triggered and determined by a smart controller attached to each IRS, which can be implemented via field-programmable gate array (FPGA). The IRS controller also acts as a gateway to communicate  with other network components (e.g.,  BSs/APs and user terminals) through wired or wireless backhaul/control links. In practice, to enhance IRS's environmental learning capability,
dedicated sensors can also be deployed in the first layer, e.g., interlaced with the reflecting elements of the IRS, for sensing the surrounding radio signals of interest to facilitate the smart controller in designing the reflection coefficients, as will be discussed  in Section IV with more details.

\begin{figure*}[!t]
\centering
\includegraphics[width=0.9\textwidth]{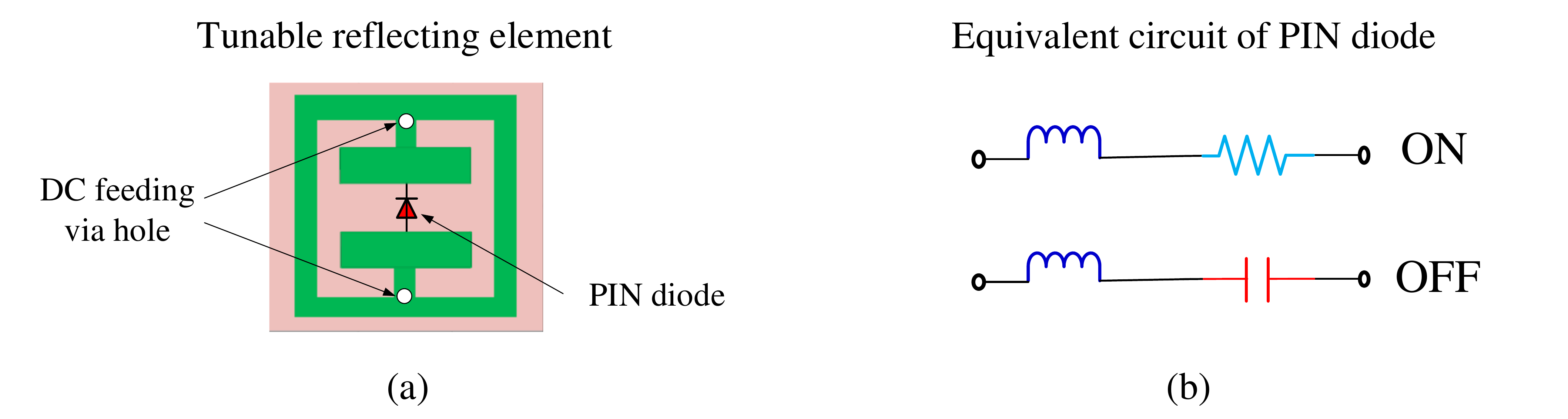}
\caption{An example of the tunable reflecting element based on PIN diode. } \vspace{-0.0cm}\label{hardware:element}
\end{figure*}

 To reconfigure IRS elements for highly controllable reflection, there are three main approaches proposed in the literature, namely, 1) mechanical actuation (via, e.g., mechanical rotation and translation), 2) functional materials (e.g., liquid crystal and  graphene), and 3) electronic devices (e.g.,  positive-intrinsic-negative (PIN) diodes, field-effect transistors (FETs), or micro-electromechanical system (MEMS) switches) \cite{nayeri2018reflectarray}. In particular, the third approach has been widely adopted in practical implementation  due its fast response time, low reflection loss as well as relatively low energy consumption and hardware cost.
Fig. \ref{hardware:element} shows an example design of the reflecting element and its equivalent circuit based on the  PIN  diode that is  loaded in the center of the element.  By applying different biasing voltages to the PIN diode via a direct-current (DC) feeding line, the PIN diode can be  switched to either ``ON'' or ``OFF'' state, which enables the element to result in  a phase-shift difference of $\pi$ in the incident signal.   According to \cite{zhang2018space}, the PIN diode switching frequency can be up to 5 megahertz (MHz), which corresponds to the switching time of 0.2 microsecond ($\mu$s). This is much smaller than the typical channel coherence time that is on the order of millisecond (ms) and thus well suited for mobile applications with time-varying channels.
Besides tuning the phase shift, additional control of the reflection amplitude of each IRS element provides more flexibility in reshaping the reflected signal to achieve various communication objectives effectively. This also offers a flexible way to trade-off between the hardware cost and reflection performance in practice, as amplitude control is generally of lower cost to implement as compared to phase control. There are various ways to achieve amplitude adjustment for IRS.
 One common way is by adjusting the load resistance/impedance in each element   \cite{yang2017design}. For example, by changing the resistance of each element, a certain portion of the incident signal energy is dissipated as heat, thus achieving a dynamic range of the reflection amplitude in $[0, 1]$. This is quite similar to the operation  of a passive radio frequency identification (RFID) tag that controls the strength of the reflected signal power by varying its load impedance for data modulation. 
In practice, it is desirable to have independent control of the amplitude and phase shift of each IRS element for optimizing the reflection design, which, however, requires more  sophisticated hardware designs (e.g., the multilayer surface design \cite{nayeri2018reflectarray}) than the above mentioned for their separate control only.

Ideally, the IRS reflection amplitude and phase shift per element can be independently and continuously tuned, thus yielding their following respective feasible sets:
\begin{align}
\mathcal{F}_{\beta}  & = \Big\{  \beta      \Big|     \beta \in [0, 1]      \Big\}, \label{eq:apd}  \\
\mathcal{F}_{\theta}  & = \Big\{  \theta      \Big|     \theta \in [0, 2\pi) \Big\}.   \label{eq:phz}
\end{align}
Alternatively, by defining   $\psi \triangleq \beta e^{\jmath \theta}$ as the reflection coefficient per element, we obtain the following equivalent feasible set as
\begin{align}
\mathcal{F} &= \Big\{  \psi \Big| |\psi| \leq 1  \Big\}.
\end{align}
Note that the above ideal feasible sets for IRS reflecting elements usually lead to theoretical performance limits of IRS-aided wireless communication systems under  practical reflection models, which are introduced as follows.

\subsubsection{Discrete reflection amplitude and phase shift}
 While tuning the reflection coefficient continuously  is beneficial for optimizing  the communication performance, it is practically difficult to implement  since higher-resolution reflecting elements require not only increased cost but also more complex hardware design. For example, to enable $8$ levels of phase shifts per IRS element,   at least $\log_28=3$ PIN diodes are required.
 This renders  the element design more challenging due to its limited size, and also requires more controlling pins at the IRS controller to control the required  PIN diodes. Although a single varactor diode can be used to achieve multi-level phase shifts, it requires a wide range of biasing voltages, and thus is more costly to implement.   As such,  for practical IRSs that usually have a large number of reflecting elements, it is more cost-effective  to implement only discrete and finite  amplitude/phase-shift levels that require only a small number of control bits for each element, e.g., two-level (reflecting or absorbing) amplitude control, and/or two-level ($0$ or $\pi$) phase-shift control \cite{cui2014coding,wu2018IRS_discrete}.

In general, let  $b_{\beta}$ and $b_{\theta}$ denote the number of bits for controlling  the corresponding number of reflection amplitude and phase-shift levels denoted by  $K_{\beta}$ and $K_{\theta}$, respectively, where  $K_{\beta}=2^{b_{\beta}}$ and $K_{\theta}=2^{b_{\theta}}$. Then  the sets of discrete reflection amplitudes and phase shifts at each element  of IRS can be respectively expressed as
\begin{align}
\mathcal{F}'_{\beta} &= \{ \bar{\beta}_1, \cdots,  \bar{\beta}_{K_{\beta}}  \},  \\
 \mathcal{F}'_{\theta} &= \{\bar \theta_1, \cdots, \bar \theta_{K_{\theta}} \},
 \end{align}
 where $0 \leq \bar{\beta}_m< \bar{\beta}_{m'}\leq 1$ for $1 \leq m<{m'}\leq K_{\beta}$ and $0\leq \bar \theta_l< \bar \theta_{l'}<2\pi$ for $1 \leq l<l'\leq K_{\theta}$.
For example, by assuming that the discrete amplitude and phase-shift values are obtained by  uniformly quantizing the interval $[0, 1]$ and $ [0, 2\pi)$, respectively,  we have
\begin{align}
\mathcal{F}'_{\beta} &= \{0, \Delta\beta, \cdots, (K_{\beta}-1) \Delta\beta  \},  \label{eq:14}\\
 \mathcal{F}'_{\theta} &= \{0,\Delta\theta, \cdots, ( {K_{\theta}}-1)\Delta\theta \},  \label{eq:15}
 \end{align}
 where $\Delta\beta= 1/(K_{\beta}-1)$ and $\Delta\theta= 2\pi/K_{\theta}$.   Compared to the ideal continuous reflection amplitude/phase-shift models in \eqref{eq:apd} and \eqref{eq:phz},  their quantized versions in \eqref{eq:14} and \eqref{eq:15}   inevitably result in coarser reflected signal amplitude/phase control and thus degraded communication performance. Furthermore, they also bring new challenges  in optimizing  the IRS reflecting elements' amplitudes/phase-shifts in practice due to the resultant discrete optimization variables that are in general more difficult to handle than their continuous counterparts,  as will be discussed in Section III with more details.

{In practice, to further reduce the  hardware cost and design complexity, only discrete phase-shift control or discrete amplitude control may be implemented, thus leading to the following two special cases of the above discrete models, namely,
\begin{itemize}
  \item IRS with discrete phase-shift control only,  where for each reflecting element, only the phase shift can be tuned while the reflection amplitude is set to its maximum value of one, i.e.,  $\mathcal{F}'_{\beta} =  \left\{  \beta      \big|     \beta =1    \right\}$ and $ \mathcal{F}'_{\theta}  = \{0,\Delta\theta, \cdots, ( {K_{\theta}}-1)\Delta\theta \}$;
  \item IRS with discrete amplitude control only,  where for each reflecting element, only the reflection amplitude can be tuned while the phase shift is set to be a constant (say, zero without loss of generality), i.e.,   $\mathcal{F}'_{\beta} = \{0, \Delta\beta, \cdots, (K_{\beta}-1) \Delta\beta  \}$ and $\mathcal{F}'_{\theta}  = \left\{  \theta      \big|     \theta =0 \right\}$.
\end{itemize}
 Generally speaking, phase-shift control (or phase beamforming)  is of higher cost to implement as compared to amplitude control (or amplitude beamforming) for IRS, while the former can achieve better passive beamforming performance than the latter given the same number of control bits/discrete levels per reflecting element.    A detailed study on their performance comparison will be given in Section III.}

\begin{figure*}[!t]
\centering
\includegraphics[width=0.55\textwidth]{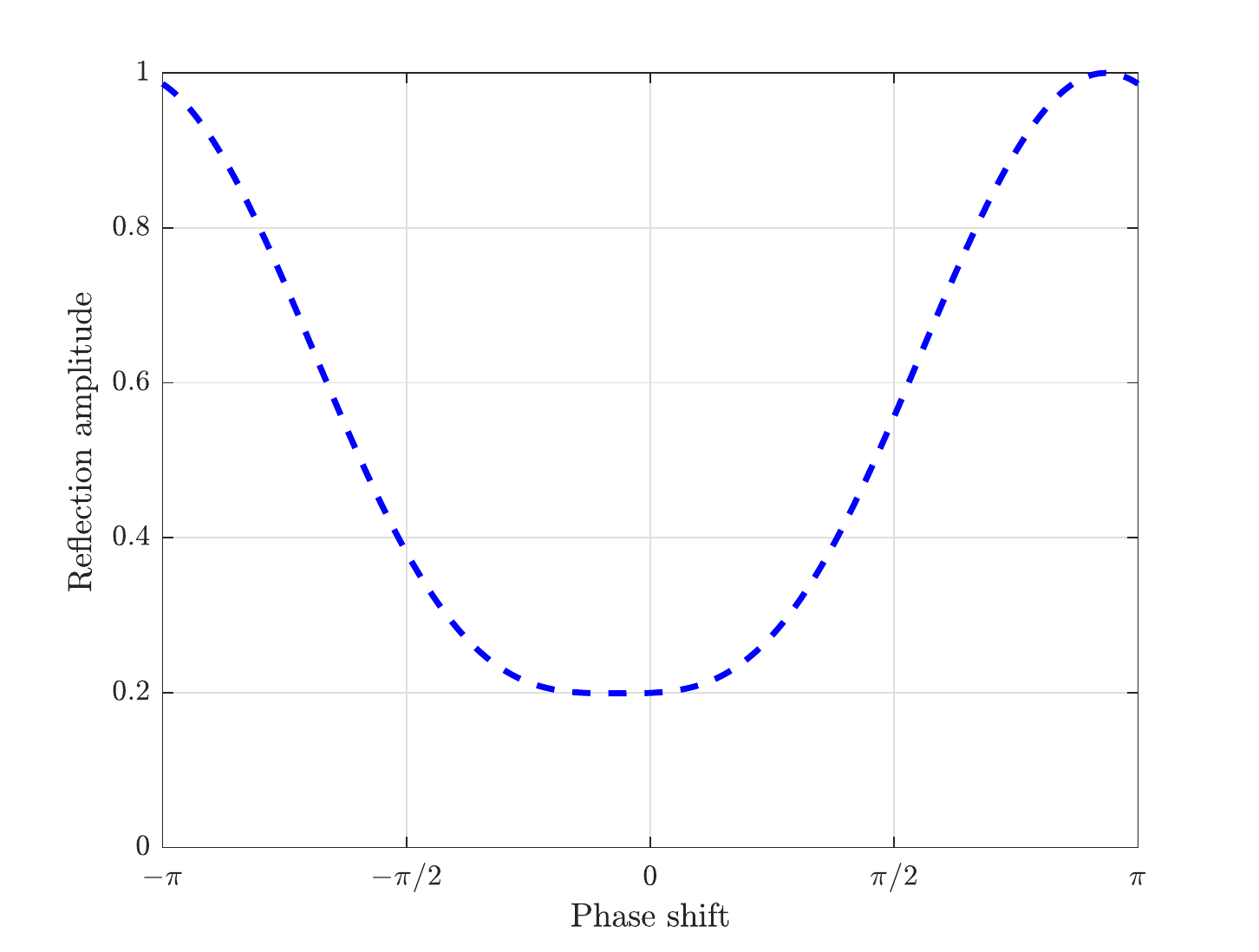}  
\caption{Reflection amplitude versus phase shift for the practical IRS reflecting element \cite{abeywickrama2019ICCintelligent,abeywickrama2019intelligent}. } \vspace{-0.0cm}\label{amplitude:depend:phaseshift}
\end{figure*}
\subsubsection{Coupled reflection amplitude and phase shift}
Although independent control of the reflection amplitude and phase shift simultaneously provides the maximum design flexibility, it imposes challenges in the element design.  Recently, a practical reflection model for IRS was proposed in \cite{abeywickrama2019ICCintelligent,abeywickrama2019intelligent} by modelling  each reflecting element as a resonant circuit with certain inductance,  capacitance, and resistance.  Based on this model, it was revealed  that the amplitude response of the reflecting element is in general non-linearly coupled with  its phase shift, which thus are not independently adjustable.  Specifically, as shown in Fig. \ref{amplitude:depend:phaseshift}, the reflection amplitude typically attains  its minimum value at the zero phase shift, but monotonically increases and asymptotically approaches the maximum value of one as the phase shift tends to  $\pi$ or $-\pi$. This is due to the fact that in each reflecting element, any phase shift is achieved by tuning its effective capacitance/resistance, which inevitably changes the reflection amplitude. To be specific, when the phase shift approaches zero, the reflective currents (also referred to as image currents)  are in-phase with the element currents, and thus the electric field as well as the current flow in the element are both enhanced, which results in maximum energy dissipation and the lowest reflection amplitude. In contrast, when the phase shift is around $\pi$ or $-\pi$, the reflective currents are out-of-phase with the element currents, and thus the electric field as well as the current flow in the element both diminish, thus leading to minimum energy loss and the highest reflection amplitude.
 This analytical reflection model of IRS is also consistent with the experimental results reported in e.g., \cite{zhu2013active}. The reflection amplitude and phase-shift  coupling has a great impact on the optimal reflection coefficient design of IRS reflecting elements, as it needs to strike an optimal balance between the signal amplitude and phase reflected by each element such that the reflected signals by all IRS elements are combined at the receiver with maximum power or achieving maximum signal-to-noise ratio (SNR).

In addition to the ideal versus practical IRS reflection models discussed in the above, IRS is theoretically passive but consumes energy in practice as well.   Although IRS only reflects signal without power amplification, it still needs a power supply to sustain the operation of reconfiguring the reflecting elements as well as its smart controller.  For example, if PIN diodes are used to tune the phase shifts of IRS reflecting elements, when the PIN diode is OFF, the element does not consume any energy, whereas when the PIN diode is ON, it consumes a certain amount of power, e.g., about 0.33 milliwatt (mW)  in \cite{tang2019wireless} and 50 $\mu$W in \cite{kaina2014shaping}. Nevertheless, compared to the power consumption of  active antennas (e.g., massive MIMO and multi-antenna relay), such power consumption is much lower and thus can be practically ignored for comparison. On the other hand, the power consumption of the smart controller will depend on the controller's circuit implementation (e.g., FPGA) and communication module used. For example, the power consumption of an IRS controller with 256 reflecting elements is only about 0.72 W \cite{tang2019wireless}, which is significantly lower than that of the active BS/relay in practice. Therefore,  IRS generally consumes substantially  lower energy consumption than the existing active relays in the wireless network. Interested readers may refer to \cite{kaina2014shaping,tang2019wireless,nayeri2018reflectarray} for more  details on IRS energy consumption modelling as well as the impact of energy consumption on its performance comparison with conventional active relays.

\subsection{Other Related Work and Future Direction}
IRS reflection modelling and practical hardware design are critical to its reflection optimization and achievable performance gains in IRS-aided wireless systems. However, their research is still in an early stage and needs to address  many interesting and important problems that are open and thus worth further investigating. In the following, we list some of these promising topics to motivate future work.

One crucial assumption for the linear channel model of IRS given in \eqref{eq:8} is the ignorance of any reflected signal coupling among  neighbouring IRS elements. In practice, increasing the number of reflecting elements given the same IRS size is generally helpful in achieving more fine-grained passive beamforming and thus enhanced performance. This, however, will reduce the element spacing and may render the mutual coupling more severe and thus no more negligible,  as nearby reflecting elements will interact with each other through their circuit coupling and thus result in coupled reflection coefficients. As such, the linear channel model of IRS given in \eqref{eq:8} will become inaccurate and more complex non-linear modelling may be needed to characterize the mutual coupling effect, which is an interesting problem to address in future work. Furthermore, it is also worth developing efficient decoupling/isolation techniques to minimize the effect of mutual coupling by e.g., using defected ground structures, parasitic scatterers, and neutralization lines \cite{zhang2019mutual,nayeri2018reflectarray}.

Another assumption adopted in the IRS channel model given in \eqref{eq:8} is that the reflection coefficient is insensitive to the incident angle of the signal  impinging on IRS. However,  recent experimental results  in \cite{tang2019wireless} revealed that the IRS reflection coefficient, particularly the phase shift, can be highly sensitive to the incident angle. Such an angle-dependent reflection model poses a new challenge in IRS reflection optimization, especially in a multi-path propagation environment since the reflection response of IRS depends on the angle-of-arrival (AoA) of each signal path to the IRS. More importantly, the widely adopted assumption of  channel reciprocity in typical time-division duplex (TDD) based wireless systems may become no longer valid when IRS is involved, which implies that the uplink training based channel estimation is not applicable to the donwlink communication. Thus, further studies on this effect are worth pursuing.


Last but not least, the IRS channel model in \eqref{eq:8}  has assumed that the phase shift of each reflecting element is constant over the signal bandwidth, $B$. Although this assumption is valid for the narrow-band system with $B\ll f_c$, it may become inaccurate for the broadband system when $B$ is much larger and thus becomes  comparable with $f_c$. As shown in \eqref{reflection}, the phase shift by IRS is resulted by delaying the input signal with a certain amount of time, which will cause a linear phase drift that increases with the signal frequency deviating from $f_c$. Such a non-uniform phase shift over frequency can cause undesired phase errors for multi-carrier (such as orthogonal frequency-division multiplexing (OFDM)) modulated signal, and thus should  be compensated by either hardware or signal processing, which needs further investigation. In addition, the effects of other IRS hardware imperfections such as frequency/time offset, phase noise, etc. on its reflection design and achievable performance remain largely unknown, which call for further investigation in future work.

\section{IRS  Reflection Optimization}\label{Sec:passivebf}
\begin{figure}[!t]
\centering
\includegraphics[width=0.7\textwidth]{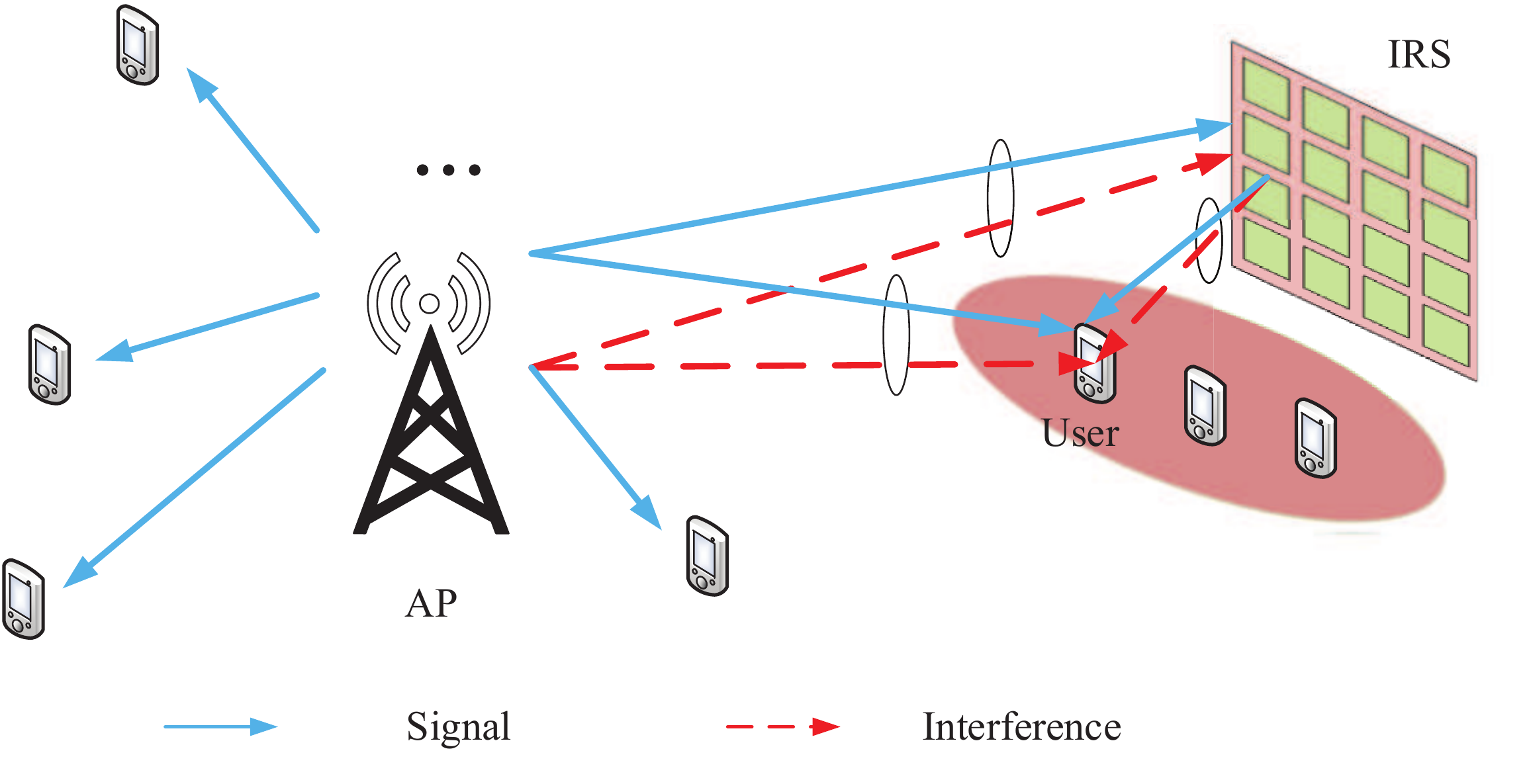}
\caption{IRS-aided  multi-user communication  system. } \label{P2Psystem:model}
\end{figure}

In this section, we study  the passive reflection optimization for IRS-aided wireless communication under various system setups, namely, from single-user to multi-user, from single-antenna to multi-antenna, from narrow-band to broadband, as well as from single-cell to multi-cell communications. For the purpose of exposition, we assume perfect knowledge of all channels considered  in this section, while the issue of IRS channel estimation will be addressed in Section IV. In Fig. \ref{P2Psystem:model}, we show a single-cell  IRS-aided multi-user system, where an IRS comprising $N$ reflecting elements is deployed to assist in the communications from the AP (or BS)  to a set of $K$ users in the downlink. The users are located arbitrarily in the cell, so some of them may not be in the vicinity of the IRS in general. It is assumed that the AP is equipped with $M_t$ antennas and each user is equipped with $M_r$ antennas without loss of generality.

\subsection{IRS-aided SISO System: Passive Beamforming Basics and Power Scaling Order}
First, we  consider the single-user single-input-single-output (SISO) case, i.e., $K=1$ and $M_t=M_r=1$, and the narrow-band system with flat-fading channels.  The baseband equivalent channels  from the AP to the IRS, from the IRS to the user, and from the AP to the user are denoted by $\bm{g}\in \mathbb{C}^{N\times 1}$, $\bm{h}^H_{r}\in \mathbb{C}^{1\times N}$, and ${h}^*_d \in \mathbb{C}$, respectively.  
Based on the IRS reflection model in (6), the signal received at the user is expressed as
\begin{align}\label{SectionII:channel}
y= ( \bm{h}^H_r\ttheta \bm{g} +  {h}^*_d)\sqrt{P_t}x + z,
\end{align}
where $x$ is the information signal modeled by an independent and identically distributed (i.i.d.) random variable with zero mean and unit variance, $P_t$ is the transmit power at the AP, and $z$ denotes additive white Gaussian noise (AWGN) at the user receiver modeled as CSCG  with zero mean and variance $\sigma^2$. Accordingly, the user receive SNR is written as
\begin{align}\label{SectionII:receivedpower}
\gamma = \frac{{P_t| \bm{h}^H_r\ttheta \bm{g}+  {h}^*_d|^2}}{\sigma^2} =\frac{{P_t| \sum_{n=1}^{N}{h}^*_{r,n}\beta_n e^{\jmath \theta_n}  {g}_n+  {h}^*_d|^2}}{\sigma^2} .
\end{align}
Thus,  the maximum achievable rate  in bits  per  second per Hertz (bps/Hz) of the considered IRS-aided point-to-point SISO link is given by $r=\log_2(1+\gamma)$.   

We aim to maximize the achievable rate  $r$ (or SNR $\gamma$ equivalently)  by optimizing the passive reflect  beamforming at the IRS.  By ignoring the constant terms and assuming continuous reflection amplitude and phase shift,  the optimization problem is formulated as
\begin{align}
\text{(P1)}: ~~\max_{\bm{\theta}, \bm{\beta}} ~~~&\left| \sum_{n=1}^{N}{h}^*_{r,n}   {g}_n \beta_n e^{\jmath \theta_n}+  {h}^*_d\right|^2  \label{eq:obj}\\
\mathrm{s.t.}~~~~&0\leq \theta_n < 2\pi,  n=1,\cdots, N, \label{phase:constraints} \\
~~~~&0\leq \beta_n \leq 1, n=1,\cdots, N, \label{apd:constraints}
\end{align}
where $\bm{\theta}= [\theta_1, \cdots, \theta_N]^T$ and $\bm{\beta}= [\beta_1, \cdots, \beta_N]^T$. For given $\bm{\beta}\geq \0$, the optimal phase-shift solutions to (P1) are given by \cite{wu2018IRS,JR:wu2018IRS}
\begin{align}\label{eq:phase:solution}
\theta^{\star}_n = {\text{mod}}[ \zeta - (\phi_n +\psi_n ),2\pi ], n=1,\cdots, N,
\end{align}
 where  $\phi_n$, $\psi_n$, and  $\zeta$ are the phases of ${h}^*_{r,n}$,  ${g}_n$, and ${h}^*_d$, respectively. Note that the solutions in \eqref{eq:phase:solution} do not depend on the values in $\bm{\beta}$, which implies that they are indeed optimal to (P1). This is expected since the optimal phase shifts should align all the signals reflected by the IRS regardless of their strength with the signal coming directly from the AP to achieve coherent combining and thus maximize the received signal power at the user. Besides, if the AP-user direct link is negligible compared to the IRS-reflected link (e.g., when the former is severely blocked) and thus can be ignored, i.e., ${h}^*_d= 0$,  the optimal solutions in \eqref{eq:phase:solution} can be multiplied with any arbitrary (common) phase shift without changing the optimal value of (P1). Thus,  we can set $\zeta$ equal to zero in \eqref{eq:phase:solution} without loss of optimality.   In general, by substituting \eqref{eq:phase:solution} into (P1), this problem  is reduced to
 \begin{align}
\text{(P1.1)}: ~~\max_{ \bm{\beta}} ~~~&\left| \sum_{n=1}^{N}   |{h}_{r,n} |   |{g}_n | \beta_n  +   |{h}_d | \right|^2  \label{P1.1eq:obj}\\
\mathrm{s.t.}~~~~&0\leq \beta_n \leq 1, n=1,\cdots, N. \label{apd2:constraints}
\end{align}
 It follows easily from \eqref{P1.1eq:obj}  that the optimal reflection amplitude solutions are  given by $\beta^{\star}_n =1, \forall n$, since maximizing the reflection amplitudes helps achieve the maximum  user receive power due to coherent combining. Another interesting observation from  (P1) is that the optimal IRS reflection design depends on the cascaded  channels via the IRS only, i.e.,  $\text{diag}(\bm{h}^H_r)\bm{g}$,  without the need of knowing the individual channels, $\bm{h}^H_r$ and $\bm{g}$. It is worth mentioning that this result usually holds for IRS-aided communication systems and can be leveraged to greatly simplify the IRS channel estimation design as will be discussed  in Section IV.


\subsubsection{Receive Power Scaling with $N$} A fundamental question regarding IRS passive beamforming performance is how the maximum receive  SNR grows with the number of IRS reflecting elements, $N$, when $N$ becomes asymptotically large. To address this problem,  we focus on the IRS-reflected link only by assuming that the AP-user direct channel ${h}^*_d$ equals to zero, since the former becomes more dominant than the latter as $N\rightarrow \infty$. Based on  \eqref{eq:phase:solution}, the user receive power, denoted by $P_r$, is given by $P_r =P_t \big|\sum_{n=1}^N |h_{r,n}||g_n| \big|^2$.  Assume i.i.d. Rayleigh fading with average power $\varrho^2_h$ and $\varrho^2_g$ for each entry in $\bm{h}^H_{r}$ and $\bm{g}$, respectively. It was shown in \cite{JR:wu2018IRS} that as $N\rightarrow \infty$, the asymptotic  receive power is approximately given by
\begin{align}\label{eq:square:power:gain}
P_r \approx  N^2\frac{P_t\pi^2\varrho^2_h\varrho^2_g}{16}.
\end{align}
This result suggests that with sufficiently large $N$, the user receive power increases {\it quadratically} with $N$, i.e., in the order of $\mathcal{O}(N^2)$  \cite{JR:wu2018IRS}. Alternatively,  we can scale down the transmit power at the AP by a factor of $1/N^2$ without compromising the receive SNR. This is because the IRS not only achieves the reflect beamforming gain of $\mathcal{O}(N)$ in the IRS-user link, but also captures an additional  aperture gain of $\mathcal{O}(N)$  in the AP-IRS link.

\begin{figure}[h]
\centering
\includegraphics[width=0.4\textwidth]{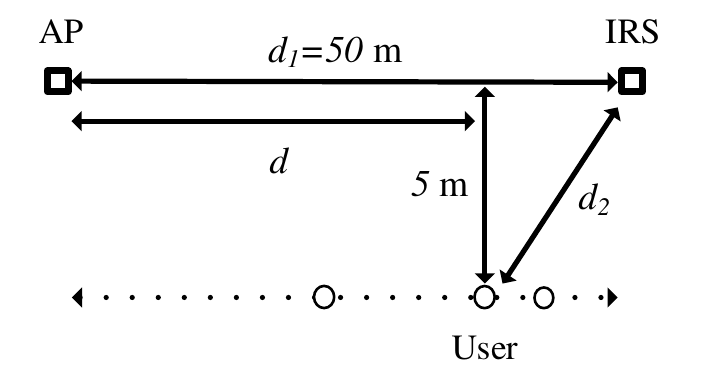}  
\caption{Simulation setup of the IRS-aided single-user  system.}
 \label{simulation:setup}
\end{figure}

\begin{figure}[h]
\centering
\includegraphics[width=0.65\textwidth]{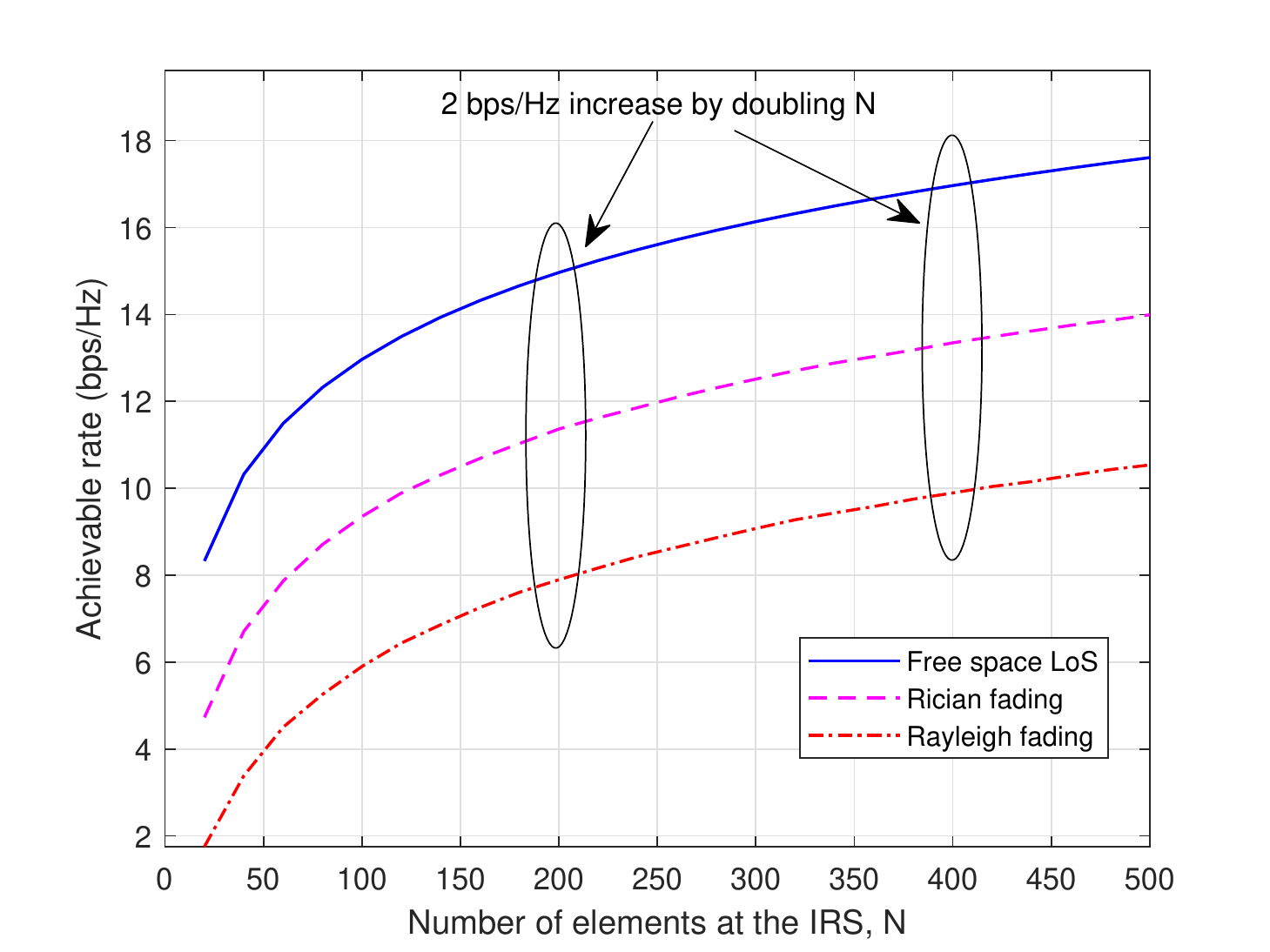}  
\caption{Achievable rate versus $N$ for IRS under different channel models.}
 \label{simulation:squared:power:gain}
\end{figure}

To illustrate the receive power scaling order, we consider a setup shown in Fig. \ref{simulation:setup}. The path loss model for both AP-IRS and IRS-user links is set as   $c_0d_0^{-a}$ where $d_0$ is the corresponding link distance in meter (m) and $a$ denotes the path loss exponent. Other parameters are set as $c_0= -30$ dB, $P_t=50$ mW and $\sigma^2=-90$ dBm.
In Fig. \ref{simulation:squared:power:gain}, we plot the achievable rate versus the number of IRS reflecting elements, $N$, with $d=d_1=50$ m. Specifically, we consider three different channel models, namely, free-space LoS channel, Rician fading channel with a Rician factor of $3$, and i.i.d. Rayleigh fading channel (a special case of the Rician fading channel without the LoS component or with zero Rician factor). The path loss exponents of the three channel models are set as $2$, $2.4$ and $2.8$, respectively.  From Fig. \ref{simulation:squared:power:gain},  it is observed that for all considered  channel models,  the achievable rate of the IRS-aided SISO system increases about $2$ bps/Hz by doubling $N$ when $N$ is sufficiently large, e.g., from $N=200$ to $N=400$, which validates the asymptotic power scaling order in \eqref{eq:square:power:gain}.

In practice, it is useful to estimate the coverage range of IRS for a given average  receive power (or SNR) requirement of the user  with a fixed transmit power at the AP.
By substituting $\varrho^2_h=c_0d_{2}^{-a}$ (with $d_2$ shown in Fig. \ref{simulation:setup})  into \eqref{eq:square:power:gain}, it follows that
\begin{align}\label{eq:pow3}
N \thickapprox d_{2}^{\frac{a}{2}}c_1,
\end{align}
where $c_1 = \sqrt{\frac{P_r}{P_tc_0}} \frac{4}{\varrho_g}$. From \eqref{eq:pow3}, it follows that as the desired IRS coverage range increases,  the number of its reflecting elements also needs to be increased to keep the same  average receive power without inuring more transmit power at the AP, e.g., by increasing  $N$ linearly with $d_{2}$  if $a=2$.

\subsubsection{Comparison with M-MIMO  and MIMO Relay}
It is worth comparing the performance of IRS with $N$ passive reflecting elements to that of M-MIMO or MIMO relay with the same number $(N)$ of active antennas. For M-MIMO, it is well known that a transmit/receive beamforming gain  of order $\OO(N)$ can be achieved \cite{Hien2013,hoydis2013massive}. {While it was shown in \cite{JR:wu2018IRS} that even with perfect self-interference cancellation, the receive SNR by using the FD AF-based MIMO relay also  increases linearly with $N$ when $N$ is asymptotically large.  This is due to the processing noise effect at the AF relay. To be specific, although the user receive signal power in the FD AF MIMO relay system scales in the same order of $N^2$  as that of the IRS-aided system, the forwarded relay noise power at the receiver scales linearly with $N$  in contrast to the noise-free signal reflection by the IRS, thus resulting in a lower receive SNR gain order of $\OO(N)$. Last, it is worth mentioning that for the HD AF MIMO relay system, its receive SNR scaling order with $N$ can be shown to be identical to  that of its  FD counterpart.}

\begin{figure}[h]
\centering
\includegraphics[width=0.65\textwidth]{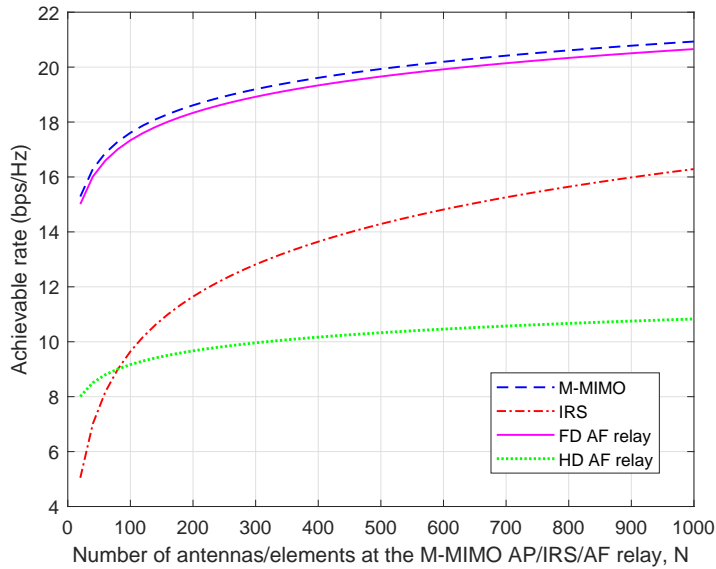}  
\caption{{Performance comparison  of M-MIMO, IRS and FD/HD AF-based MIMO relay.} }
 \label{simulation:AFrelay}
\end{figure}

In Fig. \ref{simulation:AFrelay}, we compare the achievable rates of the aforementioned  three technologies versus $N$. Rayleigh fading channel with the path loss exponent of 2.4 is assumed  for both AP-IRS and IRS-user links (with the IRS replaced by a multi-antenna relay in the MIMO relay case), and other parameters are set the same as for Fig. \ref{simulation:squared:power:gain}.  It is observed that when $N$ is small, IRS performs the worst due to the severe product-distance  path loss (see (7) in Section II) but  insufficient passive beamforming gain. However,  with the increasing number of reflecting elements, the rate gaps between the  IRS-aided system and  the M-MIMO and FD/HD MIMO relay-aided systems become smaller and IRS even outperforms the HD MIMO relay  thanks to its larger beamforming gain (i.e., $\OO(N^2)$ versus $\OO(N)$) when $N$ is large as well as its more spectral-efficient FD operation. Furthermore, since IRS's reflecting elements are passive and do not need any  transmit RF chains, their  cost is much lower as compared to that of active antennas for the M-MIMO/MIMO relay. Therefore, it may be unfair to compare their achievable rates with the same number of passive/active elements as in Fig. \ref{simulation:AFrelay}.  Instead, with a given hardware cost,  IRS can potentially offer significantly  more reflecting elements as compared to affordable antennas in  the case of M-MIMO/MIMO relay and thus achieve  more comparable or even superior performance against them in practice.

\subsection{IRS-aided  MISO System:  Joint Active and Passive Beamforming}
Next, we consider the  single-user system but  with multiple antennas at the AP, i.e., $M_t>1$, where the downlink/uplink communication becomes the multiple-input-single-output (MISO) and single-input-multiple-output (SIMO) system, respectively. In this case, the active AP transmit/receive beamforming needs to be optimized jointly with the IRS passive beamforming. For brevity,  we consider only the MISO case in this subsection while the results  given are applicable to the SIMO case in the reverse user-AP link as well.  Let $\bm{G}$ and $\bm{h}^H_d$ denote the channel matrix and channel vector from the AP to IRS and from the AP to user, respectively. Similar to (P1), the rate/SNR maximization problem in the MISO case is equivalently  formulated as\footnote{Note that  the reflection amplitudes of all IRS elements are set to one or their maximum value since with the optimal transmit beamforming vector $\bm{w}^{\star}$, the objective function of (P2) is reduced to that of  (P1), thus $\beta^{\star}_n =1, n=1,\cdots, N$, are optimal as shown in Section III-A.}
\begin{align}
\text{(P2)}: ~~\max_{\bm{w}, \bm{\theta}} ~~~&| (\bm{h}^H_r\ttheta \bm{G}+\bm{h}^H_d )\bm{w}|^2  \label{eq:obj}\\
\mathrm{s.t.}~~~~& \|\bm{w}\|^2\leq  P_t,  \\
& 0\leq \theta_n < 2\pi,  n=1,\cdots, N, \label{phase:constraints:multiantenna}
\end{align}
where $\bm{w}$ denotes the transmit beamforming vector at the AP and $P_t$ is the maximum transmit power.  Intuitively,  if the channel of the AP-user link, $\bm{h}^H_d$, is much stronger than that of the AP-IRS link, $\bm{G}$, it is preferable for the AP to beam toward the user directly, while in the opposite case, especially when the AP-user link is severely blocked, the AP should  steer its beamforming direction toward the IRS to leverage its reflected signal to serve the user. In general, the transmit beamforming  at the AP needs to be jointly designed with the phase shifts at the IRS based on all the AP-IRS, IRS-user, and AP-user channels  in order to maximize their cooperative gain.

 Unfortunately, (P2) is a non-convex optimization problem due to the non-concavity of its objective function with respect to $\bm{w}$ and $\bm{\theta}$.  Nevertheless, one can observe that  if we fix the transmit beamforming vector $\bm{w}$, (P2)  is reduced to its SISO counterpart, i.e., (P1), which thus motivates the alternating optimization (AO) approach to solve (P2) sub-optimally \cite{JR:wu2018IRS}. Specifically, with fixed $\bm{w}$, we obtain the optimal phase shifts by solving the resulted (P1); while with fixed  $\bm{\theta}$, the optimal transmit beamforming solution to (P2) is given by  the maximum-ratio transmission (MRT), i.e., $\bm{w}_{\rm MRT} = \sqrt{P_t} \frac{(\bm{h}^H_r\ttheta \bm{G}+\bm{h}^H_d  )^H}{\|\bm{h}^H_r\ttheta \bm{G} +\bm{h}^H_d \|}$. The above procedure is repeated until convergence is reached, which guarantees attaining at least a locally optimal solution to (P2) \cite{JR:wu2018IRS}.

An alternative  approach to solve (P2) is by first reducing it into  an optimization problem with phase shifts only. Specifically, by substituting $\bm{w}_{\rm MRT}$ to \eqref{eq:obj},  (P2) is reduced to  the following problem with $\bm{\theta}$ only,
\begin{align}\label{secIII:p2}
\text{(P3)}: ~~\max_{\bm{\theta}} ~~~&\|\bm{h}^H_r\ttheta \bm{G}+ \bm{h}^H_d\|^2\\
\mathrm{s.t.}~~~~& 0\leq \theta_n \leq 2\pi, n=1,\cdots, N.  \label{SecIII:phaseconstraint}
\end{align}
In contrast to (P1) for  the SISO case,  the  IRS phase shifts in (P3) for  the MISO case need to balance the channel powers from different transmit antennas, thus are more difficult to solve in general. In fact,  it is a non-deterministic polynomial-time hard (NP-hard)  problem with respect to $N$ and thus it is impossible to obtain the optimal solution to (P3) with polynomial complexity of $N$. To show this,  let $\bm{v} = [v_1, \cdots, v_N]^H$ where $v_n = e^{\jmath \theta_n}$, $\forall n$. Then, the constraints in \eqref{SecIII:phaseconstraint} are equivalent to the unit-modulus constraints:  $|v_n|^2=1, \forall n$. By applying the change of variables $\bm{h}^H_r\ttheta \bm{G} =\bm{v}^H\bm{\Phi} $ where $\bm{\Phi}=\text{diag}(\bm{h}^H_r)\bm{G} \in \mathbb{C}^{N \times M_t}$,  (P3) is equivalent to
\begin{align}\label{secIII:p4}
\text{(P4)}: ~~\max_{\bm{v}} ~~~&\bm{v}^H\bm{\Phi}\bm{\Phi}^H\bm{v} + \bm{v}^H\bm{\Phi}\bm{h}_d+\bm{h}^H_d\bm{\Phi}^H \bm{v}  + \|\bm{h}^H_d\|^2\\
\mathrm{s.t.}~~~~& |v_n|^2=1, n=1,\cdots, N. \label{P4:C9}
\end{align}
Note that (P4) is a non-convex quadratically constrained quadratic program (QCQP), which is generally NP-hard \cite{so2007approximating}.
As such, various methods have been proposed in the literature to attain  high-quality suboptimal solutions for (P4) including, e.g.,  1) semidefinite relaxation (SDR) with Guassian randomization \cite{JR:wu2018IRS}, and 2)  AO where each of the phase shifts is optimized in closed-form as in \cite{JR:wu2019discreteIRS} with the others being fixed in an iterative manner, which guarantees a locally optimal solution. In addition, the SDR approach with matrix rank  relaxation also provides a performance upper bound and thus is usually adopted for benchmarking  the performance of suboptimal solutions  \cite{JR:wu2018IRS}. Moreover, the above methods can be jointly applied, e.g.,
the solution obtained by scheme 1) can be further improved by adopting it as an initialization for scheme 2).

\begin{figure}[h]
\centering
\includegraphics[width=0.65\textwidth]{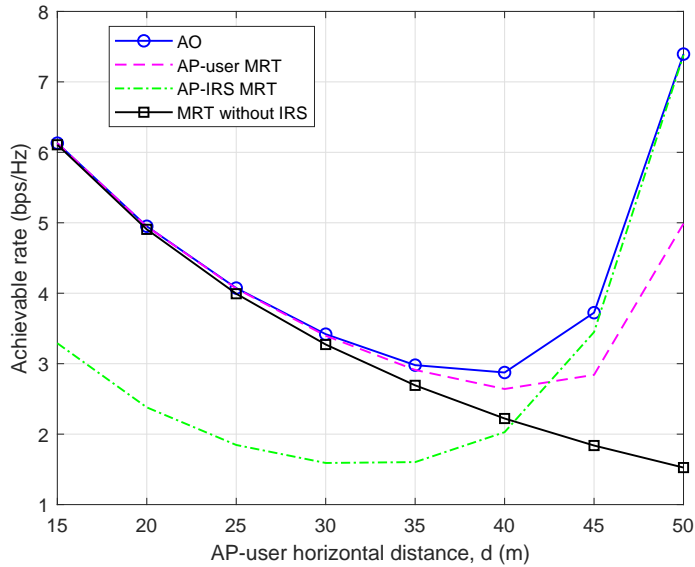}
\caption{Achievable rate of IRS-aided MISO system  versus AP-user horizontal distance, $d$. } \label{simulation:MISO:distance} \vspace{-4mm}
\end{figure}

In Fig. \ref{simulation:MISO:distance}, we show the achievable rate of the IRS-aided MISO system versus the AP-user horizontal distance, $d$, for the same setup as in Fig. \ref{simulation:setup} with $M_t=4$ and $N=40$.  Rayleigh fading channels are assumed  for both the AP-user and IRS-user links, while the free-space LoS channel is adopted for the AP-IRS link.  Four schemes are considered, namely, 1) the joint transmit and passive beamforming design by AO \cite{JR:wu2019discreteIRS}, 2) AP-user MRT by setting $\bm{w} = \sqrt{P_t}\frac{\bm{h}_d}{\|\bm{h}_d\|}$ to achieve MRT based on the AP-user direct channel, 3) AP-IRS MRT by setting $\bm{w} = \sqrt{P_t}\frac{\bm{g}}{\|\bm{g}\|}$  to achieve MRT based on the AP-IRS  rank-one channel, with $\bm{g}^H$ denoting any row in $\G$, and 4) MRT without IRS by setting  $\bm{w} = \sqrt{P_t}\frac{\bm{h}_d}{\|\bm{h}_d\|}$.   It is observed from Fig. \ref{simulation:MISO:distance} that the  AP-user MRT scheme performs close to the AO scheme when the user is nearer to the AP, while it results in  considerable rate loss when the user is nearer the IRS. This is expected since in the former case, the user received signal is dominated by the AP-user direct   link whereas the IRS-user link is dominant in the latter case. Moreover, it is observed that the AP-IRS MRT scheme behaves in the opposite way as compared to the AP-user MRT counterpart. Moreover, Fig. \ref{simulation:MISO:distance}  shows  that if the transmit beamforming is not designed properly based on all the channels, the achievable rate by using the IRS may be even worse than  the conventional MRT  without the IRS, e.g., in the case of  the AP-IRS MRT scheme for $d\leq 40$ m. This demonstrates that the joint active and passive beamforming design is essential to strike an optimal  balance between the transmission toward   the user directly and  that via the IRS reflection, to maximize  the received signal  power at the user.





\subsection{IRS-aided MIMO and OFDM Systems}
In the above, we have considered  narrow-band communication systems under frequency-flat fading channels, where the AP and/or user are equipped with a single antenna (i.e., $M_t=1$ and/or $M_r=1$). However, for MIMO systems with multiple antennas at both the AP and user (i.e., $M_t> 1$ and $M_r> 1$) and/or broadband OFDM systems subject to frequency-selective fading channels, the passive IRS reflection optimization problems need to cater to multi-antenna channels and/or multi-path channels with different delays, thus are more complicated as well as challenging to solve. In this subsection, we present the models for IRS-aided MIMO and OFDM systems, and the new methods used to optimize the IRS reflection for them.
\subsubsection{IRS-aided MIMO System}
Different from SISO/MISO/SIMO systems that can support the transmission of one single data stream only, the capacity of MIMO systems is generally achieved by parallel transmissions of multiple data streams at the same time, or \emph{spatial multiplexing}. Therefore, to achieve the capacity of an IRS-aided MIMO system, the IRS reflection needs to be jointly optimized with the transmit covariance matrix at the AP \cite{MIMO}. Specifically, let $\mv{H}_d\in \mathbb{C}^{M_r\times M_t}$ denote the direct channel from the AP to the user, $\mv{G}\in \mathbb{C}^{N\times M_t}$ denote the channel from the AP to the IRS, and $\mv{H}_r\in \mathbb{C}^{M_r\times N}$ denote the channel from the IRS to the user. The overall IRS-aided MIMO channel can  thus be expressed as $\tilde{\mv{H}}=\mv{H}_d+\mv{H}_r\mv{\Theta}\mv{G}$. Let $\mv{Q}\in \mathbb{C}^{M_t\times M_t}$ denote the transmit covariance matrix. By considering continuous phase shifts and maximum reflection amplitudes at all IRS elements (i.e., $\beta_n=1,\forall n$),\footnote{Note that the unit-amplitude assumption here is mainly motivated by the practical difficulty to jointly tune the amplitude and phase shift of each IRS reflecting element at the same time, while considering flexible amplitude design within $[0,1]$ may lead to improved capacity for IRS-aided MIMO systems in general \cite{MIMO}.} the capacity optimization problem is formulated as
\begin{align}
\mbox{(P5):}\quad \underset{\mv{\Theta},\mv{Q}}{\max}\quad &\log_2\det\left(\mv{I}_{M_r}+\frac{1}{\sigma^2}\tilde{\mv{H}}\mv{Q}\tilde{\mv{H}}^H\right)\\
\mathrm{s.t.}\quad & 0\leq \theta_n< 2\pi,\quad n=1,\cdots,N,\\
& \beta_n=1, \quad n=1,\cdots,N,\\
& \mathrm{tr}(\mv{Q})\leq P_t,\\
& \mv{Q}\succeq \mv{0}.
\end{align}

Note that (P5) is a non-convex optimization problem due to the non-concavity of the MIMO channel capacity with respect to $\mv{\Theta}$ and $\mv{Q}$ jointly; moreover, (P5) is more difficult to solve as compared to (P1) and (P2) for the SISO and MISO cases as the MIMO channel capacity is a more complicated log-determinant function. Nevertheless, an efficient AO-based algorithm was proposed in \cite{MIMO} to solve (P5), by iteratively optimizing one IRS element's phase shift or the transmit covariance matrix at each time, with  the other variables being fixed. Specifically, with given transmit covariance matrix and any $N-1$ IRS phase shifts, the optimal solution of the remaining phase shift to (P5) can be obtained in closed-form \cite{MIMO}; on the other hand, with given IRS reflection matrix $\mv{\Theta}$ and consequently the overall MIMO channel $\tilde{\mv{H}}$, the optimal solution of the transmit covariance matrix is given by the well-known eigenmode transmission with water-filling based spatial power allocation \cite{tse2005fundamentals,goldsmith2005wireless}. The above AO-based algorithm is guaranteed to converge to at least a locally optimal solution to (P5), with only polynomial complexity in terms of $N$, $M_t$, or $M_r$ \cite{MIMO}.

\begin{figure}[t]
	\centering
	\includegraphics[width=8cm]{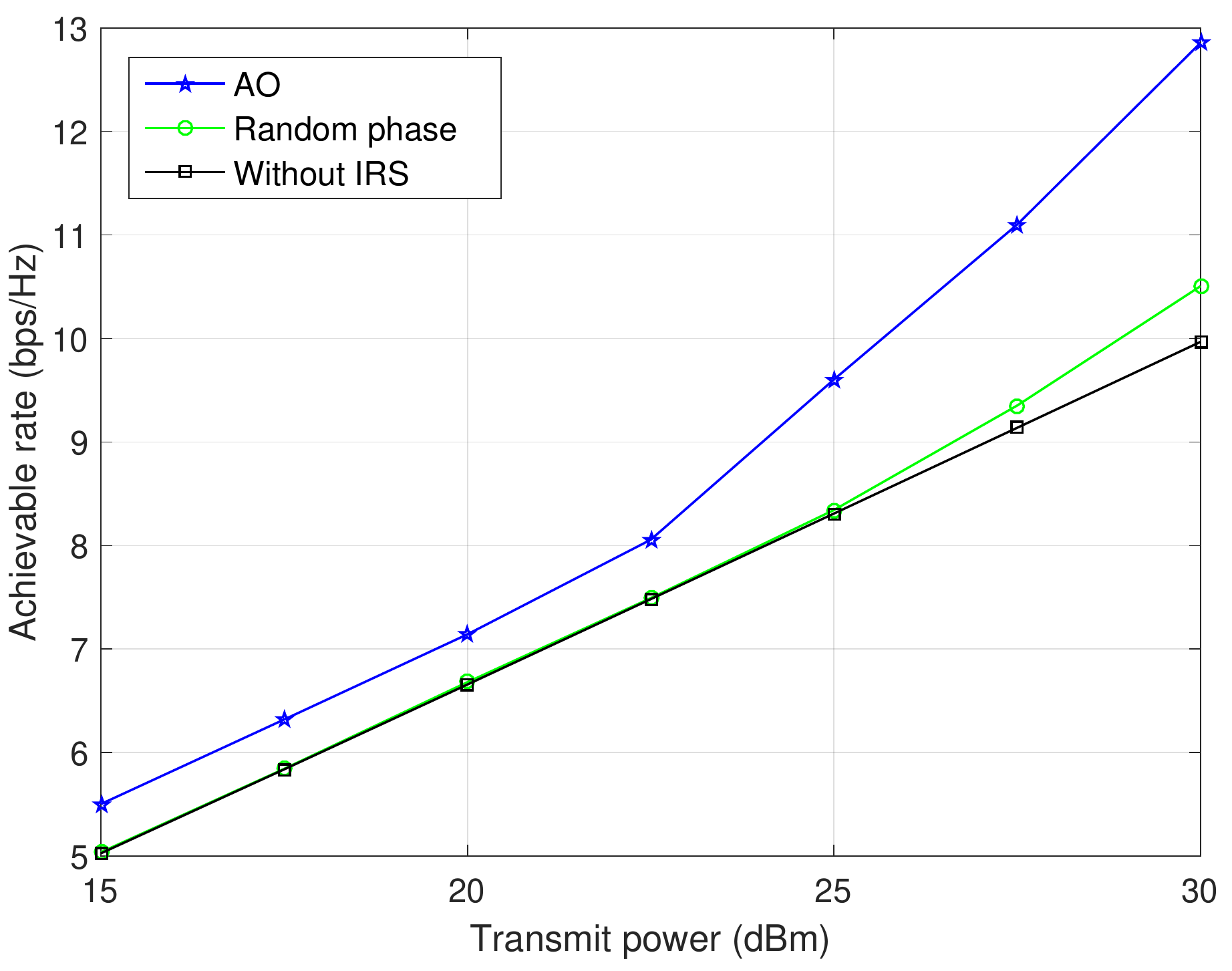}
	\vspace{-3mm}
	\caption{Achievable rate versus transmit power for IRS-aided MIMO system.}\label{MIMO}
	\vspace{-5mm}
\end{figure}

It is worth noting that by properly designing the IRS reflection coefficients, various key parameters of the IRS-enhanced MIMO channel can be significantly improved, including the channel total power, condition number, rank, etc \cite{MIMO}. Particularly, for practical scenarios where the direct channel between the AP and user is of low rank or even rank-one (e.g., LoS-dominant channel, key-hole channel), the AO-based algorithm results in a higher-rank MIMO channel with larger \emph{spatial multiplexing gain} as compared to the MIMO channel without IRS, thus substantially boosting the channel capacity in the high-SNR regime. For illustration, we show in Fig. \ref{MIMO} the achievable rate versus transmit power $P_t$ for an IRS-aided MIMO system with $M_t=M_r=4$ and $N=30$. The system setup is similar to that in Fig. \ref{simulation:setup}, but with $d_1=d=4000$ m and $d_2=1$ m. We assume that the AP-user direct channel follows the free-space LoS model with path loss exponent $2$ (thus being rank-one); the AP-IRS channel follows the Rician fading model with Rician factor $0.5$ and path loss exponent $2.1$; and the user-IRS channel follows the Rayleigh fading model with path loss exponent $2.1$. It is observed from Fig. \ref{MIMO} that the achievable rate of the AO-based algorithm is significantly larger than the MIMO channel capacity without IRS, as well as the achievable rate of a benchmark scheme with the phase shift of each IRS reflecting element independently and uniformly generated in $[0,2\pi)$ and the corresponding optimal transmit covariance matrix.

\subsubsection{IRS-aided OFDM System}
Next, we extend the IRS reflection design to OFDM systems under frequency-selective fading \hbox{channels.} Despite the broadband system, we assume that the system bandwidth is practically much smaller than the carrier frequency, i.e., $B\ll f_c$ for simplicity. To illustrate the new signal and channel models in this case, we first consider that both the AP and user are equipped with a single antenna, i.e., $M_t=M_r=1$. Let $L_0$ denote the number of delayed taps in the AP-user direct channel; $L_1$ and $L_2$ denote that in the AP-IRS and IRS-user channels, respectively. Let $\mv{h}^d=[h^d_0,\cdots,h^d_{L_0-1}]^T\in \mathbb{C}^{L_0\times 1}$ denote the baseband equivalent time-domain channel for the AP-user direct link; $\mv{h}_n^{r_1}=[h_{n0}^{r_1},\cdots,h_{n(L_1-1)}^{r_1}]^T\in \mathbb{C}^{L_1\times 1}$ and $\mv{h}_n^{r_2}=[h_{n0}^{r_2},\cdots,h_{n(L_2-1)}^{r_2}]^T\in \mathbb{C}^{L_2\times 1}$ denote that for the AP-IRS and IRS-user links associated with IRS reflecting element $n$, respectively. Similar to Section II,  let $x(t)$ denote the baseband equivalent transmit signal at the AP. The passband signal impinging on IRS element $n$ is thus given by
\begin{align}
y_{in,n}(t)=\mathrm{Re}\left\{\left[\sum_{l=0}^{L_1-1}h^{r_1}_{nl}x(t-\tau_{l}(t))\right]e^{\jmath2\pi f_ct} \right\},
\end{align}
where $\tau_l(t)=l/B$ denotes the time delay of the $l$th tap. The reflected signal by IRS element $n$ is then expressed as
\begin{align}
y_{out,n}(t)=\beta_ny_{in,n}(t-t_n)
=&\mathrm{Re}\left\{\beta_n\left[\sum_{l=0}^{L_1-1}h^{r_1}_{nl}x(t-t_n-\tau_{l}(t))\right]e^{\jmath2\pi f_c(t-t_n)} \right\}\nonumber\\
\approx& \mathrm{Re}\left\{\beta_ne^{\jmath\theta_n}\left[\sum_{l=0}^{L_1-1}h^{r_1}_{nl}x(t-\tau_{l}(t))\right]e^{\jmath2\pi f_ct} \right\},
\end{align}
where $t_n\in [0,1/f_c]$ denotes the time delay at IRS element $n$; $x(t-t_n-\tau_l(t))\approx x(t-\tau_l(t))$ holds since $t_n\leq 1/f_c\ll 1/B$; and $\theta_n= 2\pi-2\pi f_ct_n\in [0,2\pi)$ denotes the phase shift at IRS element $n$. After undergoing the IRS-user channel, the passband signal at the user receiver via IRS element $n$'s reflection is given by
\begin{align}\label{signal_OFDM}
y_{r,n}(t)=\mathrm{Re}\Big\{\Big[\sum_{l'=0}^{L_2-1}h^{r_2}_{nl'}\beta_ne^{\jmath\theta_n}\sum_{l=0}^{L_1-1}h^{r_1}_{nl}x(t-\tau_{l}(t)-\tau_{l'}(t))\Big]e^{\jmath2\pi f_ct} \Big\}.
\end{align}
Accordingly, the baseband signal model of (\ref{signal_OFDM}) can be expressed as
\begin{align}
y_n(t)=(\mv{h}_n^{r_2}\circledast\beta_ne^{\jmath\theta_n}\circledast\mv{h}_n^{r_1})x(t)\overset{\Delta}{=}{\mv{h}}_n^rx(t),
\end{align}
where ${\mv{h}}_n^r=\mv{h}_n^{r_2}\circledast\beta_ne^{\jmath\theta_n}\circledast\mv{h}_n^{r_1}=\beta_ne^{\jmath\theta_n}(\mv{h}_n^{r_2}\circledast\mv{h}_n^{r_1})\in \mathbb{C}^{L^r\times 1}$ denotes the cascaded AP-IRS-user  channel via IRS element $n$, with $L^r=L_1+L_2-1$. Note that the cascaded channel is the \emph{convolution} of the AP-IRS multi-path channel, the (single-tap)  IRS reflection coefficient, and the IRS-user multi-path channel.

We consider an OFDM system where the total bandwidth is equally divided into $Q\geq 1$ orthogonal sub-bands each with a different central sub-carrier frequency. For ease of exposition, let $\tilde{\mv{g}}_n^r=[(\mv{h}_n^{r_2}\circledast\mv{h}_n^{r_1})^T,0,\cdots,0]^T\in \mathbb{C}^{Q\times 1}$ denote the zero-padded convolved AP-IRS and IRS-user time-domain channel via IRS element $n$, and define $\tilde{\mv{G}}^r=[\tilde{\mv{g}}_1^r,\cdots,\tilde{\mv{g}}_N^r]\in \mathbb{C}^{Q\times N}$. The cascaded AP-IRS-user channel can thus be expressed as $\tilde{\mv{h}}^r=\tilde{\mv{G}}^r\mv{\theta}$, where $\mv{\theta}=[\beta_1e^{\jmath\theta_1},\cdots,\beta_Ne^{\jmath\theta_N}]^T$. Let $\tilde{\mv{h}}^d=[\mv{h}^{d^T},0,\cdots,0]^T\in \mathbb{C}^{Q\times 1}$ denote the zero-padded time-domain AP-user direct channel. The superposed effective channel impulse response (CIR) is thus given by
\begin{align}
\tilde{\mv{h}}=\tilde{\mv{h}}^d+\tilde{\mv{h}}^r=\tilde{\mv{h}}^d+\tilde{\mv{G}}^r\mv{\theta}.
\end{align}
Note that the number of delayed taps in the effective AP-user channel (or non-zero entries in $\tilde{\mv{h}}$) is $L=\max(L_0,L_1+L_2-1)$. By further assuming that the cyclic prefix (CP) length of each OFDM symbol is no smaller than $L$ (so that the inter-symbol interference can be eliminated), the channel frequency response (CFR) at each sub-carrier $q$ can be expressed as
\begin{align}\label{CFR}
c_q=\mv{f}_q^H\tilde{\mv{h}}^d+\mv{f}_q^H\tilde{\mv{G}}^r\mv{\theta},\quad q=1,\cdots,Q,
\end{align}
where $\mv{f}_q^H$ denotes the $q$th row of the $Q\times Q$ discrete Fourier transform (DFT) matrix. It is worth noting from (\ref{CFR}) that the IRS phase-shift values in $\mv{\theta}$ impact the CFR at each OFDM sub-carrier identically, i.e., \emph{without frequency selectivity}. By ignoring the rate loss due to the CP insertion, the achievable rate of the IRS-aided OFDM system in bps/Hz is given by
\begin{align}\label{rate_OFDM}
r=\frac{1}{Q}\sum_{q=1}^Q\log_2\left(1+\frac{p_q|\mv{f}_q^H\tilde{\mv{h}}^d+\mv{f}_q^H\tilde{\mv{G}}^r\mv{\theta}|^2}{\bar{\sigma}^2}\right),
\end{align}
where $p_q$ denotes the transmit power allocated to sub-carrier $q$ with $\sum_{q=1}^Qp_q\leq P_t$, and $\bar{\sigma}^2$ denotes the average receiver noise power at each sub-carrier.

\begin{figure}[t]
	\centering
	\includegraphics[width=8cm]{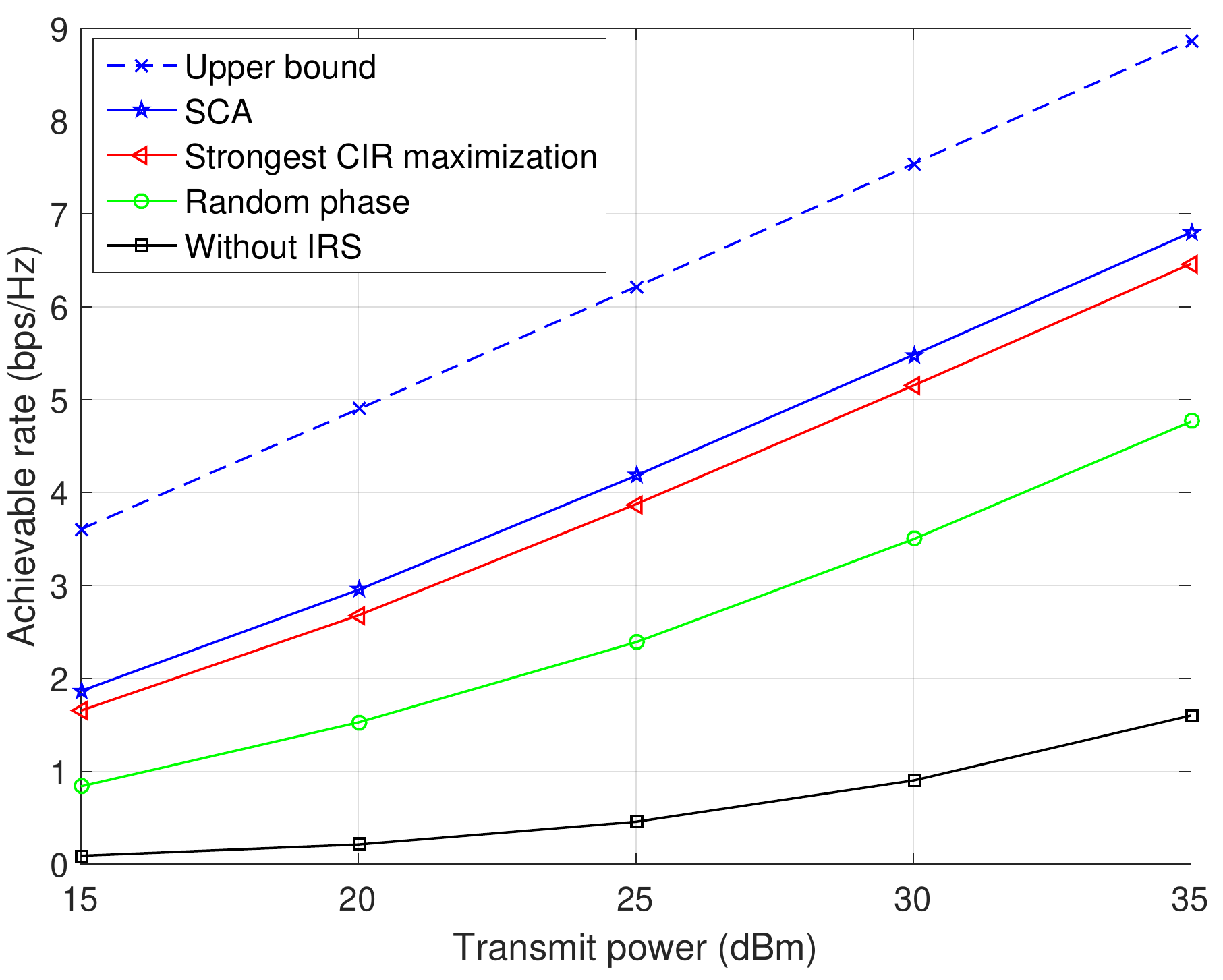}
	\vspace{-3mm}
	\caption{Achievable rate versus transmit power for IRS-aided OFDM system.}\label{OFDM}
	\vspace{-5mm}
\end{figure}

To maximize the achievable rate in (\ref{rate_OFDM}), the IRS phase shifts in $\mv{\theta}$ need to cater to the frequency-varying channels at different sub-carriers, or equivalently, the time-domain channels at different delayed taps. Moreover, $\mv{\theta}$ needs to be jointly optimized with the transmit power allocations over the $Q$ sub-carriers, $\{p_q\}_{q=1}^Q$, thus rendering the resultant optimization problem more difficult to solve as compared to (P1) in the narrow-band case. To tackle this problem, an efficient successive convex approximation (SCA) based algorithm was proposed in \cite{OFDM_Protocol} by approximating the non-concave rate function in (\ref{rate_OFDM}) using its concave lower bound based on the first-order Taylor expansion. The SCA-based algorithm is guaranteed to converge to a stationary point of the joint IRS reflection and transmit power optimization problem, and requires only  polynomial complexity over $N$ or $Q$ \cite{OFDM_Protocol}. To further lower the complexity, \cite{OFDM_BX} proposed a simplified algorithm where the IRS phase shifts are designed to only align with the time-domain channel with strongest path power, thus termed as the ``strongest CIR maximization''. In Fig. \ref{OFDM}, we consider an IRS-aided OFDM system with $M_t=M_r=1$, $N=30$, $Q=64$, and show the achievable rate versus transmit power $P_t$ for various schemes. The system setup is similar to that in Fig. \ref{simulation:setup}, but with $d_1=200$ m, $d_2=\sqrt{26}$ m, and $d=199$ m. All the channels involved are assumed to follow the Rician fading model. The Rician factors for the direct AP-user channel, AP-IRS channel, and IRS-user channel are set as $0$, $3$ dB, and $-20$ dB, respectively; while the corresponding path loss exponents for the three channels are set as $3.5$, $2.8$, and $2.2$, respectively. The number of delayed taps in the AP-user direct channel is set as $L_0=16$, while those in the AP-IRS channel and IRS-user channel are set as $L_1=4$ and $L_2=13$, respectively. We also set $\bar{\sigma}^2=-108$ dBm. It is observed from Fig. \ref{OFDM} that the SCA-based algorithm proposed in \cite{OFDM_Protocol} achieves significantly higher rate as compared to the OFDM system without IRS and that with random IRS phase shifts and the corresponding optimal transmit power allocation. Moreover, the strongest CIR maximization algorithm achieves close performance to the SCA-based algorithm, thus being an efficient alternative with lower complexity. Furthermore, we show an achievable rate upper bound by assuming that (ideally) different IRS reflection coefficients can be designed for different sub-carriers, thus making the IRS reflection design ``frequency-selective''. It is observed that this rate upper bound outperforms the SCA-based solution with the practical frequency-flat (non-selective) IRS reflection quite substantially, and the rate gap increases with the number of sub-carriers. This thus reveals that a fundamental limitation of IRS-aided OFDM systems lies in the lack of frequency-selective IRS reflection due to its passive operation.

Finally, it is worth noting that for the more general IRS-aided MIMO-OFDM systems where the AP and/or user are equipped with multiple antennas, the IRS reflection design for rate maximization is more involved, due to the need of catering to more channels in both space and frequency; moreover, the IRS reflection needs to be jointly optimized with multiple transmit covariance matrices at different sub-carriers. To resolve this problem, \cite{MIMO} proposed an efficient AO-based algorithm by extending that in the narrow-band MIMO case and leveraging the convex relaxation technique. The results in \cite{MIMO} showed that despite the lack of frequency selectivity, IRS is still effective in improving the rate of MIMO-OFDM systems with properly designed IRS reflection coefficients over the conventional system without IRS.

\subsection{IRS-aided Multi-user System}
Next, we consider the general case with multiple users in the IRS-aided system (see, e.g., Fig. \ref{P2Psystem:model}). For the purpose of exposition, we focus on the following two narrow-band multi-user systems under the SISO and MISO setups, respectively. In particular, for the SISO case, we compare the performance of orthogonal multiple access (OMA) such as time division multiple access (TDMA) and  frequency division
multiple access (FDMA) with that of non-orthogonal multiple access (NOMA); while for the MISO case, we consider the  spatial division multiple access (SDMA) where the multi-antenna AP serves multiple single-antenna users simultaneously in the same frequency band. Our main objective is to highlight the main differences in system design and performance optimization with IRS versus traditional systems without IRS.

\subsubsection{OMA versus NOMA}
\begin{figure}[t]
	\centering
	\includegraphics[width=8cm]{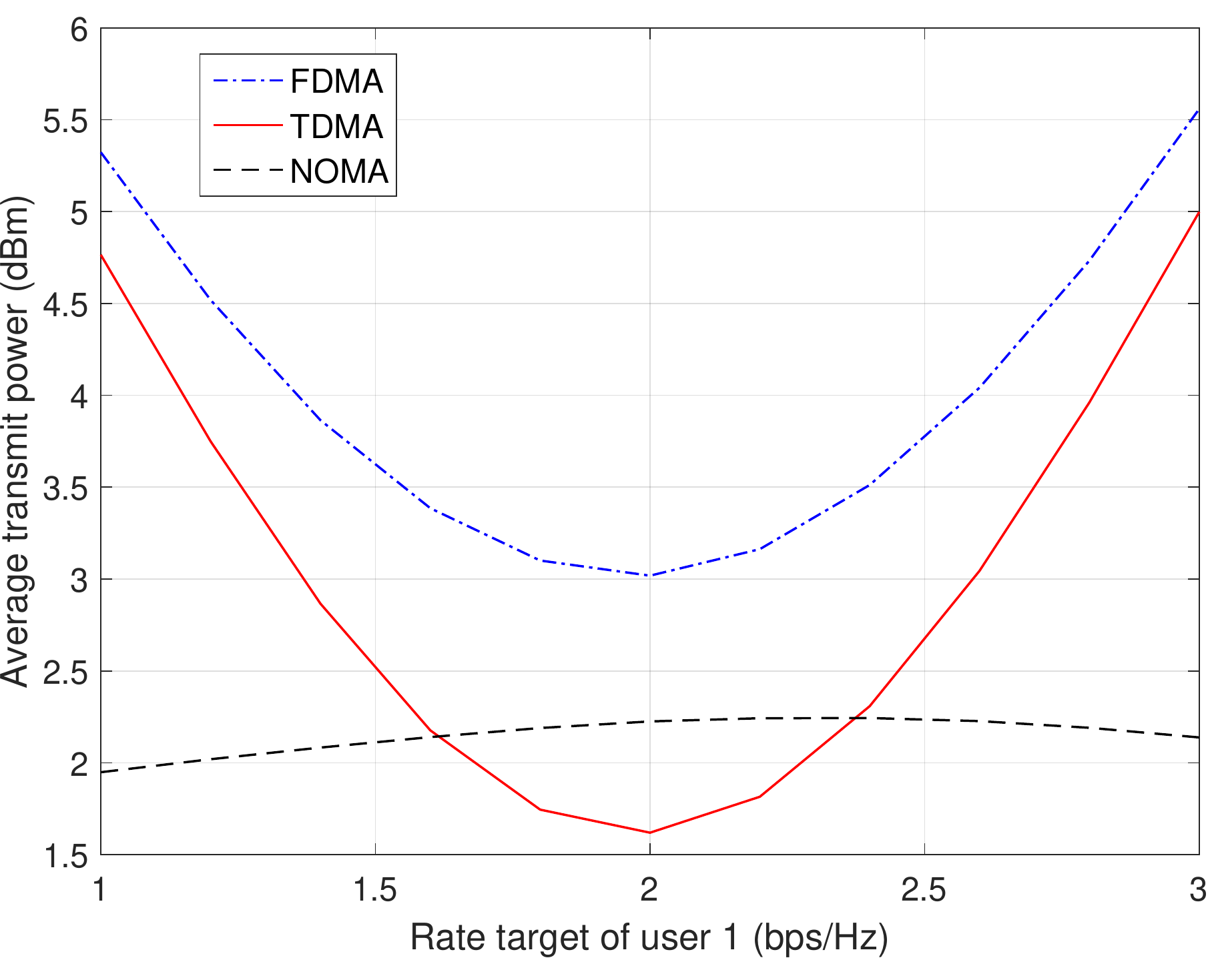}
	\vspace{-3mm}
	\caption{Minimum transmit power versus user 1's  rate target for an IRS-aided two-user system with the sum-rate fixed as  4 bps/Hz.}\label{NOMA}
	\vspace{-5mm}
\end{figure}

In TDMA, the users are served in orthogonal time slots, thus different IRS reflection coefficients can be applied over time to maximize the rate of each user, by exploiting the ``time selectivity'' of  IRS. In contrast, in FDMA and NOMA where the users are simultaneously served in orthogonal or common frequency bands, the IRS reflection design needs to cater to the channels of all users, thus is more challenging compared to the TDMA case. To investigate which multiple access scheme is more favorable for IRS-aided systems, \cite{NOMA_BX} considered a two-user downlink communication system with $K=2$, and compared the average transmit power at the AP required by TDMA, FDMA, and NOMA for achieving the same rate targets of the two users. Under this setup, it was shown that TDMA outperforms FDMA due to the lack of frequency-selective IRS reflection in the latter case; while,  surprisingly, although NOMA outperforms OMA (i.e., both TDMA and FDMA) due to successive interference cancellation  in traditional multi-user systems without the IRS, TDMA may perform better than NOMA for near-IRS users with symmetric rate targets, thanks to the IRS adaptive reflections based on users' individual channels in TDMA. For illustration, we consider an IRS-aided two-user system where the locations of AP, IRS, and two users are set as $[0,0]$ m, $[50,0]$ m, $[48,1]$ m, and $[48,-1]$ m, respectively, in a two-dimensional plane, and show in Fig. \ref{NOMA} the minimum average transmit power required versus the rate requirement of user $1$ with the sum-rate of the two users being fixed as $4$ bps/Hz. Similar to \cite{NOMA_BX}, the number of IRS reflecting elements is set as $N=100$, which are divided into $5$ sub-surfaces; the reflection coefficient at each element is assumed to have unit amplitude and discrete phase shift with $8$ levels; all the channels are modeled by i.i.d. Rayleigh fading with the path loss exponents of the AP-user, AP-IRS, and IRS-user links being $3.2$, $2.5$, and $2.6$, respectively. The average noise power at the user receivers is set as $\sigma^2=-80$ dBm. It is observed from Fig. \ref{NOMA} that both TDMA and NOMA outperform FDMA in terms of the minimum transmit power required. Moreover, when the user rates are symmetric, NOMA requires even larger transmit power than TDMA; while the performance of NOMA is more robust (less sensitive) to the users' rate disparity as compared to that of TDMA.

For the more general multi-user case with $K> 2$, the IRS reflection designs for FDMA and NOMA become more difficult, due to the increased number of user channels that need to be catered to at the same time, which also limits the passive beamforming gain of IRS for each user. To overcome this difficulty, \cite{OFDMA} proposed a novel \emph{dynamic passive beamforming} scheme for an IRS-aided orthogonal frequency division multiple access (OFDMA) system, where each channel coherence interval is divided into multiple time slots each for serving only a \emph{subset} of selected users, and the IRS passive beamforming is dynamically adjusted over different time slots to induce artificial (yet properly-tuned) channel fading for exploiting the IRS's time-selective reflection as well as multi-user channel diversity in resource allocation optimization. By jointly designing the IRS passive beamforming over time with the OFDMA time-frequency resource allocation, it was shown in \cite{OFDMA} that improved performance can be achieved over the benchmark scheme with optimized but fixed IRS reflection coefficients for all users in each channel coherence interval.

\subsubsection{SDMA} 
In SDMA, the multi-antenna AP employs different linear precoding/combining vectors to serve different users simultaneously in the same frequency band for the downlink/uplink transmissions, thus significantly improving the spectral efficiency over the single-antenna AP. However, conventional SDMA does not perform well for users close to each other (e.g., in a hot-spot scenario), especially when they are located at the cell edge. This is because users in this case are likely to lie in similar directions from their serving  AP, which induces high correlation among their channels that is detrimental to the achievable spatial multiplexing gain due to the more severe multi-user co-channel interference. As such,  conventional beamforming techniques such as  zero-forcing (ZF) to null/suppress their mutual interference  become inefficient.

However, by properly deploying IRS in such scenarios, the above  issue can be efficiently solved by leveraging  its spatial interference nulling/cancellation capability \cite{JR:wu2018IRS}. Specifically,  IRS can effectively reduce the undesired channel correlation among these users by optimizing its reflection coefficients to provide additional controllable signal paths.  Besides, the users near the IRS are expected to be able to tolerate more  interference from the AP as compared to those farther away from the IRS. This thus provides more flexibility in designing the transmit beamforming at the AP for  serving other users outside the coverage of any IRS. As a result,  the signal-to-interference-plus-noise ratio (SINR)  performance of all users in the network can be significantly improved, regardless of whether they are aided directly by any IRS or not \cite{JR:wu2018IRS}.

 Despite the above benefits, the active and passive beamforming designs in the multi-user case are closely coupled and usually lead to more complicated optimization problems as compared to that in the single-user case. As such, AO has been widely adopted to obtain high-quality suboptimal  solutions, by iteratively optimizing one of the transmit and reflect beamforming vectors/coefficients with the other being fixed, until the convergence is reached \cite{JR:wu2018IRS,zhao2019intelligent,zhao2020exploiting}.  However, such AO-based algorithms may become inefficient as the number of  QoS constraints increases since they are prone to getting  trapped at undesired suboptimal   solutions due to the more stringent coupling among  the variables. To address this issue, a new and more efficient penalty-based algorithm was proposed in \cite{wu2019jointSWIPT} for a more general system setup, which guarantees attaining at least a locally optimal solution. Specifically,  the QoS constraints are first decoupled by introducing a set of auxiliary variables and the reformulated problem is then solved by jointly applying the penalty-based method and block coordinate descent (BCD)  method \cite{wu2019jointSWIPT}.


\subsection{Passive Beamforming with Discrete Reflection Amplitude and Phase Shift}
In the preceding subsections, we have mainly considered continuous phase shifts of IRS with the maximum reflection amplitude, while as discussed in Section II-B, practical IRSs have discrete phase shifts and/or discrete reflection amplitudes, which may result in signal misalignment at designated receivers and thus degraded communication performance as compared to the ideal case of continuous phase shifts with flexible  reflection amplitudes \cite{wu2018IRS_discrete,JR:wu2019discreteIRS,di2019hybrid,zhang2019reconfigurable,xu2019discrete,zhao2020exploiting}. Besides, such constraints  also complicate the IRS reflection design and render the corresponding optimization problems more difficult to solve as compared to their continuous counterparts. For example, exhaustive search over all possible discrete phase-shift values is needed to solve the design problem optimally, which, however, incurs prohibitively high complexity in practice with large values of $N$ \cite{wu2018IRS_discrete}.
Although some optimization techniques such as branch-and-bound (BB) method can be used to obtain the optimal solution to such problems with reduced complexity on average, the computational complexity in the worst case is still exponential over $N$ and thus the same as that for the exhaustive search.
In practice, one heuristic  approach is to firstly relax such constraints and  solve the problem with  continuous amplitude/phase-shift values,  then quantize the obtained solutions to their nearest values in  the corresponding discrete sets.  While this approach is generally able to reduce the computational complexity  significantly to polynomial orders of $N$, it may suffer various losses in performance as compared to the continuous solution due to quantization/round-off  errors, depending on the number of discrete  levels as well as $N$, and is also generally suboptimal for the original discrete optimization problem.  To further improve the performance  of the above  approach, AO can be applied to iteratively optimize the discrete amplitude/phase-shift of each element by fixing those of the others at each time   \cite{wu2018IRS_discrete}.

To draw useful insight into the performance loss of employing discrete phase shifts for IRS as compared to continuous phase shifts,   we consider a modified problem of (P1) by ignoring the AP-user direct link and replacing the phase-shift constraints in \eqref{phase:constraints} by their discrete counterparts. As a result,  the user receive power (or achievable rate) maximization problem with discrete phase shifts can be formulated as
\begin{align}
\text{(P1-DP)}: ~~\max_{\bm{\theta}} ~~~&\left| \sum_{n=1}^{N}{h}^*_{r,n}   {g}_n e^{\jmath \theta_n}\right|^2  \label{eq:obj11}\\
\mathrm{s.t.}~~~~& \theta_n \in  \mathcal{F}'_{\theta},  n=1,\cdots, N, \label{phase:constraints:discrete}
\end{align}
where $ \mathcal{F}'_{\theta}$ is defined in (15). To facilitate our analysis, we adopt the quantization approach to solve (P1-DP) by first obtaining the optimal continuous phase shifts in \eqref{eq:phase:solution} and then quantizing them independently to the nearest values in  $\mathcal{F}'_{\theta}$. The corresponding objective value is denoted by $P_r(b_{\theta})$. In general, $P_r(b_{\theta})$ serves as a lower bound of the maximum objective value of (P1), whereas when $b_{\theta}\rightarrow \infty$,   $P_r(b_{\theta})$ converges to its maximum value in the continuous phase-shift case. As such, the performance loss of IRS with $b_{\theta}$-bit phase shifters (or $2^{b_{\theta}}$ uniform phase-shift levels) per reflecting element
 as compared to the ideal case with continuous phase shifts at all reflecting elements can be characterized by a ratio given by $\eta (b_{\theta}) \triangleq  {P_r(b_{\theta})}/{P_r(\infty)}$.
Under the same channel assumption for deriving the asymptotic result given in \eqref{eq:square:power:gain}, it was shown in \cite{JR:wu2019discreteIRS} that
\begin{align}\label{ratio}
\eta (b_{\theta})=  \Big(\frac{2^{b_{\theta}}}{\pi}\sin\left(\frac{\pi}{2^{b_{\theta}}}\right)\Big)^2.
\end{align}
It is observed from \eqref{ratio} that when $N$ is sufficiently large,  the power ratio $\eta (b_{\theta})$ depends only on the number of discrete phase-shift levels, $2^{b_{\theta}}$,  but is regardless of $N$. In other words,  using a practical IRS  even with discrete phase shifts can still achieve the same asymptotic receive power scaling order of $\mathcal{O}(N^2)$ given in \eqref{eq:square:power:gain}  as in the case of continuous phase shifts.  As such,  the design of  IRS hardware and control module can be greatly simplified by using discrete phase shifters, without compromising the performance significantly in the large-$N$ regime.  Since $\eta (1)  =-3.9$ dB while $\eta (2)= -0.9$ dB, using 2-bit phase shifters is practically sufficient to achieve close-to-optimal  performance within $1$ dB.

In \cite{arun2019rfocus}, a practical binary/1-bit reflection amplitude control was considered for IRS where the phase shifts of all the elements are set to zero and only the reflection amplitude of each IRS element is optimized  as either  0 or 1. Similar to (P1-DP),  the receive power maximization problem in this case can be formulated as
\begin{align}
\text{(P1-DA)}: ~~\max_{\bm{\beta}} ~~~&\left| \sum_{n=1}^{N}{h}^*_{r,n}   {g}_n \beta_n \right|^2  \label{eqDA:obj}\\
\mathrm{s.t.}~~~~& \beta_n \in  \{0, 1\},  n=1,\cdots, N, \label{apd:constraints}
\end{align}
where $\bm{\beta}= [\beta_1, \cdots, \beta_N]^T$. To solve (P1-DA), a heuristic method was proposed in \cite{arun2019rfocus} where $\beta_n=1$ if the phase of ${h}^*_{r,n}   {g}_n$ lies in $[-\pi/2, \pi/2]$;  otherwise, $\beta_n=0,  n=1,\cdots, N$.
 It was shown that the user receive power with this scheme is reduced  by a factor of $\pi^2$, i.e., incurring about 10 dB loss, as compared to the case with continuous  phase shifts \cite{arun2019rfocus}.

  \begin{figure}[h]
\qquad
 \includegraphics[width=0.95\textwidth]{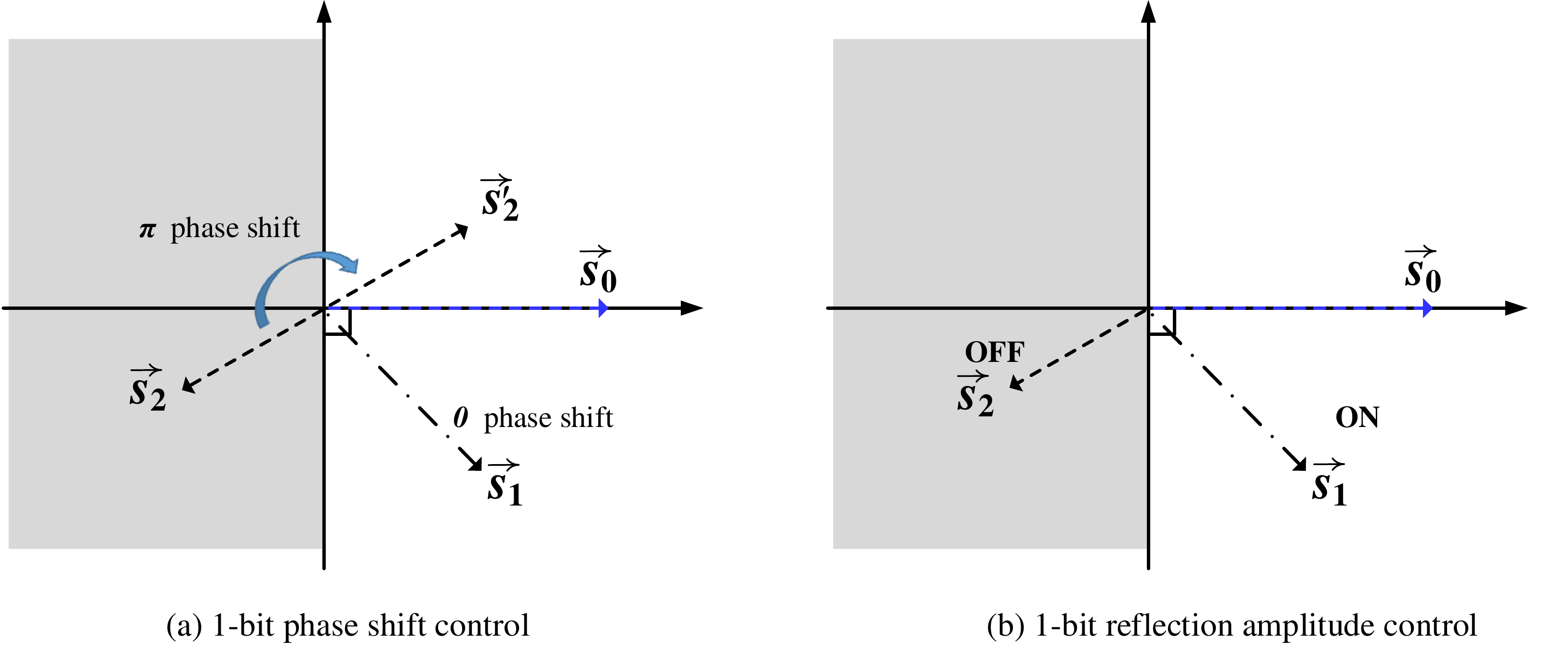}
\caption{Illustration of 1-bit phase shift control versus  1-bit reflection amplitude control in the complex signal plane,  where the former utilizes  both {$ {\protect \overrightarrow{\mv S_1}}$} and $ {\protect \overrightarrow{\mv S_2}}$ while the latter only uses  $ {\protect \overrightarrow{\mv S_1}}$.   } \label{system:1bit:phz:rotation}
\end{figure}

  \begin{figure}[h]
\centering
\includegraphics[width=0.65\textwidth]{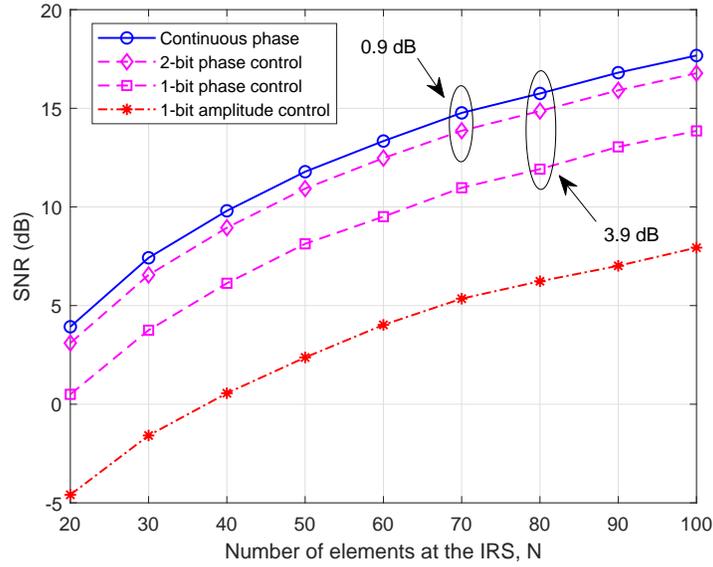}
\caption{Receive  SNR versus $N$ for different discrete amplitude/phase-shift designs  of IRS. } \label{simulation:discrete} 
\end{figure}

To compare the two schemes intuitively, we illustrate them  in the complex signal plane in  Fig. \ref{system:1bit:phz:rotation}. Suppose that $\overrightarrow{\mv {S_{0}}}$ indicates the desired signal  to be aligned with,  while $\overrightarrow{ {\mv {S_{1}}}}$ and $\overrightarrow{\mv {S_2}}$ denote the reflected signals by any two IRS elements without applying any phase shift/amplitude control yet. For the 1-bit phase shift control shown in Fig. \ref{system:1bit:phz:rotation} (a) \cite{JR:wu2019discreteIRS,wu2018IRS_discrete}, it is intuitive  that $\overrightarrow{\mv {S_1}}$  can help improve the user receive power by simply setting $\theta_1=0$ since it has an acute angle with $\overrightarrow{\mv{S_0}}$, i.e., in the right halfplane. On the other hand,  setting $\theta_2= \pi$ will  rotate $\overrightarrow{\mv{S_2}}$ into the right halfplane, which helps improve the combined signal amplitude as well.  In contrast, for the 1-bit amplitude control shown in Fig.  \ref{system:1bit:phz:rotation} (b)  \cite{arun2019rfocus}, since $\overrightarrow{\mv{S_1}}$ and $\overrightarrow{\mv{S_2}}$ cannot be rotated, the optimal strategy is to turn ON and OFF their corresponding reflecting elements, by setting $\beta_1=1$ and $\beta_2=0$. By comparing Figs. \ref{system:1bit:phz:rotation} (a) and (b), it is observed that  both $\overrightarrow{\mv{S_1}}$ and $\overrightarrow{\mv{S_2}}$ are utilized in 1-bit phase shift control whereas only $\overrightarrow{\mv{S_1}}$ is used in 1-bit amplitude control, which implies that the former generally outperforms the latter or equivalently the latter needs to deploy more reflecting elements (i.e., larger IRS) to achieve the same performance as the former.

Considering the same simulation setup as in Fig. \ref{simulation:setup}, we plot the maximum receive SNR versus $N$ under  different discrete amplitude/phase-shift cases of IRS in Fig. \ref{simulation:discrete} with $M_t=1$, $P_t=5$ mW, and $\sigma^2=-100$ dBm.  It is observed that when $N$ is sufficiently large, the SNR loss of using IRS with 1-bit or 2-bit phase shift control approaches a constant, i.e., 3.9 dB or 0.9 dB, which is consistent with the theoretical result in \eqref{ratio}.  Besides,   one can observe that compared with 1-bit phase shift control, 1-bit amplitude control suffers substantial power/SNR loss since the IRS reflected signals cannot be fully utilized with the simple  ON/OFF control.

\subsection{Other Related Work  and Future Direction}
In the last subsection, we overview other related works on IRS reflection design and point out promising directions for future work.

\begin{figure}[!t]
\centering
\subfigure[Downlink transmission]{\includegraphics[width=0.7\textwidth]{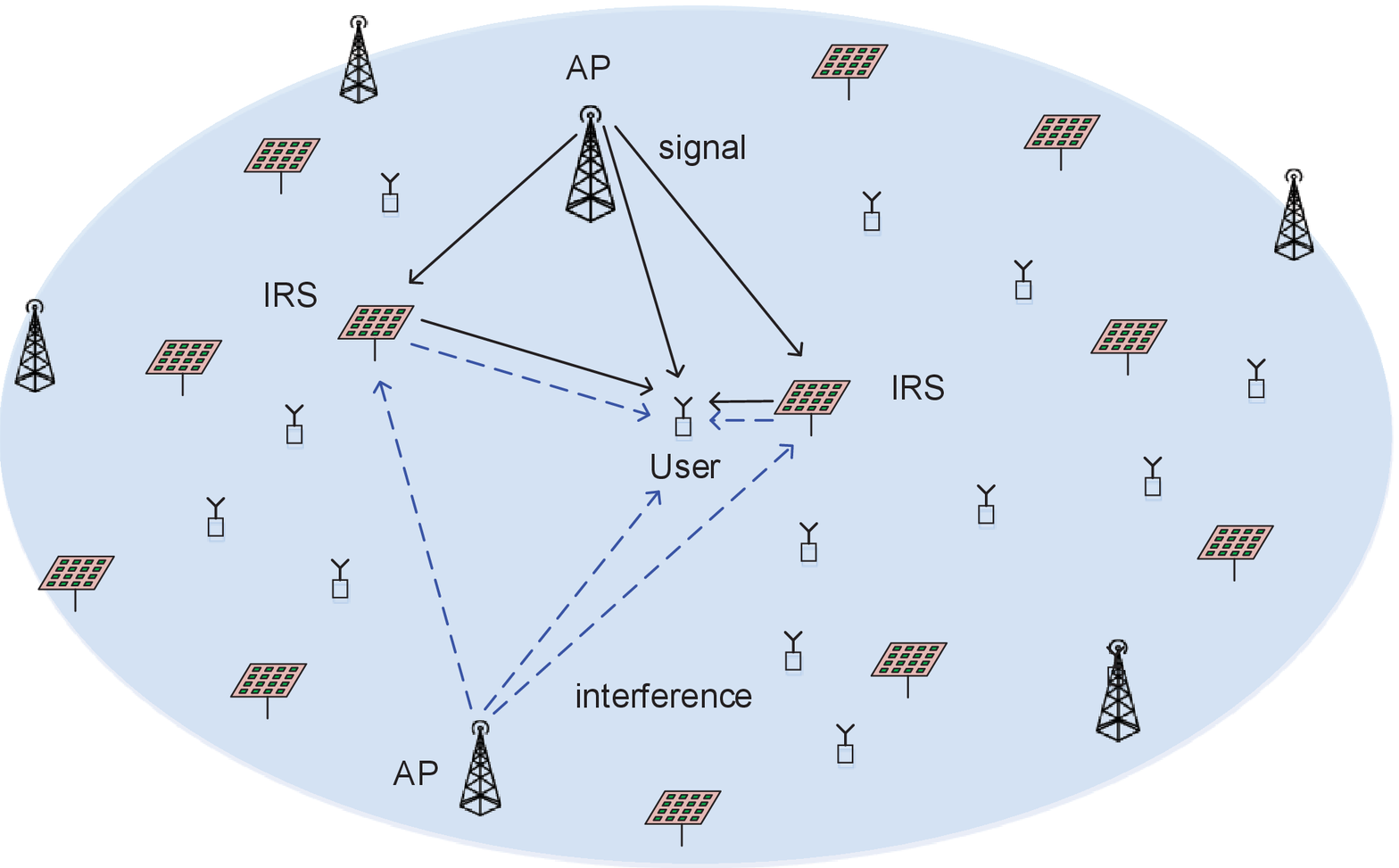} } 
\subfigure[Uplink transmission]{\includegraphics[width=0.7\textwidth]{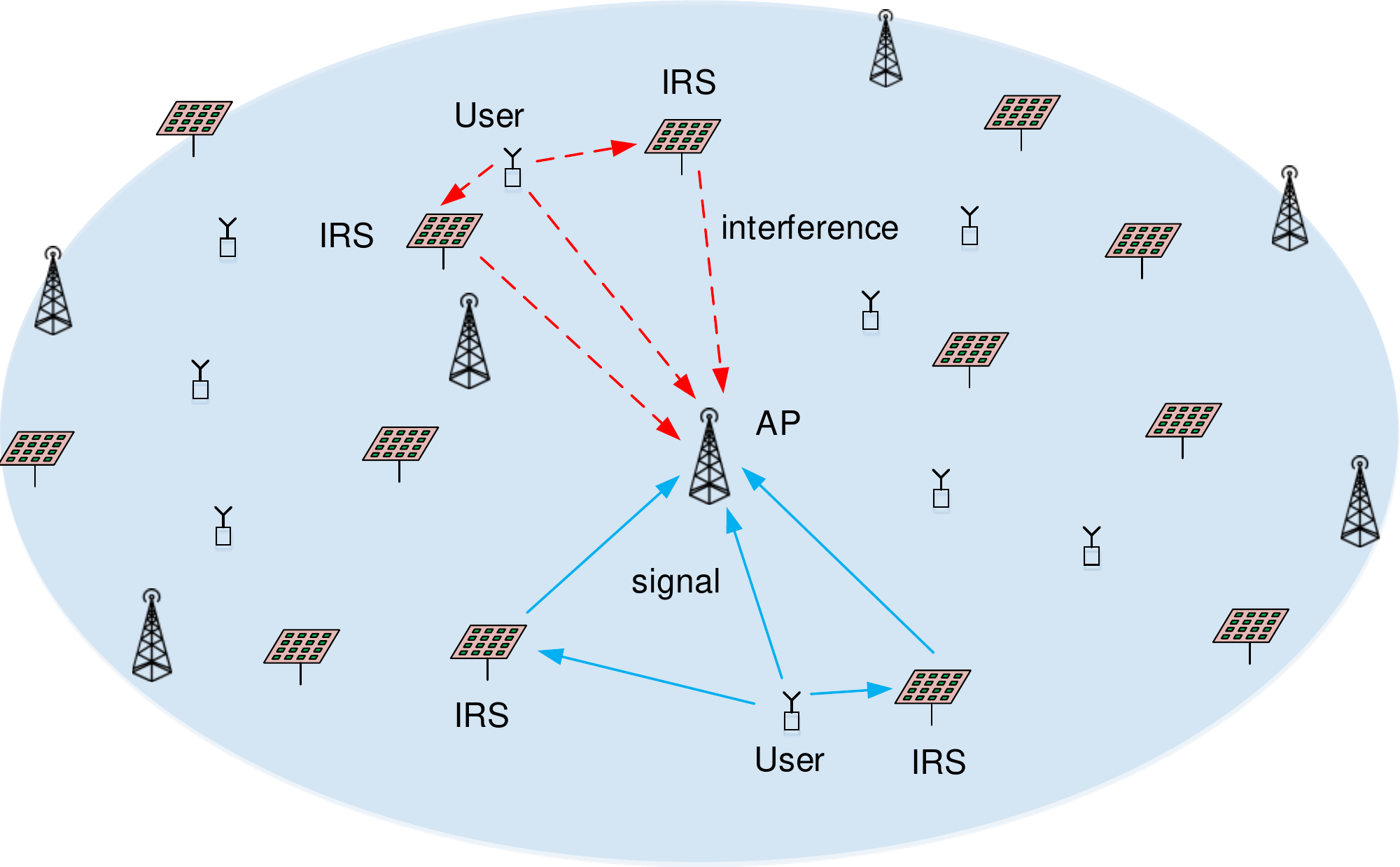} }
\caption{IRS-aided  multi-cell wireless network.  }  \label{P2Psystem:model:multicell}
\end{figure}

\begin{table*}[!t]
\centering
{{\caption{ A summary of representative works on IRS reflection optimization.}\label{table11}
\vspace{-0.3cm}
\small
{\begin{tabular}{|m{1.8cm}|m{2.3cm}|m{3cm}|m{3.3cm}|m{2.7cm}|}
  \hline
{\bf Reference}  &{\bf System setup} &{\bf Design objective } &  {\bf Optimization  techniques } & {\bf  CSI assumption} \\ \hline
{ \cite{wu2018IRS,JR:wu2018IRS} }          & SU/MU MISO        &Power minimization  &  { AO, SDR }  & Instantaneous CSI   \\ \hline
{ \cite{feng2020deep} }          & SU MISO      &Power minimization  &  { Deep learning  }  & Instantaneous CSI   \\ \hline
{ \cite{OFDM_Protocol,OFDM_GC}}  & SU OFDM  & Rate maximization  &  {AO, duality}  &  Instantaneous CSI     \\ \hline
{ \cite{MIMO,MIMO_ISIT} }          &SU MIMO       &  Rate maximization  & {AO} &   Instantaneous CSI   \\ \hline
{\cite{huang2018largeRIS} }  & MU MISO  &  Energy efficiency \textcolor{white}{xx} maximization&  {AO, gradient descent,  majorize-minimization (MM)} &   Instantaneous CSI     \\ \hline
{\cite{wu2019jointSWIPT} }  & MU MISO  & Power minimization &  {Penalty method } & Instantaneous CSI      \\ \hline
{\cite{zhao2019intelligent} }  & MU MISO  & Rate maximization  &  {Penalty method, \textcolor{white}{xxxx}  stochastic optimization} & Instantaneous and statistical CSI      \\ \hline
{\cite{yangliu2020} }  & MU MISO  &  Min SINR \textcolor{white}{xxxxxx} maximization &   Penalty method,  Alternative direction method of multipliers (ADMM) &  Instantaneous CSI        \\ \hline
{\cite{kammoun2020asymptotic} }  & MU MISO  &  Min SINR \textcolor{white}{xxxxxx} maximization &  { Projected gradient \textcolor{white}{xx} ascent} & Statistical CSI      \\ \hline
{\cite{NOMA_BX} } & MU SISO NOMA  & Power minimization  &  { AO} &   Instantaneous CSI      \\ \hline
\cite{guo2019weighted}  & MU MIMO    & Weighted sum-rate \textcolor{white}{x} maximization  &{AO, SCA, fractional \textcolor{white}{x} programming  } &   Instantaneous CSI   \\ \hline
\cite{MIMO_Pan}  & Multi-cell MIMO  & Rate maximization  & AO, MM &   Instantaneous CSI    \\ \hline
\end{tabular} } } }
\end{table*}

Although this section focuses  on the single-cell system with one IRS,  it is worthy of extending the results  to more general setups with multiple APs and/or multiple IRSs in an IRS-aided wireless network as shown in Fig. \ref{P2Psystem:model:multicell}. As such, the reflection coefficients of IRSs need to be jointly designed with the transmissions of multiple APs to not only improve the desired signal power but also mitigate the  intra-cell interference as well as inter-cell interference (ICI)   \cite{hua2020intelligent,MIMO_Pan}. In Table \ref{table11}, we summarize the representative works on IRS reflection optimization based on  their considered system setups, adopted optimization techniques, etc. Furthermore, besides discrete reflection amplitudes/phase shifts,   the performance of IRS-aided systems under other hardware constraints, such as coupled reflection amplitude and phase shift shown in Section II-B, is also worth investigating. In particular, it was shown in \cite{abeywickrama2019intelligent} that the asymptotic power scaling order, i.e., $O(N^2)$ unveiled in \cite{JR:wu2018IRS} under the ideal phase shift model, still holds for the practical case with phase-shift dependent non-uniform IRS reflection amplitude.  On the other hand, from the optimization perspective, it is also important to develop more advanced and computationally efficient algorithms such as machine learning based methods for IRS reflection design, especially for practically large IRSs \cite{feng2020deep,yu2020optimal,yang2020intelligent}.

Besides SINR/rate maximization, another line of research has aimed to study other performance metrics of IRS-aided systems such as outage probability and average bit error rate (BER)\cite{zhang2019analysis}.
In \cite{kammoun2020asymptotic}, random matrix theory was leveraged to study the asymptotic max-min SINR in the single-cell MISO downlink system, which characterizes the effect of channel large-scale fading parameters on the SINR performance. Moreover, IRS reflection design and spatial throughput analysis for large hybrid active and passive wireless networks is also an interesting topic to pursue \cite{lyu2020hybrid,SpatialThroughput}.


Last but not the least, while significant performance gains offered by IRS have been shown in this section, they are based on the assumption of perfect CSI for all channels considered. In practice,  the acquisition of accurate CSI of the IRS-reflected links is crucial, which, however, is a difficult task due to the passive operation  of  IRS as well as  its large number of reflecting elements. Thus, how to efficiently estimate the channels in IRS-aided wireless systems and optimize the IRS reflection with imperfect/partial CSI (see, e.g., \cite{zhao2019intelligent,zhao2020exploiting}) is practically important, as will be addressed in the next section.


\section{IRS Channel Estimation}

\subsection{Problem Description and Challenges}
To fully achieve various performance gains brought by IRS, the acquisition of accurate CSI is crucial, which, however, is practically challenging.
To be specific, considering the IRS-aided uplink multi-user MIMO communication in a narrow-band system over flat-fading channels, the received signal at the $M_B$-antenna BS from $K$ users (each of which is equipped with $M_u$ antennas) can be expressed as
\begin{align}\label{signal}
{\bm y}=\sum_{k=1}^K\left({\bm G}^T {\bm \Theta}  {\bm H}_{r,k}+{\bm H}_{d,k}\right) {\bm x}_k +{\bm z},
\end{align}
where ${\bm G}\in {\mathbb{C}^{N\times  M_B}} $, ${\bm H}_{r,k}\in {\mathbb{C}^{N\times M_u }}$ and ${\bm H}_{d,k}\in {\mathbb{C}^{M_B\times M_u }}$ denote the IRS-BS, user $k$-IRS, and user $k$-BS direct
channels, respectively, with $k=1,\ldots,K$; ${\bm \Theta}=\mathrm{diag} \left( e^{\jmath \theta_1},e^{\jmath \theta_2},\ldots, e^{\jmath \theta_N} \right)$ represents the diagonal phase-shift matrix of one or more IRSs comprising $N$ reflecting elements in total, with the reflection amplitude of each element set to one or its maximum value for simplicity;
${\bm x}_k\in {\mathbb{C}^{M_u\times 1}}$ is the transmit  signal of user $k$;
and ${\bm z} \in {\mathbb{C}^{M_B\times 1}}$ is the AWGN vector at the BS.
Accordingly, the uplink CSI includes ${\bm G}$, $\left\{{\bm H}_{r,k}\right\}_{k=1}^K$, and $\left\{{\bm H}_{d,k}\right\}_{k=1}^K$,
and thus the total number of uplink channel coefficients consists of two parts:
\begin{itemize}
	\item The number of channel coefficients (equal to $K\times{N M_u }+{M_B N }$) for the links to/from the (equivalent single)  IRS (i.e., $\left\{{\bm H}_{r,k}\right\}_{k=1}^K$ and ${\bm G}$), which are newly introduced due to the employment of IRS;
	\item The number of channel coefficients (equal to $K\times M_B M_u$) for the direct links (i.e., $\left\{{\bm H}_{d,k}\right\}_{k=1}^K$), which exist in conventional  communication systems without  IRS.
\end{itemize}
Note that the total number of channel coefficients may be different for TDD and frequency-division duplexing (FDD) systems.
In particular, the FDD system requires to estimate twice the number of channel coefficients in \eqref{signal} due to the generally non-symmetric uplink and downlink channels; while in contrast, the TDD system may only need to acquire either the uplink or downlink channel coefficients by exploiting the uplink-downlink channel reciprocity.
Furthermore, in broadband communication systems over frequency-selective fading channels, more channel coefficients are induced for both user-BS direct channels as well as user-IRS-BS reflected channels due to the multi-path delay spread and the resultant convolution of time-domain impulse responses of user-IRS and IRS-BS multi-path channels, which
makes the channel acquisition problem even more challenging \cite{zheng2020intelligent,OFDM_BX,OFDM_Protocol}.

\begin{figure*}
	\centering
	\subfigure[Semi-passive IRS.]{
		\begin{minipage}[b]{0.45\textwidth}
			\includegraphics[width=1\textwidth]{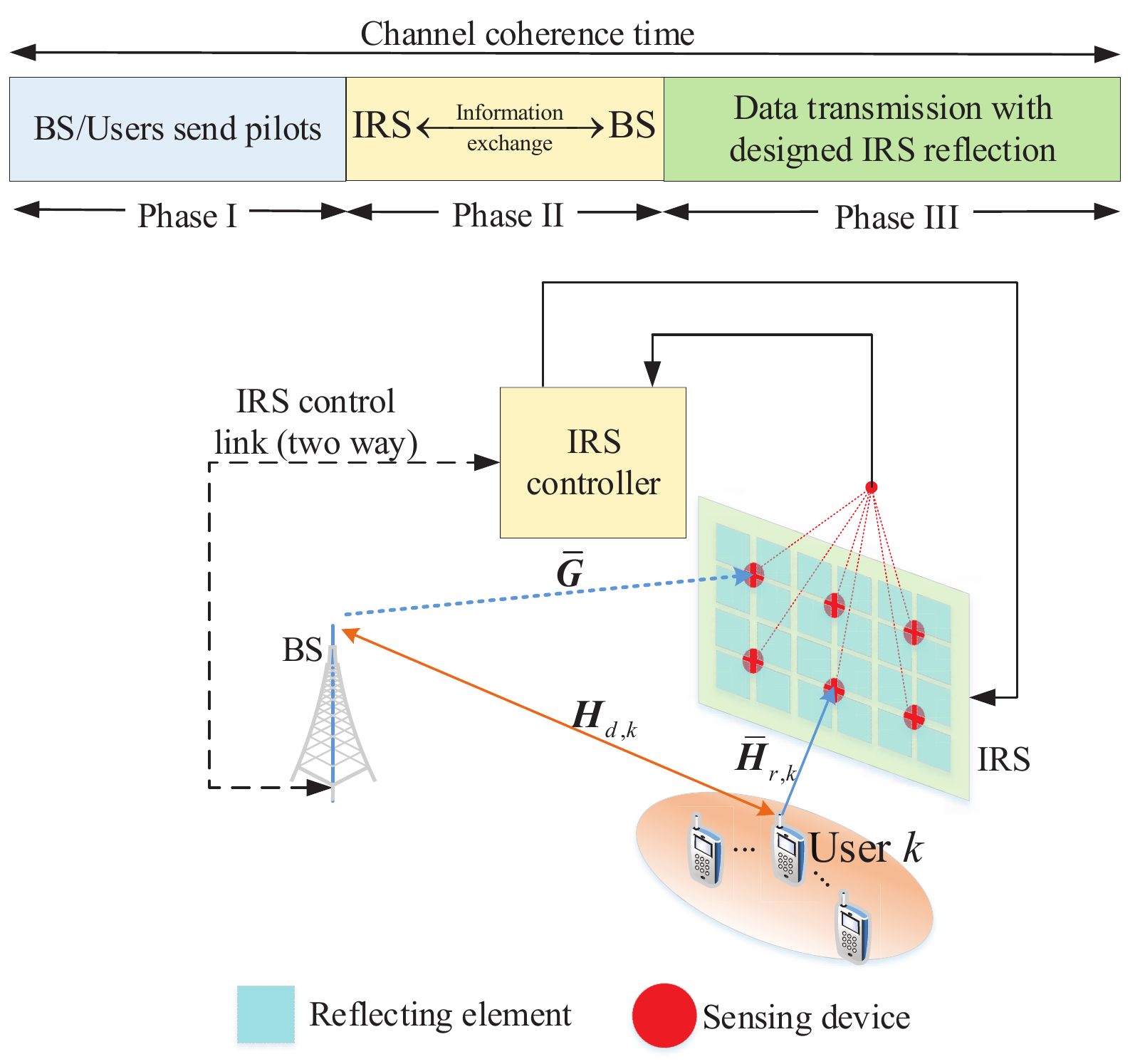}
		\end{minipage}\label{SemiIRS}
	}
	~~~~~~~~~~~~~~\subfigure[Passive IRS (in the uplink).]{
		\begin{minipage}[b]{0.5\textwidth}
			\includegraphics[width=1\textwidth]{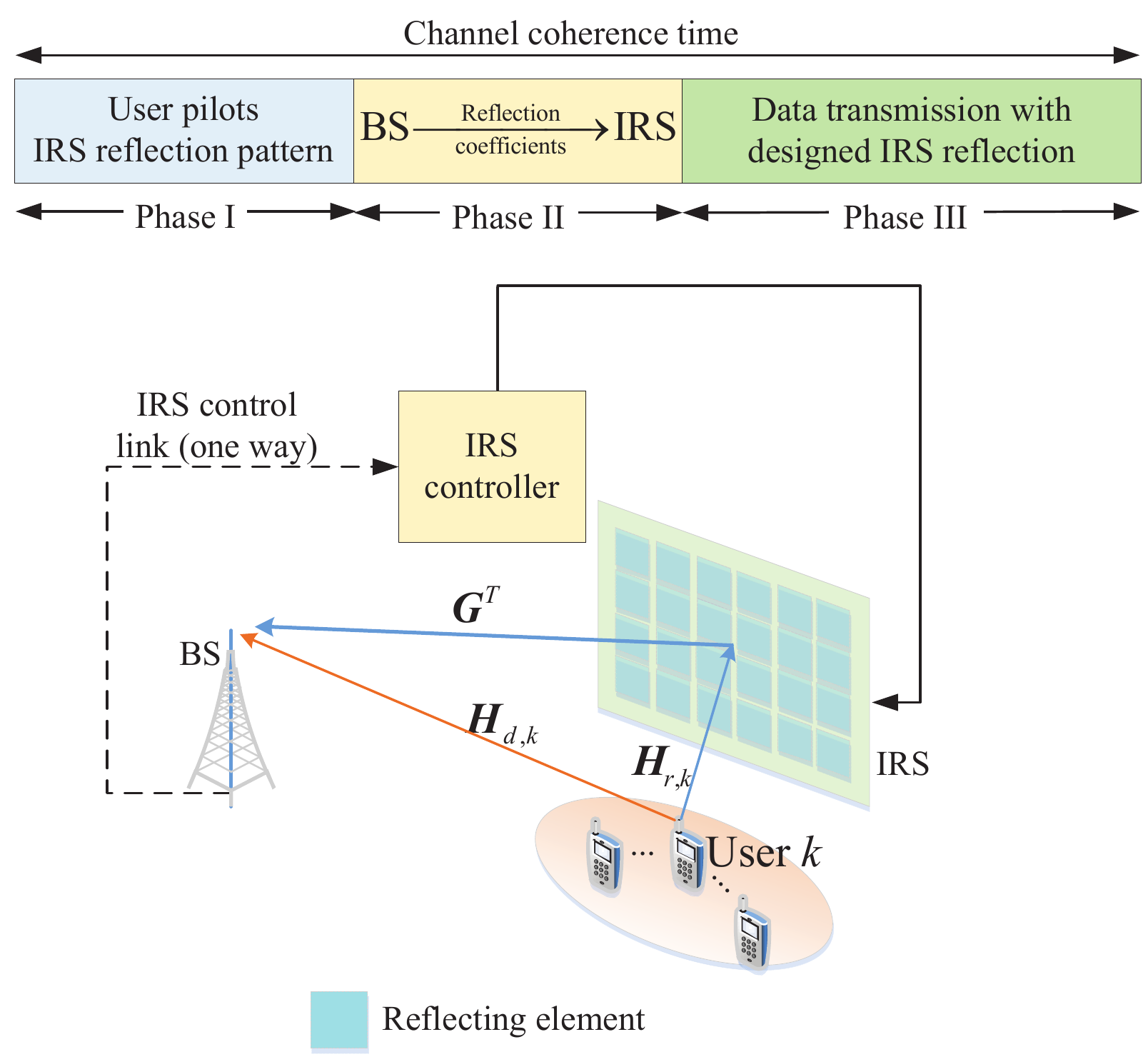}
		\end{minipage}\label{FullyIRS}
	}
	\caption{Two practical IRS configurations  and their respective transmission protocols.} \label{architecture}
\end{figure*}

Besides the substantially more IRS-induced channel coefficients as compared to the conventional system without IRS, another challenge in IRS channel estimation arises from its low-cost reflecting elements that do not possess any active RF chains and thus cannot transmit pilot/training signals to facilitate channel estimation, which is in sharp contrast to the active BSs/user terminals in conventional wireless systems.
In the existing literature, there are two main approaches for IRS channel estimation based on two different IRS configurations, depending on whether it is mounted with sensing devices (receive RF chains) or not, termed as semi-passive IRS and (fully) passive IRS, respectively, as shown in Fig.~\ref{architecture}. In the following, we present these two IRS configurations  in detail, discuss the state-of-the-art results on  IRS channel estimation based on them, respectively, and finally highlight the remaining important issues that need to be tackled in future work.

\subsection{Semi-Passive IRS Channel Estimation}\label{semi}
To endow the IRS with sensing capability for channel estimation, additional sensing devices (such as low-power sensors) need to be  integrated into IRS, e.g., interlaced with IRS reflecting elements, as shown in Fig.~\ref{architecture} (a), each equipped with a low-cost receive RF chain (e.g., low-resolution analog-to-digital converter (ADC)) for processing the sensed signal.
As such, the semi-passive IRS generally operates in one of the following two modes alternately over time:
\begin{itemize}
	\item Channel sensing mode: With all the reflecting elements turned OFF,
	the sensors are activated to receive the pilot signals from the BS/users in the downlink/uplink for estimating their respective channels to IRS;
	\item Reflection mode: With the sensors deactivated, the IRS reflecting elements are turned ON to reflect the data signals from the BS/users for enhancing the downlink/uplink communication, respectively.
\end{itemize}

Accordingly, a general transmission protocol for semi-passive IRS is illustrated in Fig.~\ref{architecture}~(a), where each channel coherence interval  is divided into three phases. In the first phase, the BS/users send their pilot signals in the downlink/uplink to estimate the BS-user direct channels as in the conventional wireless system without IRS, while the IRS operates in the channel sensing mode to estimate the CSI from the BS/users based on the signals received by its sensors. After that, in the second phase, the CSI is exchanged between the IRS controller and BS, based on which the active and passive beamforming coefficients are jointly designed at the IRS controller or BS and then sent to the other via the separate wired/wireless backhaul link between them.
Finally, in the third phase, the IRS switches to the reflection mode to assist
data transmission between the BS and users with the designed active/passive beamforming coefficients set at the BS/IRS.
It is worth pointing out that for semi-passive IRS, only the downlink/uplink CSI of BS/users$\rightarrow$IRS links can be estimated by the IRS sensors; whereas that of their corresponding reverse links can be obtained only in the TDD system by leveraging the channel reciprocity. However, such CSI is unavailable in the FDD system, thus making channel estimation in FDD systems infeasible for semi-passive IRS.

As shown in Fig.~\ref{architecture} (a), ${\bar{\bm G}}\in {\mathbb{C}^{ N_s \times M_B}} $ and ${\bar{\bm H}}_{r,k}\in {\mathbb{C}^{ N_s\times M_u}}$ denote the channels from the BS and user $k$ to the $N_s$ IRS sensors, respectively, with $N_s < N$ in general for reducing the IRS cost and energy consumption.
It should be noted that the channels from the BS/users to the IRS sensors are not identical to those from them to the IRS reflecting elements given in \eqref{signal}, while they are usually correlated due to close proximity.
As such, the essential challenge for semi-passive IRS channel estimation is how to construct
the \emph{high-dimensional} channels ${\bm G}$ and ${\bm H}_{r,k}$ in \eqref{signal}
from the estimated  CSI on the \emph{low-dimensional} channels ${\bar{\bm G}}$ and ${\bar{\bm H}}_{r,k}$.
To resolve this problem, advanced signal processing tools, such as compressed sensing, data interpolation, and machine learning, can be applied to construct the CSI of BS/users$\rightarrow$IRS links from the estimated CSI via the IRS sensors by exploiting their inherent spatial correlation.
Moreover, it is worth noting that the channel estimation accuracy for semi-passive IRS is generally limited by the number of available sensors, their finite (e.g., $1$-bit) ADC resolution, and the channel sensing (downlink/uplink training) time.
Intuitively, installing more sensors provides more channel-sensing measurements for reducing the IRS CSI construction error in general, applying higher-resolution ADCs can reduce the quantization error, and increasing the channel sensing time can help average out the sensing noise more effectively.

However, a systematic study on the fundamental limits, practical algorithms, and their cost-performance trade-offs for semi-passive IRS channel estimation is still lacking in the literature, although a handful of   preliminary works \cite{alexandropoulos2020hardware,Tahaa2019enabling,Taha2019Deep,taha2020deep,Han2020deep} on addressing some of these  aspects have recently appeared. Specifically, an AO approach using random spatial sampling and analog combining techniques was proposed in \cite{alexandropoulos2020hardware} for semi-passive IRS channel estimation under the narrow-band beamspace-based channel model, where its mean-squared error (MSE) performance was evaluated with respect to the channel sensing time.
In \cite{Tahaa2019enabling,Taha2019Deep,taha2020deep},
semi-passive IRS channel estimation algorithms based on compressed sensing and deep learning techniques were proposed for the IRS-aided SISO system, where the impact of both the number of IRS sensors and channel sensing time on the achievable rate was evaluated by simulations.
In \cite{Han2020deep}, the direction-of-angle estimation problem was investigated in the IRS-aided M-MIMO system, for which a deep neural network-based method was proposed to effectively reduce the quantization error induced by the low-resolution ADCs.

\subsection{Passive IRS Channel Estimation}\label{fully}
When there are no sensors mounted on the IRS for low-cost implementation, IRS becomes fully passive and thus it is generally infeasible to acquire the CSI between the IRS and BS/users directly.
In this (perhaps more practical yet challenging) case, an alternative approach is to estimate the
cascaded user-IRS-BS channels at the BS/users in the uplink/downlink, respectively. Note that this approach, unlike the case of semi-passive IRS, applies to both TDD and FDD systems, although in the TDD case, only the uplink or downlink cascaded channels need to be estimated thanks to the uplink-downlink channel reciprocity.
Without loss of generality, we consider the IRS-aided uplink multi-user MIMO system given in \eqref{signal}, where the cascaded user-IRS-BS channel is the transposed Khatri-Rao product of the user-IRS and IRS-BS channels, with each channel coefficient given by $\left[{\bm G}\right]_{n,m_B} \left[{\bm H}_{r,k}\right]_{n,m_u}, \forall n=1,\ldots,N, m_B=1,\ldots,M_B, m_u=1,\ldots,M_u$, and $ k=1,\ldots,K$.
As such, the total number of channel coefficients for the cascaded user-IRS-BS links is $K\times N M_B M_u$, which is in general (much) larger than that of channel coefficients for the separate user-IRS and IRS-BS links (equal to $K\times{N M_u } +{M_B N }$). This indicates that there is an inherent redundancy in the cascaded channels, which is due to the fact that all users' cascaded channels involve the common IRS-BS channel ${\bm G}^T$, which, however, is difficult to be resolved from any user's estimated cascaded channel. On the other hand, it is worth noting that from the perspective of designing the joint passive/active beamforming, the CSI on the cascaded links is generally sufficient and without loss of optimality as compared to that on their corresponding  links, as shown in Section III.
Moreover, it is worth pointing out that different from the semi-passive IRS case where the IRS channels need to be constructed from the estimated CSI via IRS sensors, the cascaded user-IRS-BS channels in the case of passive IRS can be directly estimated at the BS/users without the need of sophisticated channel reconstruction.

In Fig.~\ref{architecture} (b), we illustrate the transmission protocol for the uplink with the passive IRS, where each channel coherence interval  is divided into three consecutive phases.
First, users transmit orthogonal pilots to the BS and meanwhile the IRS varies its reflection coefficients according to a pre-designed reflection pattern, based on which the BS estimates both the user-BS direct channels and the cascaded user-IRS-BS channels.
Second, based on the estimated CSI, the IRS reflection coefficients for data transmission are designed at the BS jointly with its receive beamforming and then sent to the IRS controller through the backhaul link.
Third, the IRS controller sets the reflection coefficients accordingly for assisting independent data transmissions from the users to BS.
Note that for TDD system, the estimated uplink CSI can also be used to design IRS reflection for data transmissions in the downlink from the BS to users; while for FDD system, the transmission protocol in Fig. \ref{architecture} (b) is still applicable for the downlink, with the only modification that the roles of the BS and users are swapped in the first phase, i.e., the direct/cascaded channels are estimated at the users based on the pilot signals sent by the BS, then the users need to feed back their estimated CSI to the BS for the joint optimization of IRS reflection and BS transmit beamforming.

For the (uplink) passive IRS channel estimation, the key problem is the joint design of the pilot sequences, IRS reflection pattern, and signal processing algorithm at the receiver to accurately estimate both the direct user-BS and cascaded user-IRS-BS channels with minimum training overhead (number of pilot symbols). This problem has not been addressed before in channel estimation for conventional wireless systems without IRS and thus is new and also non-trivial to solve in general. For the purpose of exposition, we consider first the simple IRS-aided single-user system (i.e., $K=1$) with flat-fading channels, where both the BS and user are equipped with one single antenna (i.e., the SISO case with $M_B=M_u=1$).
In this case, one practical method for IRS channel estimation is by employing an ON/OFF-based IRS reflection pattern \cite{Mishra2019Channel}, i.e., each one of the IRS elements is turned ON sequentially with the others set OFF at each time, thereby the user-BS direct channel and the cascaded channels associated with different IRS elements are estimated separately. Note that this method requires at least $N+1$ pilot symbols for estimating the total $N+1$ channel coefficients in this system. Albeit being simple to implement, the ON/OFF-based IRS reflection pattern incurs substantial reflection power loss as only
one element is switched ON at each time and thereby the reflected signal is rather weak.
To overcome this power loss issue and improve the channel estimation accuracy,
the all-ON IRS reflection pattern can be employed with orthogonal reflection coefficients over time, such as those drawn from different columns of an $(N+1)\times (N+1)$ DFT matrix  \cite{OFDM_BX,jensen2019optimal}, whereby all the $N+1$ channel coefficients can be estimated  over $N+1$ pilot symbol durations.
Moreover,
to reduce the training overhead for IRS with practically large number of reflecting elements as well as simplify IRS reflection design for data transmission, an efficient approach is to group adjacent IRS elements (over which the IRS channels are usually spatially correlated) into a sub-surface \cite{OFDM_BX,OFDM_Protocol}, referred to as {\it IRS element grouping}; as a result, only the effective cascaded user-IRS-BS channel associated with each sub-surface needs to be estimated, thus greatly reducing the training overhead. In addition, based on the estimated effective cascaded channels for the sub-surfaces, their reflection coefficients (each set identical for all the IRS elements in the same sub-surface) can be optimized more efficiently for data transmission. Thus, the element grouping strategy provides a flexible trade-off between training/design overhead/complexity  and IRS passive beamforming gain in practice \cite{OFDM_BX,OFDM_Protocol}.

For the passive IRS-aided single-user MIMO/MISO system (i.e., $M_B>1$ and $M_u\ge1$),
it is required to estimate more channel coefficients due to the increased channel dimensions and the matrix multiplication of the user-IRS and IRS-BS channels in the cascaded user-IRS-BS channel. By adopting the orthogonal pilots over different transmit antennas at the BS/user for the downlink/uplink training, the total training overhead is increased by $M_B$ or $M_u$ times as compared to the single-user SISO case, which can be practically prohibitive if the number of transmit antennas $M_B$ or $M_u$ is large.
To tackle this difficulty, certain IRS channel properties (such as low-rank, sparsity, and spatial correlation) can be exploited to facilitate the cascaded channel decomposition as well as reduce the training overhead \cite{wang2019compressed,he2019cascaded,de2020parafac,mirza2019channel,he2020channel}.\footnote{The cascaded channel decomposition exploiting certain IRS channel properties can also be readily applied to the multi-user case by treating multiple users as one equivalent user with $K M_u$ transmit antennas in the uplink or estimating the channels of different users in parallel in the downlink \cite{chen2019channel,hu2019two,liu2019matrix,wei2020parallel,wan2020broadband}.}
Furthermore, to accelerate the training process, deep learning and hierarchical searching algorithms have also been developed for the channel estimation in passive IRS-aided MIMO/MISO systems \cite{khan2019deep,elbir2020deep,ning2019channel,ning2019channelMU}.




For channel estimation with passive IRS serving multiple users (i.e., $K>1$),
a straightforward method is by adopting the single-user channel estimation design to estimate the channels of $K$ users separately over consecutive time \cite{nadeem2019intelligent}, which, however, increases the total training overhead by $K$ times as compared to the single-user case and thus is practically prohibitive if $K$ is very large.
Recall that all the users in fact share the same (common) IRS-BS channel ${\bm G}^T$ in \eqref{signal} in their respective cascaded user-IRS-BS channels. By exploiting this fact, the training overhead for IRS channel estimation in the multi-user case can be significantly reduced. For example, a user can be selected as the reference user of which the cascaded channel is first estimated. Then, based on this reference CSI, the cascaded channels of the remaining $K-1$ users can be efficiently estimated by exploiting the fact that
these cascaded channels are scaled versions of the reference user's cascaded channel and thus only the low-dimensional
scaling factors, rather than the whole high-dimensional cascaded channels, need to be estimated.
In particular, for the IRS-aided  multi-user MISO system (i.e., $M_u=1$), it was shown in \cite{wang2019channel} that the minimum uplink training overhead is $K+N+\max\left(K-1,\left\lceil \frac{(K-1)N}{M_B}\right\rceil\right)$, which decreases with an increasing $M_B$.
Note that this result exploits the redundancy of receive antennas at the BS for training overhead reduction and thus is in sharp contrast to the conventional multi-user channel estimation without IRS for which the minimum training overhead is independent of the number of receive antennas at the BS.


For broadband systems with frequency-selective fading channels, it is necessary to estimate more channel coefficients due to the multi-path delay spread and the resultant convolution of the user-IRS and IRS-BS multi-path channels in each cascaded user-IRS-BS channel. Moreover, although the channels are frequency-selective, the IRS reflection coefficients are frequency-flat (as shown in Section III), which thus cannot be flexibly designed for different frequencies (e.g., different sub-carriers in OFDM-based systems). Due to the above reasons, the IRS channel estimation for broadband frequency-selective fading channels is even more challenging as compared to that for narrow-band flat-fading channels, under all the SISO/MISO/MIMO as well as single-user/multi-user setups.
Fortunately, since the number of OFDM sub-carriers is typically much larger than the maximum number of delayed paths in practical systems, there exists great redundancy for channel estimation, which can be exploited for designing OFDM-based pilot symbols to efficiently estimate the channels of multiple users at the same time \cite{zheng2020intelligent}. In addition, the previously discussed techniques for narrow-band IRS channel estimation such as DFT-based IRS reflection pattern, IRS element grouping, compressed sensing, IRS-BS common channel exploitation, etc., can be jointly applied with customized OFDM-based pilots to achieve efficient broadband IRS channel estimation. For example, by exploiting the common IRS-BS channel and LoS dominant user-IRS channels, it was shown in \cite{zheng2020intelligent} that only $N+1$ OFDM symbols are required for estimating the cascaded channels of up to $\left\lfloor \frac{(N+1)(Q-L)}{N+L}\right\rfloor+1$ users at the same time, with $Q$ and $L$ denoting the number of OFDM sub-carriers and the maximum number of delayed paths, respectively.
Note that different from \cite{wang2019channel} that exploits the redundancy of receive antennas at the BS for training overhead reduction, \cite{zheng2020intelligent} exploits the redundancy of OFDM sub-carriers to support more users for concurrent channel estimation and thus improve the training efficiency, which is applicable to all the SISO/MISO/MIMO setups with training overhead independent of the number of receive antennas at the BS.
Therefore, it is also possible to fully exploit the redundancy of both receive antennas and OFDM sub-carriers to further improve the training efficiency for the general IRS-aided broadband multi-user MIMO system.

Most of the existing works (e.g., \cite{jensen2019optimal,wang2019compressed,he2019cascaded,de2020parafac,mirza2019channel,he2020channel}) on passive IRS channel estimation have assumed continuous phase shifts of IRS reflecting elements. However, as shown in Section II, the number of phase-shift levels in practice needs to be finite considering the hardware implementation.
This thus gives rise to another issue on how to design orthogonal/near-orthogonal  IRS reflection coefficients over time with discrete phase-shift levels, which is not an issue  in the case of continuous phase shifts with the DFT-based orthogonal training reflection design.
One efficient way for tackling  this problem  is to construct near-orthogonal reflection coefficients using proper quantization techniques \cite{you2019intelligent,you2019progressive}.
However, the resultant
channel estimation errors are generally higher than the case with orthogonal reflection coefficients, and also correlated over the reflecting elements. As a result, their induced  interference is correlated in general  during data transmission, which  needs to be properly taken into account in the design of IRS passive/reflect beamforming for data transmission \cite{you2019intelligent,you2019progressive}. Moreover, existing IRS channel estimation works have mostly focused on the design of ``all-at-once" channel estimation, i.e., all the cascaded channel coefficients of all IRS reflecting elements are estimated at once. However, this requires long training time which may not be practically available in some scenarios such as short-packet transmissions with insufficient pilot length and/or fast fading channels with short coherence time. Thus, it is desirable to develop flexible training protocols with adjustable pilot durations to adapt to different  transmission requirement and channel coherence time.
In \cite{you2019intelligent}, a novel hierarchical training reflection pattern was proposed to progressively resolve the cascaded CSI of the IRS reflecting elements based on the element grouping concept, which also leads to successively refined passive beamforming designs for data transmission that effectively enhance the system rate performance in a block-by-block manner.

\begin{figure}[!t]
	\centering
	\includegraphics[width=3.5in]{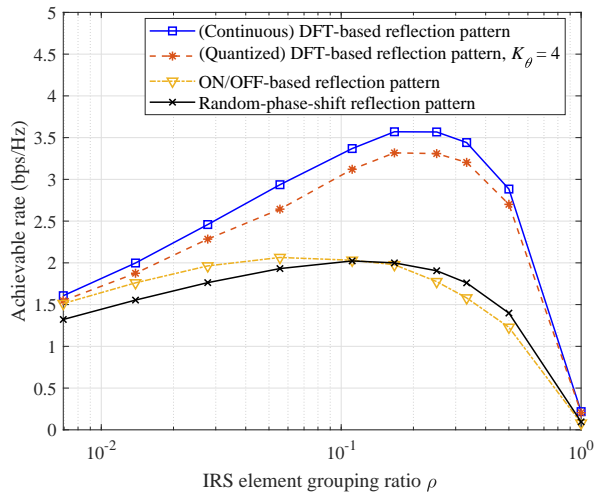}
	\caption{Achievable rate versus IRS element grouping ratio $\rho={\bar N}/N$ with $N=12\times 12=144$, SNR $=5$ dB, and channel coherence time $T_c=150$ channel uses.}
	\label{tutorial_result}
\end{figure}
It is worth noting that for IRS channel estimation under different setups, there exists a general trade-off between the training overhead for channel estimation and passive beamforming gain for data transmission.
Specifically, too little training leads to inaccurate/insufficient CSI which results in inefficient reflection design and thus degrades the passive beamforming gain for data transmission, while
too much training renders less time for data transmission,
both causing reduced achievable rate.
To show this fundamental trade-off in IRS channel estimation, Fig.~\ref{tutorial_result} plots the achievable rate versus the IRS element grouping ratio $\rho={\bar N}/N$ with ${\bar N}$ denoting the number of IRS sub-surfaces (which is proportional to the training overhead) for a passive IRS-aided single-user narrow-band system with different channel estimation schemes.
It is observed that given a channel coherence time, the achievable rate with each scheme first increases
with the element grouping ratio, due to the enhanced passive beamforming gain as a result of more accurate CSI,
while it then decreases as the element grouping ratio exceeds a certain value,
due to the reduced time for data transmission.
Moreover, compared to the optimal DFT-based orthogonal  IRS reflection pattern with continuously adjustable phase shift at each reflecting element, some performance loss is observed for the IRS reflection pattern design under the practical constraint of discrete phase shifts with $K_\theta$ uniformly quantized levels in $[0, 2\pi)$ \cite{you2019intelligent,you2019progressive}.
On the other hand, both DFT-based reflection patterns (with continuous/discrete phase shifts) achieve much larger rates than the ON/OFF-based reflection pattern that does not fully utilize the large aperture
of IRS as well as the heuristic  random-phase-shift reflection pattern where the IRS independently generates multiple sets of reflection coefficients and the set that achieves the largest effective channel gain is selected for data transmission.

\begin{table*}[!t]
	\begin{center}\caption{A Summary of Representative Works on IRS Channel Estimation}\label{Table_est}
		\resizebox{\textwidth}{!}{%
			\begin{tabular}{|c|c|c|l|}
				\hline
				\bf{IRS configuration} &
				\bf{System setup} &
				\begin{tabular}[c]{@{}c@{}}\bf{Phase-shift}\\  \bf{model}\end{tabular} &
				\qquad\qquad\qquad\qquad\qquad\qquad\qquad\bf{Representative work and main contributions}  \\ \hline
				\multirow{2}{*}{Semi-passive IRS} &
				\begin{tabular}[c]{@{}c@{}}Single-user, \\ narrow-band\end{tabular} &
				- &
				\begin{tabular}[c]{@{}l@{}}Alternating optimization approach for channel estimation using random spatial sampling and analog combining \cite{alexandropoulos2020hardware}\\ Direction-of-angle estimation based on deep neural network for the IRS-aided M-MIMO system \cite{Han2020deep}\end{tabular} \\ \cline{2-4}
				&
				\begin{tabular}[c]{@{}c@{}}Single-user, \\ broadband\end{tabular} &
				- &
				Channel estimation based on compressed sensing and deep learning for the IRS-aided SISO system \cite{Tahaa2019enabling,Taha2019Deep,taha2020deep} \\ \hline
				\multirow{12}{*}{(Fully) Passive IRS} &
				\multirow{2}{*}{\begin{tabular}[c]{@{}c@{}}Single-user, \\ narrow-band\end{tabular}} &
				Continuous &
				\begin{tabular}[c]{@{}l@{}}DFT-based IRS reflection for the IRS-aided SISO system \cite{jensen2019optimal}\\ Compressed-sensing-based channel estimation with the cascaded channel sparse representation \cite{wang2019compressed}\\ Cascaded channel decomposition/factorization and matrix completion for the IRS-aided MIMO system \cite{he2019cascaded,de2020parafac}\\ Iterative algorithms for channel estimation of the IRS-assisted MIMO system \cite{mirza2019channel,he2020channel}
					\\ Beam training with hierarchical search codebook design for the IRS-aided MIMO system \cite{ning2019channel}
					\\ Deep-learning-aided channel acquisition for the IRS-aided MISO system  \cite{khan2019deep}\end{tabular} \\ \cline{3-4}
				&
				&
				Discrete &
				\begin{tabular}[c]{@{}l@{}}ON/OFF-based IRS reflection for the IRS-aided SISO system \cite{Mishra2019Channel}\\ Progressive channel estimation with discrete phase shifts for the IRS-aided SISO system \cite{you2019intelligent,you2019progressive}\end{tabular} \\ \cline{2-4}
				&
				\multirow{2}{*}{\begin{tabular}[c]{@{}c@{}}Single-user, \\ broadband\end{tabular}} &
				Continuous &
				DFT-based IRS reflection with IRS element grouping for the IRS-aided SISO system \cite{OFDM_BX} \\ \cline{3-4}
				&
				&
				Discrete &
				ON/OFF-based IRS reflection with IRS element grouping for the IRS-aided SISO system \cite{OFDM_Protocol} \\ \cline{2-4}
				&
				\multirow{2}{*}{\begin{tabular}[c]{@{}c@{}}Multi-user, \\ narrow-band\end{tabular}} &
				Continuous &
				\begin{tabular}[c]{@{}l@{}}Reference-user-based channel estimation by exploiting the common IRS-BS channel for the IRS-aided MISO system \cite{wang2019channel}\\ Compressed-sensing-based channel estimation with the common channel sparse representation \cite{chen2019channel}\\ Dual-link pilot transmission scheme with a full-duplex BS for the IRS-aided MISO system \cite{hu2019two}\\ Cascaded channel decomposition/factorization for the IRS-aided MIMO/MISO system \cite{liu2019matrix,wei2020parallel} 		
					\\ Beam training with hierarchical search codebook design for the IRS-aided MIMO system \cite{ning2019channelMU}		
					\\ Deep-learning-aided downlink channel acquisition for the IRS-aided MISO system \cite{elbir2020deep}\end{tabular} \\ \cline{3-4}
				&
				&
				Discrete &
				User-by-user successive ON/OFF-based IRS reflection for the IRS-aided SISO system \cite{nadeem2019intelligent} \\ \cline{2-4}
				&
				\begin{tabular}[c]{@{}c@{}}Multi-user, \\ broadband\end{tabular} &
				\begin{tabular}[c]{@{}c@{}}Continuous\end{tabular} &
				\begin{tabular}[c]{@{}l@{}}Reference-user-based channel estimation by exploiting the common IRS-BS channel and optimal training design \cite{zheng2020intelligent}\\ Compressed-sensing-based downlink channel estimation for the IRS-aided MIMO system \cite{wan2020broadband}\end{tabular} \\ \hline
			\end{tabular}
		}
	\end{center}
\end{table*}

\subsection{Other Related Work and Future Direction}
In Table~\ref{Table_est}, we summarize the representative works on IRS channel estimation according to their considered IRS configurations  and system setups.
It is noted that there has been very limited work on semi-passive IRS channel estimation, and it is still unclear how this approach is compared with the passive IRS channel estimation in terms of performance and cost. In addition, most of the existing works consider narrow-band channel estimation, while channel estimation for the more prevalent broadband communication requires more investigation. Furthermore, most of the works assume TDD systems and/or continuous phase shifts, while channel estimation for FDD systems (e.g., downlink channel training with passive IRS) and with discrete phase shifts is also practically important as well as  challenging to design, which deserves further studies in the future.

\section{IRS Deployment}
In the preceding sections, we have shown the effectiveness of IRS in serving as a local hub to improve the communication performance of nearby users via optimized signal reflection, where the IRS is assumed to be deployed at a given location. However, with a total number of reflecting elements, there are various IRS deployment strategies by placing these elements at different locations, e.g., near the BS/AP/users, or dividing them into smaller-size IRSs that are deployed in the network in a distributed manner. Note that the deployment strategy for IRS reflecting elements has a significant impact on the realizations/distributions of all IRS-reflected channels in an IRS-aided system and hence its fundamental performance limit. On the other hand, from the implementation perspective, IRS deployment also needs to take into account various practical factors such as deployment/operational cost, user demand/distribution, space constraint, as well as the propagation environment.

It is worth noting that the deployment strategy for IRSs is generally different from that for active communication nodes such as BSs/APs or relays, which  has been thoroughly investigated in the literature (see, e.g., \cite{BS_Shen,BS_Youssef,Relay_Yu,Relay_Tapolcai}), due to the following reasons. Firstly, since IRS-reflected channels suffer the severe product-distance/double path-loss due to IRS's passive reflection  without signal amplification/regeneration, IRS needs to be placed in close vicinity to the transmitter or receiver to minimize the path-loss. This is in sharp contrast to the deployment of active BSs/APs that need to be geographically well separated for coverage maximization \cite{BS_Shen,BS_Youssef}, or active relays with signal processing and amplification capabilities that are usually deployed in the middle of the transmitter and receiver for balancing the SNRs of the two-hop links \cite{Relay_Yu,Relay_Tapolcai}. Secondly, thanks to the significantly lower cost of IRSs as compared to active BSs/APs/relays, they can be much more densely deployed in the network so as to effectively alter the signal propagation in the network. However, this results in a much larger-scale deployment optimization problem with drastically increased complexity to solve. Thirdly, one challenging issue in deploying active communication nodes arises from their mutual interference, which greatly complicates the deployment optimization problem. In contrast, since IRSs are passive, their reflected signals decay in power rapidly over distance; as a result, as long as IRSs are deployed sufficiently far apart from each other, their mutual interference is practically negligible, which greatly simplifies their deployment design.

Motived by the above, in this section, we study the new IRS deployment problem to draw essential and useful insights for practical design. For the purpose of exposition, we consider the basic single-/multi-user communication system setups and focus on characterizing their maximum achievable rates (or capacity/capacity region) by optimizing their corresponding IRS deployment. In particular, we divide our discussion into two subsections, which address  the \emph{link-level/single-user} design and \emph{network-level/multi-user} design, respectively. Moreover, we discuss related works on IRS deployment under other/more general setups, as well as promising directions for future work.

\begin{figure}[t]
	\centering
	\subfigure[Single IRS]{
		\includegraphics[width=7cm]{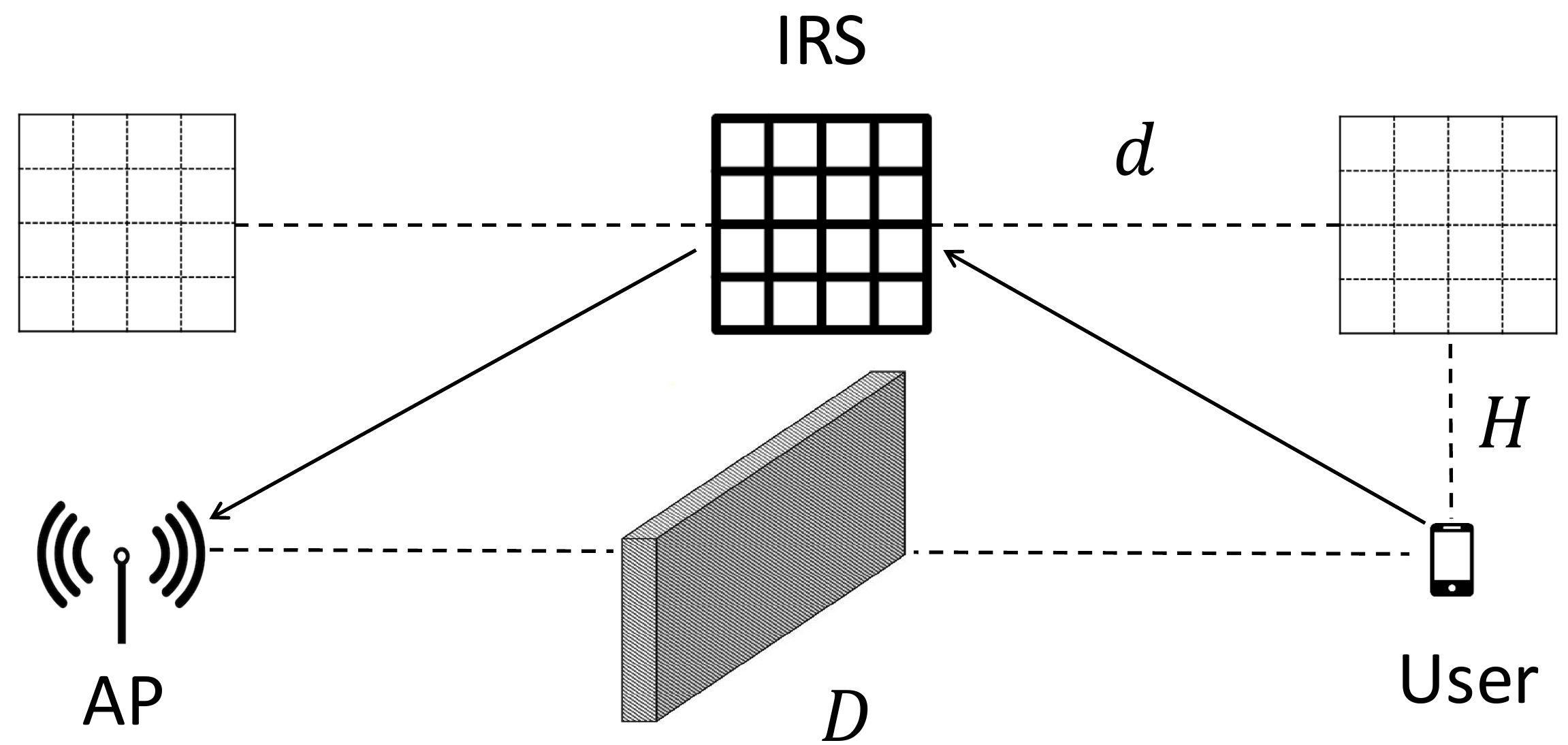}}
	\subfigure[Two cooperative IRSs]{
		\includegraphics[width=7cm]{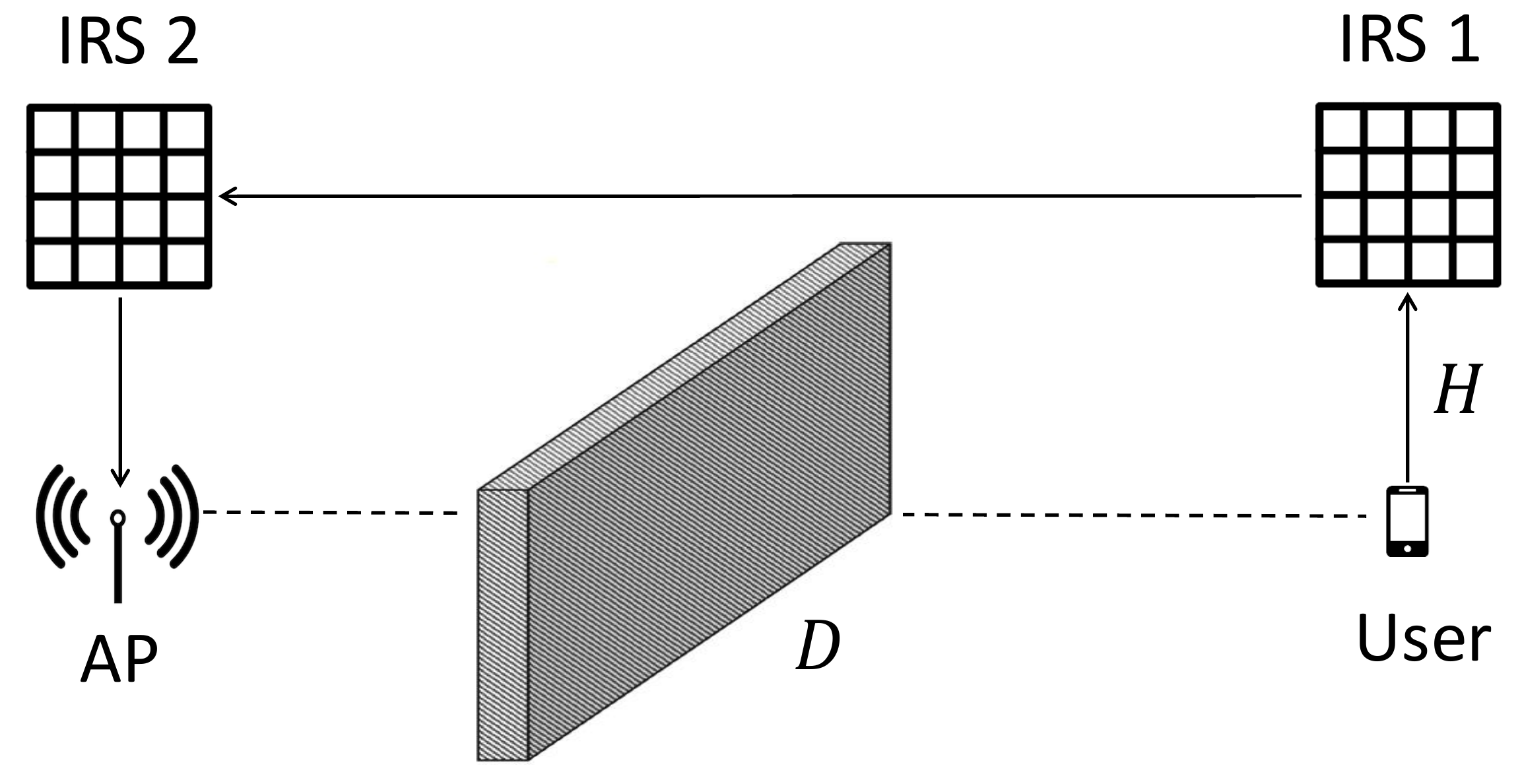}}
	
	\vspace{-1mm}
	
	\caption{IRS deployment in a point-to-point communication system.}\label{P2P}
	\vspace{0mm}
\end{figure}

\subsection{IRS Deployment Optimization at the Link Level}
First, we consider a point-to-point communication link aided by IRS and discuss three key aspects of IRS deployment for enhancing its performance at the link level in this subsection. For the purpose of exposition, we consider the deployment for a total of $N$ IRS reflecting elements to assist the communication from a single user to an AP, as illustrated in Fig. \ref{P2P}.\footnote{Note that the discussions in this subsection are also applicable to the communication from the AP to user, as well as that between the AP and a group of users located close to each other, where the IRS deployment influences the channels of all users  in the same cluster similarly.}

\begin{figure}[t]
	\centering
	\includegraphics[width=7cm]{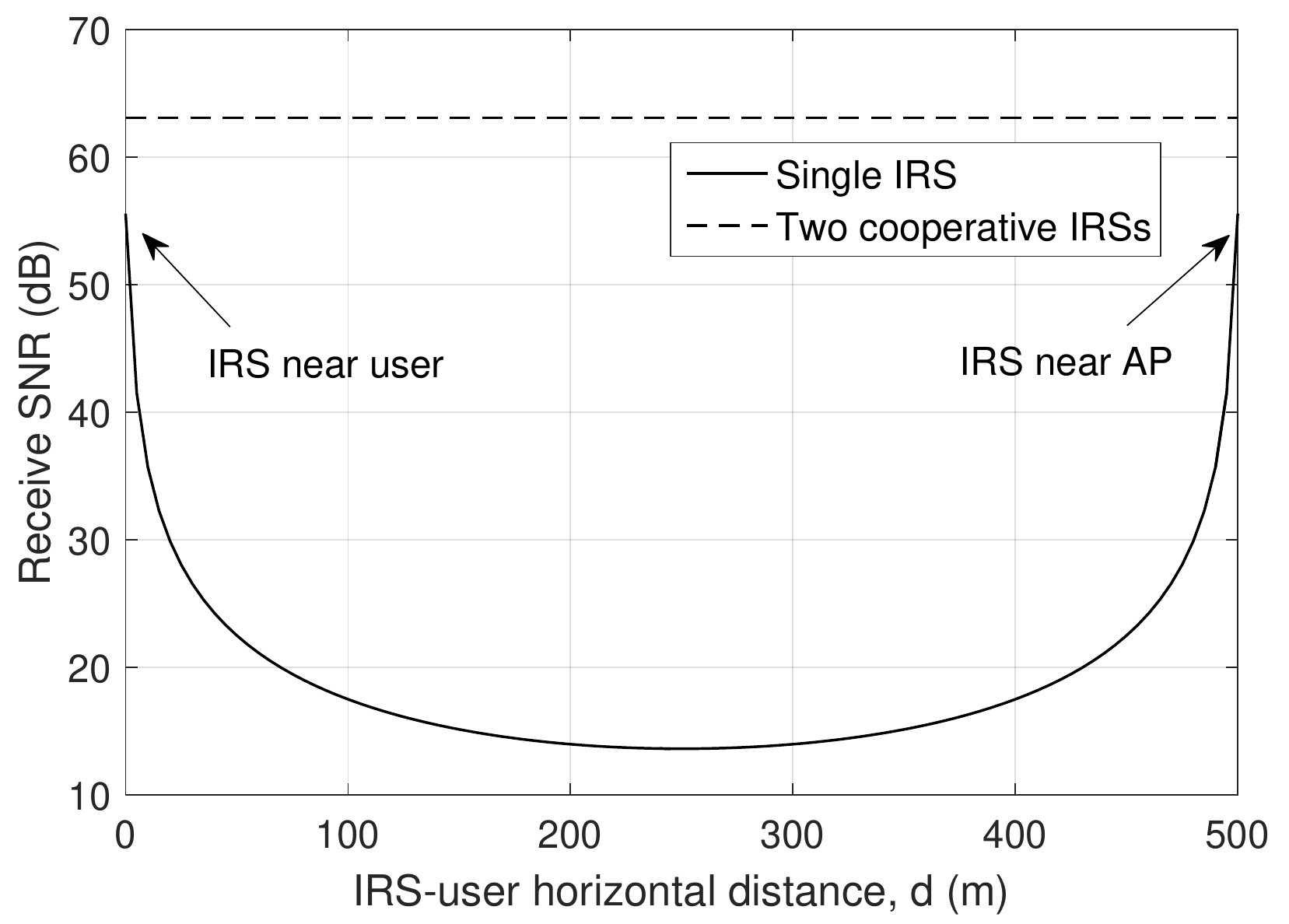}
	\vspace{-3mm}
	\caption{Receive SNR versus IRS-user horizontal distance.}\label{distance}
	\vspace{0mm}
\end{figure}

\subsubsection{Optimal Deployment of Single IRS}\label{sec_single}
To start with, we consider the simplest case where all the $N$ reflecting elements form one \emph{single}  IRS, as illustrated in Fig. \ref{P2P} (a). Moreover, we assume that both the AP and user are equipped with a single antenna, thus the achievable rate is solely determined by the receive SNR at the user. To focus our analysis on the effect of IRS location, we consider the scenario where the direct user-AP link is blocked, and the other two links to/from the IRS follow the free-space LoS channel model.

Note that despite the $\mathcal{O}(N^2)$ passive beamforming gain brought by a single IRS as shown in Section III, the received signal power at the user suffers the \emph{double path-loss} proportional to the \emph{product} of the  distances of the user-IRS and IRS-AP links. To illustrate this effect, we consider a simplified 2D system setup shown in Fig. \ref{P2P} (a) where the AP and user are located on a line with horizontal distance $D$ m, and the IRS can be flexibly deployed on a line above the AP and user by $H$ m with $H\ll D$. In this case, the receive SNR with the optimal IRS passive beamforming (see Section III-A) is given by
\begin{align}\label{rhoS}
\rho_{\mathrm{S}}=\frac{P\beta_0^2 N^2}{(d^2+H^2)((D-d)^2+H^2)\sigma^2},
\end{align}
where $P$ denotes the transmit power; $\beta_0$ denotes the path-loss at reference distance of $1$ m; $d$ denotes the horizontal distance between the IRS and user; and $\sigma^2$ denotes the average receiver noise power. Note that the user-IRS distance $\sqrt{d^2+H^2}$ increases while the IRS-AP distance $\sqrt{(D-d)^2+H^2}$ decreases as $d$ increases from $0$ to $D$; thus it can be easily shown from (\ref{rhoS}) that $\rho_{\mathrm{S}}$ is maximized {when the distance of the user-IRS or IRS-AP link is minimized}, i.e., when the IRS is placed right above the user ($d=0$) or the AP ($d=D$). With this optimal deployment strategy, the maximum receive SNR is given by  $\rho_{\mathrm{S}}^\star=\frac{P\beta_0^2 N^2}{H^2(D^2+H^2)\sigma^2}\approx \frac{P\beta_0^2 N^2}{H^2D^2\sigma^2}$ as $H\ll D$. For illustration, we show the receive SNR $\rho_{\mathrm{S}}$ versus the IRS-user horizontal distance $d$ in Fig. \ref{distance}, under a setup with $N=300$, $D=500$ m, $H=1$ m, $P=30$ dBm, $\sigma^2=-90$ dBm, and $\beta_0=-30$ dB. It is observed that placing the IRS near the user/AP yields the largest SNR, which is consistent with the above result; while placing it around the middle between the user and AP (usually optimal in the case of an active relay instead of IRS) leads to the smallest SNR.

\subsubsection{Single IRS versus Multiple Cooperative IRSs}
With a given number of IRS reflecting elements, forming them as one single IRS as discussed above is not the only strategy. Generally speaking, the IRS reflecting elements can form \emph{multiple smaller-size}  IRSs (e.g., as illustrated in Fig. \ref{P2P} (b) for the case of two IRSs), which, however, has both pros and cons as compared to the single-IRS deployment. On one hand, the increased number of IRSs in the user-AP link results in more inter-IRS reflections, thus causing even higher path-loss; while on the other hand, multiple IRSs open up the opportunity of harvesting larger multiplicative beamforming gains by performing \emph{cooperative passive beamforming} over them. Hence, it is not straightforward to conclude whether splitting the reflecting elements into multiple cooperative IRSs is advantageous over combining them as a single large IRS or not.

To answer this question, \cite{DoubleIRS} made the initial attempt by considering the scenario where two IRSs each with $N/2$ reflecting elements are deployed right above the user and the AP, respectively, which is illustrated in Fig. \ref{P2P} (b), under the same AP/user setup as in Fig. \ref{P2P} (a); moreover, all the links are assumed to be blocked except for the user-IRS 1-IRS 2-AP link, with all the involved channels following the free-space LoS model. It was shown in \cite{DoubleIRS} that as long as the distance between IRS 1 and IRS 2 satisfies a far-field condition such that the inter-IRS channel is of rank one, a passive beamforming gain of order $\mathcal{O}((N/2)^4)$ can be achieved by properly aligning the passive beamforming directions of the two IRSs. As a result, the receive SNR at the AP is given by \cite{DoubleIRS}
\begin{align}\label{rhoD}
\rho_{\mathrm{D}}=\frac{P\beta_0^3 N^4}{16H^4D^2\sigma^2}.
\end{align}
Based on (\ref{rhoD}), it can be shown that the receive SNR with two cooperative IRSs is larger than that with a single IRS (i.e., $\rho_{\mathrm{D}}> \rho_{\mathrm{S}}^\star$) if the total number of reflecting elements is larger than a threshold given by
\begin{align}\label{condition_N}
N>\frac{4H}{\sqrt{\beta_0}}.
\end{align}
This result is expected since the beamforming gain by two cooperative IRSs increases with $N$ more significantly as compared to that of one single IRS (i.e., in the order of $\mathcal{O}((N/2)^4)$ versus $\mathcal{O}(N^2)$), while the additional path-loss in the two-IRS case is fixed with given $H$. For comparison, we also show in Fig. \ref{distance} the receive SNR with two cooperative IRSs deployed as in Fig. \ref{P2P} (b). It is observed that the two-IRS case achieves substantial SNR gain as compared to the maximum SNR in the single-IRS case. Moreover, according to (\ref{rhoD}), this gain is anticipated to be more pronounced as $N$ increases, which makes it more appealing for practical systems with large $N$. Thus, it is worth exploiting the partition of the available reflecting elements into multiple cooperative IRSs for practical deployment.

\subsubsection{LoS versus Non-LoS (NLoS)}\label{sec_LoS}
As shown in the above, one main design objective for IRS deployment with single-antenna AP and user is to minimize their end-to-end path-loss via IRS reflection(s) and thus achieve the maximum receive SNR. Besides the link product-distance, the path-loss is also critically dependent on the \emph{path-loss exponent}, which characterizes the average channel attenuation over distance. Therefore, it is practically desirable to place IRS in LoS conditions with AP/user as well as other cooperative IRSs to attain the smallest (free-space) path-loss exponent (i.e., $2$) that leads to the minimum path-loss under a given link distance. This is particularly important in mmWave/THz  communications since the NLoS paths are usually much weaker than the LoS path due to severe signal blockage and penetration losses \cite{Infocom_Sun}.

However, on the other hand, LoS channels are typically of low-rank, which limits the achievable \emph{spatial multiplexing gains} of IRS-aided multi-antenna/multi-user systems with multi-stream or SDMA transmissions and thus results in their low capacity at the high-SNR regime \cite{MIMO,JR:wu2018IRS}. While deploying IRSs in locations with comparable NLoS (random) and LoS (deterministic) channel components with AP/users/other cooperating IRSs can be beneficial to improve the end-to-end MIMO channel's spatial power distribution (in terms of rank, condition number, etc.) and hence its capacity. Therefore, there is in general an interesting trade-off in IRS deployment for choosing IRS locations to achieve balanced LoS versus NLoS propagation as well as passive beamforming versus spatial multiplexing gains, so as to maximize the system capacity.

\subsection{IRS Deployment at the Network Level: Distributed or Centralized?}
\begin{figure}[t]
	\centering
	\subfigure[Distributed IRS deployment]{
		\includegraphics[width=6cm]{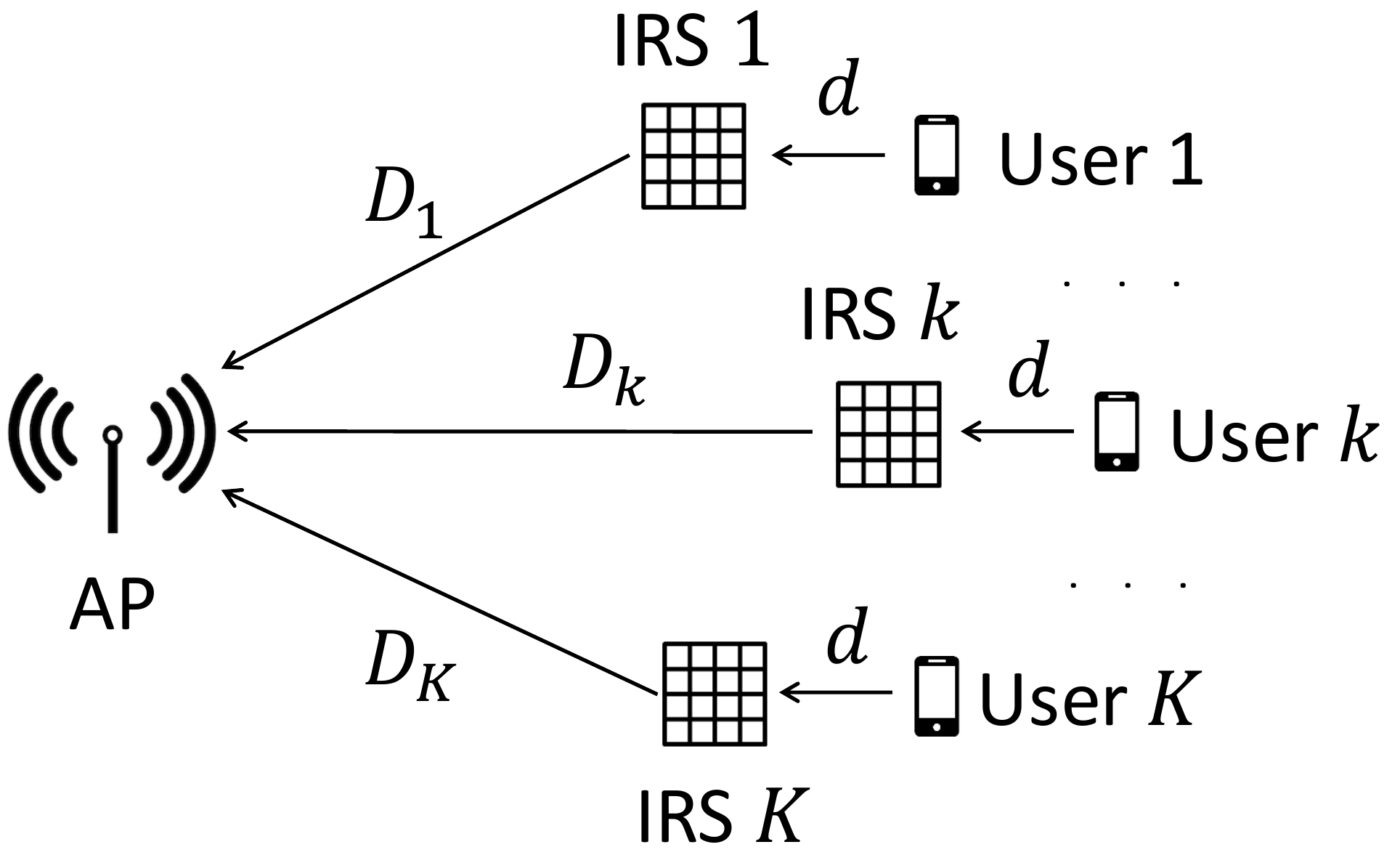}}
	\subfigure[Centralized IRS deployment]{
		\includegraphics[width=6cm]{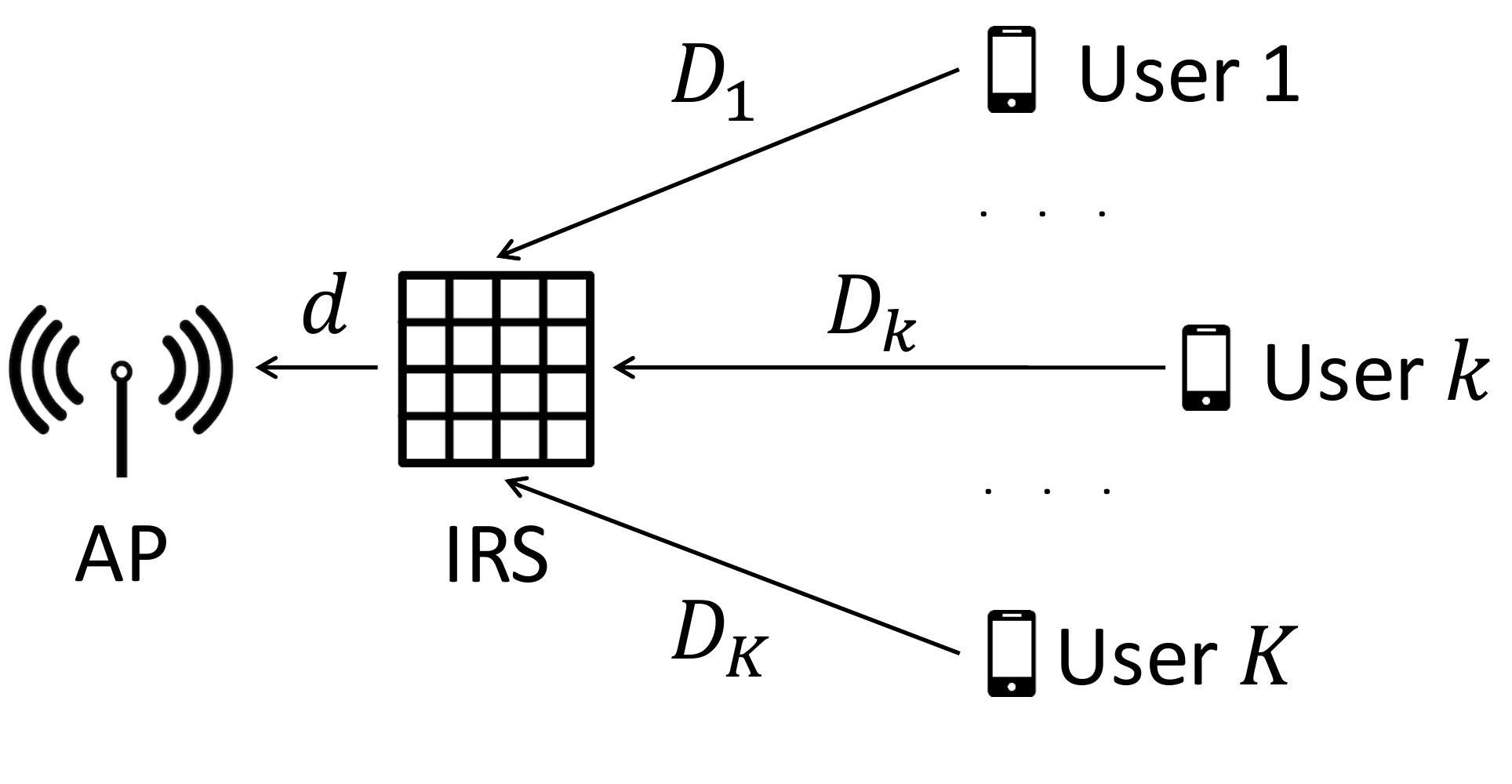}}
	
	\vspace{-2mm}
	
	\caption{IRS deployment in a multi-user communication network.}\label{Fig_System}
	\vspace{-6mm}
\end{figure}

Next, we address the IRS deployment for the more general multi-user network with one AP communicating with $K> 1$ users (or $K$ groups of nearby users) that are located sufficiently far apart from each other. In this case, there are two different strategies to deploy $N$ IRS reflecting elements in the network: \emph{distributed deployment} where the reflecting elements form multiple distributed IRSs each located near one user, as illustrated in Fig. \ref{Fig_System} (a); or \emph{centralized deployment} where all the reflecting elements form one large IRS located in the vicinity of the AP, as illustrated in Fig. \ref{Fig_System} (b) \cite{MAC}. Note that for the single-user case with $K=1$, the above two deployment strategies are equivalent since they both achieve the maximum received signal power at the AP as shown in Fig. \ref{distance}. However, for the multi-user case with $K>1$, the two deployment strategies lead to distinct channels between the users and AP in general. Specifically, under centralized deployment, all users can be served by $N$ reflecting elements; while under distributed deployment, each user is served by its nearest IRS with only a fraction of the $N$ reflecting elements, since the signals reflected by other (farther-away)  IRSs are too weak due to the much larger path loss. In the following, we compare the achievable rates (or rate/capacity region) of the $K$-user multiple access channel (MAC) in the uplink communication\footnote{The discussions for the downlink communication in the broadcast channel (BC) are similar and can be found in \cite{Deployment_Journal}.} with the above two IRS deployment strategies, under different multiple access schemes. For the purpose of exposition, we consider that the AP and users are all equipped with a single antenna, and similar to Fig. \ref{P2P}, we ignore the direct links between the users and AP. In addition, for fair comparison of the two deployment strategies, we assume that their respective user-IRS-AP channels are akin to each other (the so-called ``twin'' channels in \cite{MAC}),\footnote{Specifically, for each IRS reflecting element, its channels with AP and assigned user $k$ under distributed deployment are equal to those with user $k$ and AP under centralized deployment, respectively.} as illustrated in Fig. \ref{Fig_System}.

First, we consider TDMA where the $K$ users communicate with the AP in orthogonal time slots. For ease of exposition, we assume that $N$ reflecting elements are equally divided into $K$ IRSs, each located near a different user under the distributed deployment, which yields an $\mathcal{O}((N/K)^2)$ passive beamforming gain per user. In contrast, under the centralized deployment, each user can enjoy a passive beamforming gain of $\mathcal{O}(N^2)$ by utilizing all the $N$ reflecting elements. Thus, it can be shown that the achievable rate region with centralized deployment contains that with distributed deployment under the twin channel condition \cite{Deployment_Journal}. In particular, by assuming equal time allocations over the users and free-space LoS model for all channels involved, it can be shown that the achievable sum-rates of the $K$ users under centralized/distributed deployment are given by $R_{\mathrm{cen}}=\frac{1}{K}\sum_{k=1}^K \log_2\left(1+\frac{\bar{P}\beta_0^2N^2}{d^2D_k^2\bar{\sigma}^2}\right)$ and $R_{\mathrm{dis}}=\frac{1}{K}\sum_{k=1}^K \log_2\left(1+\frac{\bar{P}\beta_0^2(N/K)^2}{d^2D_k^2\bar{\sigma}^2}\right)$, respectively, where $\bar{P}$ and $\bar{\sigma}^2$ denote the transmit  power at each user and noise power at the AP receiver, respectively. Thus, when $N\rightarrow \infty$, it can be shown that their asymptotic rate difference is given by
\begin{align}
R_{\mathrm{cen}}-R_{\mathrm{dis}}\rightarrow2\log_2 K.
\end{align}
Therefore, centralized deployment has more pronounced rate gains over distributed deployment with increasing $K$, which is intuitive since each user is assigned with a smaller number of IRS reflecting elements and thus reduced passive beamfoming gain in the latter case.

Next, we consider NOMA and FDMA where the users communicate simultaneously with the AP in the same frequency band and over orthogonal frequency bands, respectively. Unlike TDMA, for centralized IRS deployment, the $N$ reflecting elements need to cater to the channels of all users at the same time in both NOMA and FDMA, which thus results in smaller passive beamforming gain per user than $\mathcal{O}(N^2)$. In contrast, for distributed IRS deployment, the reflecting elements of each distributed IRS are designed similarly as in the TDMA case, i.e., to maximize the passive beamforming gain of its nearby user. Nevertheless, it was proved in \cite{MAC,Deployment_Journal} that under the twin channel condition, centralized deployment outperforms distributed deployment in terms of multi-user capacity region/achievable rate region for the case of NOMA and FDMA, respectively. This is because the larger number of reflecting elements at the centralized IRS provides more flexibility to trade-off users' individual achievable rates as compared to distributed IRSs each helping one user only.

\begin{figure}[t]
	\centering
	\includegraphics[width=7cm]{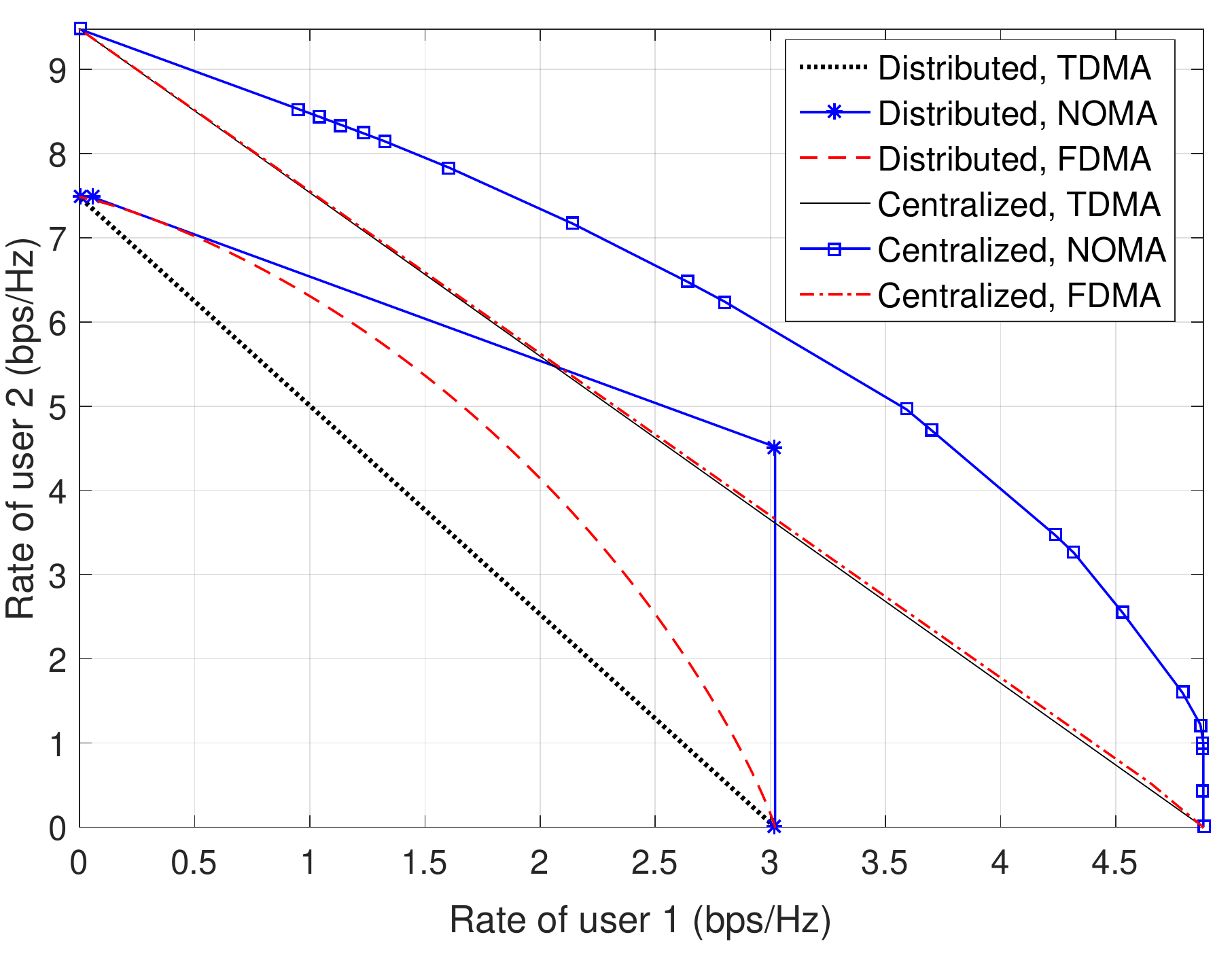}
	\vspace{-3mm}
	\caption{Performance comparison of different IRS deployment strategies and multiple access schemes in a two-user MAC.}\label{Fig_Region}
	\vspace{-6mm}
\end{figure}

In Fig. \ref{Fig_Region}, we compare the achievable rates of TDMA, NOMA, and FDMA under the two deployment strategies in a two-user MAC with $d=1$ m, $D_1=1000$ m, $D_2=200$ m, $\bar{P}=15$ dBm, $\bar{\sigma}^2=-90$ dBm, and $\beta_0=-30$ dB. All channels involved are assumed to follow the free-space LoS model for the purpose of exposition. It is observed that the capacity region/achievable rate regions under centralized IRS deployment constitute their counterparts under distributed IRS deployment in each case of TDMA, NOMA, and FDMA. Furthermore, the gains are observed to be more pronounced when the rates of the two users are \emph{asymmetric}. It is also observed that compared to distributed deployment, centralized deployment achieves substantial rate improvement for the user farther away from the AP, which thus helps deal with the ``near-far'' issue and achieve more balanced rates for the two users.

Despite that centralized IRS deployment is generally more favorable than its distributed counterpart in terms of achievable rates, it is worth noting that other practical factors may also need to be taken into account for their implementation. Firstly, distributed deployment requires more IRSs and thus more backhaul links between the AP and IRS controllers for exchanging information, which thus results in more overhead for the network. Secondly, it may not be always feasible to deploy a large centralized IRS near the AP due to site/space constraint, while deploying multiple distributed IRSs at the user side is generally more flexible. Thirdly, the performance gains of centralized against distributed IRS deployment is under the assumption of twin channel condition \cite{MAC,Deployment_Journal}, while in practice, the channel statistics under the two deployment strategies may vary drastically, in terms of LoS probability, NLoS fading distribution, channel correlation, etc., which may lead to distinct comparison results under different scenarios (e.g., MIMO, multiple users served by each distributed IRS).

\subsection{Other Related Work and Future Direction}
Besides the above discussions which have unveiled some essential trade-offs/considerations for IRS deployment in practical systems, there are other related works on this topic in the literature under various different setups, which are briefly discussed in this subsection, together with some promising directions for future work.

First, it is worthwhile to extend the results in \cite{DoubleIRS,MAC,Deployment_Journal} discussed above to more general system setups, such as multiple cooperative IRSs, other multi-user/MIMO channel models, hybrid distributed and centralized IRS deployment, etc. Furthermore, it is also important to study IRS deployment in the general multi-cell network with the inter-cell interference. In \cite{MIMO_Pan}, the authors considered a simplified setup with one IRS serving two adjacent cells and showed that the IRS should be deployed at the intersection of their edges to maximally benefit both cells. However, if multiple IRSs are available for serving multiple cells, it is still unknown of the optimal IRS deployment/association strategy in general.

Moreover, for large-size wireless network with massively deployed IRSs, how to optimally deploy them is a computationally formidable task. Therefore, new mathematical tools such as stochastic geometry \cite{SpatialThroughput,lyu2020hybrid,Exploiting} and machine learning \cite{Deployment_YW} need to be invoked to perform system-level analysis and efficient deployment design. It is expected that these new tools will play an important role in the design of IRS deployment in future wireless networks with hybrid active/passive components as well as heterogeneous user QoS requirements.

\section{Extensions}
In this section, we overview other relevant topics on IRS which are important and  deserve further investigation.

\subsection{Physical-Layer Security}
Besides cryptographic algorithms and secure communication protocol-layer designs, physical-layer (PHY-layer) security  techniques can enhance the wireless  network security by altering  the distributions of random noise, interference,  and/or fading channels in the legitimate and eavesdropping links, together with the use of Wyner's wiretap coding
\cite{leung1978gaussian}.  As a new means to smartly reconfigure the wireless propagation,  IRS is highly promising to further boost the secrecy rate of traditional wireless networks (see, e.g., \cite{cui2019secure,yang2020intelligent,yu2019enabling,chen2019intelligent,lu2020robust,guan2019intelligent,xu2019resource,yu2019robust,yang2020deep}), even for the challenging scenario when the conventional PHY-layer security techniques are ineffective \cite{JR:wu2019IRSmaga,cui2019secure}.
   Specifically, if the eavesdropping channel is stronger than the legitimate channel and/or they are highly correlated (e.g., when their channels are along the same direction from the transmitter), the legitimate receiver  can hardly achieve a positive secrecy rate  using conventional techniques (such as transmit beamforming and artificial noise (AN)/jamming). However, this issue can be efficiently addressed by deploying IRSs in the vicinity of the legitimate/eavesdropping  user and properly designing IRS passive beamforming to increase/reduce  the achievable rate of the legitimate/eavesdropping  user, thus significantly enhancing the achievable secrecy rate  \cite{cui2019secure}. Moreover, AN/jamming by the transmitter can be jointly designed with IRS passive beamforming to further reduce the  eavesdropper's rate without compromising  the rate performance of the legitimate user \cite{guan2019intelligent,xu2019resource}.

Despite the above advantages, several practical issues in the IRS-aided secrecy  communication system need to be addressed. First, exploiting IRS passive beamforming for reducing the eavesdroppers' rates requires CSI of the links between the AP and eavesdroppers as well as between the IRS and eavesdroppers, which is practically difficult to obtain since 1) IRS usually has a massive number of passive reflecting elements, and 2) the eavesdroppers may intentionally remain covert or their CSI may be coarse/outdated if their  signal leakage is utilized for channel estimation. This thus calls for new channel estimation designs as well as robust/secure IRS passive beamforming by taking into account the incomplete/imperfect CSI of the eavesdroppers \cite{yu2019robust}. Moreover, in a large-scale secrecy communication network with many  legitimate users and eavesdroppers as well as densely deployed IRSs, IRS deployment is a key to maximize the network secrecy throughput,
 which deserves further studies.
\subsection{Wireless Power Transfer}
RF-transmission enabled wireless power transfer (WPT) is envisioned as a promising technology to prolong the battery lives of Internet-of-Things (IoT)  devices in future wireless networks. To overcome the high power loss over long distance, various techniques have been proposed to be implemented at the energy transmitter and/or receiver for improving the WPT efficiency, such as
waveform design, energy transmission  and scheduling, energy  beamforming, rectifiers with high energy conversion efficiency, and so on \cite{zeng2017communications}. However, the limited size of IoT devices results in small receiving antenna aperture, which fundamentally constrains their received power level in WPT systems.
Fortunately, this issue can be efficiently alleviated  by deploying IRSs in the proximity of IoT devices to create a highly efficient  energy charging zone for them (see, e.g., \cite{wu2019jointSWIPT,wu2019weighted,pan2019intelligent,tang2019joint}). By properly deploying IRSs in LoS with the energy transmitters and receivers as well as exploiting their large aperture and high passive beamforming gains, the received power of nearby IoT devices can be substantially enhanced.
 Moreover, the improvement of wireless charging efficiency in turn helps reduce the transmit power as well as provide more flexibility in designing the transmit beamforming for information receivers and thus improves both the rate and energy performance in SWIPT systems \cite{wu2019jointSWIPT,wu2019weighted,pan2019intelligent,tang2019joint}.

 In practice, the benefits of IRS passive beamforming in WPT/SWIPT critically depend on the available CSI at the energy transmitter/AP, which needs to be acquired at the cost of time and energy resources. Specifically,  given a limited channel coherence time, the uplink pilot-assisted IRS channel estimation (see Section IV) needs to balance the trade-off between channel estimation  and downlink  power transfer, since too little channel training will result in inaccurate/insufficient CSI which in turn reduces the energy beamforming gain, while too much training will consume substantial energy at the energy receiver   and also  leave less time for energy harvesting. In addition, IRS deployment (e.g., single versus multiple cooperative IRSs, centralized versus distributed IRS as discussed in Section V) also greatly impacts  the WPT/SWIPT system performance, which is thus worthy of further investigation.

\subsection{UAV Communications}
Most of the existing research on IRS has assumed terrestrial IRSs at fixed locations, which, however, has the intrinsic limitation that the IRS can only serve nearby users usually located in half of the space, i.e., both the transmitter and receiver must lie on the same side of the IRS. To enlarge the service region, a new network architecture called aerial IRS (AIRS) was proposed in \cite{lu2020enabling}, where the IRS is mounted on an unmanned aerial vehicle (UAV) to realize over-the-air intelligent reflection. Compared to terrestrial IRS,  AIRS is more likely to establish strong LoS links with the ground nodes due to the relatively higher altitude of  UAV, thus reducing the blockage probability with them.
Besides, instead of the half-space reflection only by terrestrial IRS, AIRS can realize $360^{\circ}$ panoramic full-range reflection, i.e., one AIRS can assist in  the communication between any pair of ground nodes within its coverage area, thus greatly increasing the number of serviceable  users as compared to terrestrial IRS.
 Moreover, the coverage  area of AIRS can be further expanded by exploiting the high mobility of UAV to move closer to multiple geographically  separated users sequentially to improve their communication performance by exploiting short-range LoS channels as well as minimizing the product distance of the IRS-reflected link.  On the other hand, terrestrial IRS can also be employed to help enhance the UAV-ground communication performance via optimized signal reflections \cite{li2020reconfigurable,ma2020enhancing}.

Compared to terrestrial IRS, AIRS brings  new challenges. First, for the general case of employing multiple AIRSs to aid the communications among multiple ground nodes, it is crucial to design the three dimensional (3D) placement/trajectories  of AIRSs jointly with their passive beamforming and user association for maximizing  the average rates of all users under the practical UAV-ground channel models \cite{zeng2019accessing} (such as the elevation-angle dependent Rician fading channel \cite{you20193d} and the probabilistic LoS channel \cite{you2020hybrid}).
 In addition, practical low-complexity channel estimation/tracking methods are required for  acquiring the AIRS-users channels over their 3D trajectories, which is more challenging as compared to the case of terrestrial IRSs that are at fixed locations.


\subsection{MmWave Communications}
Due to  the enormous bandwidth available in mmWave frequencies (i.e., $30$--$300$ GHz), mmWave communications are promising  to support gigabit-per-second data rates for enabling rate-demanding wireless  applications (augmented/virtual reality (AR/VR), online gaming, etc.) in the future; whereas  they also incur significantly higher energy consumption and hardware cost due to the increasing number of active antennas/RF chains required operating at much higher frequencies as compared to sub-$6$ GHz wireless systems. Moreover, the mmWave communication channels are more susceptible  to blockage and subjected to higher propagation loss in general. These critical  issues can be efficiently resolved by properly deploying IRSs in mmWave systems to create virtual LoS channels between the BS/AP and users through smart reflections of the  IRSs, which can bypass the main obstructors between them, thus achieving both high spectral and energy efficiencies \cite{Infocom_Sun,wang2019intelligent,jamali2019intelligent,di2019hybrid}.

To fully reap the IRS passive beamforming gain in the IRS-aided mmWave communication system,
it is necessary to conduct initial  beam training/alignment at the BS/AP as well as at the IRS to establish high-SNR links before executing channel estimation and data transmission, due to the limited scattering of mmWave channels. Compared to traditional mmWave systems, the beam training/alignment in the IRS-aided mmWave system is much more involved due to the following reasons. First, the large number of IRS reflecting elements generate pencil-like IRS beam with narrow beam-width, thus incurring higher training overhead for scanning the beam space. This issue becomes more severe when the IRS needs to serve multiple users, since the overhead of the straightforward user-by-user successive beam training linearly scales with the number of users. Second, the 3D IRS passive beam training should be jointly designed with BS/AP's active beam training to adapt to the time-varying BS/AP-user and IRS-user channels due to user mobility.
 Although some initial attempts have been made to design the IRS beam training (e.g., \cite{ning2019channelMU}), the training overhead is still overwhelming, thus calling for new approaches for designing more efficient  IRS beam training for mmWave systems \cite{you2020fast}.


\subsection{Mobile Edge Computing}

Recent years have witnessed the computing paradigm shift from centralized cloud computing to distributed edge computing, which pushes abundant computational resources to the edges of networks (e.g., BS and AP) to help offload the computation tasks from mobile devices for reducing their computation latency and energy consumption \cite{mao2017mobile2}. Such computing performance gains are expected to be further enhanced if IRS is deployed for mobile edge computing (MEC) systems \cite{cao2019intelligent,liu2020intelligent,bai2019latency}.
Specifically, for a single user, as IRS passive beamforming can improve its wireless link capacity with the edge server, it is able to offload computation-intensive tasks to the edge server for reducing its computation latency without incurring high transmission energy consumption. Furthermore, for a large-size  MEC network with a massive number of users and edge servers,  IRS passive beamforming can have a great  impact on the offloading decisions and policies (i.e., amount of offloaded data bits and radio-and-computational resources) of all users by properly manipulating their offloading channels to different edge servers, thus reshaping the distribution of offloaded computation loads over the entire MEC network for  achieving better computational resource utilization and computation-latency reduction. For example, instead of congesting the same nearby server that would cause high  computation latency, some users can offload their computation tasks to the servers farther away  that are of light computation burdens, if their corresponding links can be greatly improved by the IRS passive beamforming.

As different users usually have asynchronous computation task arrivals and completion deadlines, it is necessary to jointly design the time-varying IRS passive beamforming for serving different users with the offloading scheduling order and multi-user time allocation based on their arrival-and-deadline information as well as channel conditions, which is  worthy of further investigation in the future.

\subsection{Other Miscellaneous Topics}
In addition to the above mentioned ones, there are also other interesting and promising topics on IRS that deserve further studies, including  IRS-aided cognitive radio network \cite{guan2020joint,xu2020resource}, IRS-aided sensing and localization \cite{wymeersch2019radio,bourdoux20206g}, IRS-aided THz communication \cite{chaccour2020risk}, simultaneous IRS reflection and data transmission \cite{basar2020reconfigurable}, etc., which help significantly broaden the IRS research scope as well as application horizon.

\section{Conclusion}
In this paper, we have provided  a comprehensive tutorial on the new IRS technology as a promising enabler for smart and reconfigurable wireless communication environment.  It is shown that IRS-aided wireless communication has led to a fundamental paradigm shift in the design of wireless system/network from the traditional one with active components only to a new hybrid architecture comprising both active and passive components that co-work in an intelligent way.
 Despite that the research on IRS-aided wireless communication is still in its infancy, this paper overviews  its fundamentals, state-of-the-art results on addressing the  main communication  challenges, as well as promising directions for further investigation. It is hoped that this paper will serve as  a useful and inspiring resource for future research on IRS to unlock its full potential in future-generation  (B5G/6G) wireless communications.


\begin{thebibliography}{1}

\bibitem{boccardi2014five}
F.~Boccardi, R.~W. Heath, A.~Lozano, T.~L. Marzetta, and P.~Popovski, ``Five
  disruptive technology directions for {5G},'' \emph{IEEE Commu. Mag.},
  vol.~52, no.~2, pp. 74--80, Feb. 2014.

\bibitem{shafi20175g}
M.~Shafi, A.~F. Molisch, P.~J. Smith, T.~Haustein, P.~Zhu, P.~De~Silva,
  F.~Tufvesson, A.~Benjebbour, and G.~Wunder, ``{5G: A} tutorial overview of
  standards, trials, challenges, deployment, and practice,'' \emph{IEEE J. Sel.
  Areas Commun.}, vol.~35, no.~6, pp. 1201--1221, Jun. 2017.

\bibitem{zhang2016fundamental}
S.~Zhang, Q.~Wu, S.~Xu, and G.~Y. Li, ``Fundamental green tradeoffs:
  Progresses, challenges, and impacts on {5G} networks,'' \emph{{IEEE} Commun.
  Surveys Tuts.}, vol.~19, no.~1, pp. 33--56, First Quarter 2017.

\bibitem{wu2016overview}
Q.~Wu, G.~Y. Li, W.~Chen, D.~W.~K. Ng, and R.~Schober, ``An overview of
  sustainable green {5G} networks,'' \emph{IEEE Wireless Commun.}, vol.~24,
  no.~4, pp. 72--80, Aug. 2017.

\bibitem{tse2005fundamentals}
D.~Tse and P.~Viswanath, \emph{Fundamentals of wireless communication}.\hskip
  1em plus 0.5em minus 0.4em\relax Cambridge university press, 2005.

\bibitem{goldsmith2005wireless}
A.~Goldsmith, \emph{Wireless communications}.\hskip 1em plus 0.5em minus
  0.4em\relax Cambridge university press, 2005.

\bibitem{wu2018IRS}
Q.~Wu and R.~Zhang, ``Intelligent reflecting surface enhanced wireless network:
  {J}oint active and passive beamforming design,'' in \emph{Proc. IEEE
  GLOBECOM}, Dec. 2018.

\bibitem{JR:wu2018IRS}
Q.~Wu and R.~Zhang, ``Intelligent reflecting surface enhanced wireless network via joint
  active and passive beamforming,'' \emph{{IEEE} Trans. Wireless Commun.},
  vol.~18, no.~11, pp. 5394--5409, Nov. 2019.

\bibitem{JR:wu2019IRSmaga}
Q.~Wu and R.~Zhang, ``Towards smart and reconfigurable environment: {Intelligent}
  reflecting surface aided wireless network,'' \emph{IEEE Commun. Mag.},
  vol.~58, no.~1, pp. 106--112, Jan. 2020.

\bibitem{hu2017beyond}
S.~Hu, F.~Rusek, and O.~Edfors, ``Beyond massive {MIMO}: The potential of data
  transmission with large intelligent surfaces,'' \emph{{IEEE} Trans. Signal
  Process.}, vol.~66, no.~10, pp. 2746--2758, May 2018.

\bibitem{lyu2020hybrid}
J.~Lyu and R.~Zhang, ``Hybrid active/passive wireless network aided by
  intelligent reflecting surface: System modeling and performance analysis,''
  \emph{arXiv preprint arXiv:2004.13318}, 2020.

\bibitem{basar2019reconfigurable}
E.~Basar and I.~F. Akyildiz, ``Reconfigurable intelligent surfaces for doppler
  effect and multipath fading mitigation,'' \emph{arXiv preprint
  arXiv:1912.04080}, 2019.

\bibitem{latva2019key2}
M.~Latva-Aho and K.~Lepp{\"a}nen, ``Key drivers and research challenges for
  {6G} ubiquitous wireless intelligence (white paper),'' \emph{6G Flagship,
  University of Oulu, Oulu}, 2019.

\bibitem{rajatheva2020white}
N.~Rajatheva \emph{et~al.}, ``White paper on broadband connectivity in {6G},''
  \emph{arXiv preprint arXiv:2004.14247}, 2020.

\bibitem{subrt2012intelligent}
L.~Subrt and P.~Pechac, ``Intelligent walls as autonomous parts of smart indoor
  environments,'' \emph{IET commun.}, vol.~6, no.~8, pp. 1004--1010, May 2012.

\bibitem{Infocom_Sun}
X.~Tan, Z.~Sun, D.~Koutsonikolas, and J.~M. Jornet, ``{Enabling indoor mobile
  millimeter-wave networks based on smart reflect-arrays},'' in \emph{Proc.
  IEEE INFOCOM}, Apr. 2018.

\bibitem{di2019smart}
M.~Di~Renzo \emph{et~al.}, ``Smart radio environments empowered by {AI}
  reconfigurable meta-surfaces: {A}n idea whose time has come,'' \emph{EURASIP
  J. Wireless Commun. Netw.}, {M}ay 2019.

\bibitem{basar2019wireless}
E.~Basar, M.~Di~Renzo, J.~de~Rosny, M.~Debbah, M.-S. Alouini, and R.~Zhang,
  ``Wireless communications through reconfigurable intelligent surfaces,''
  \emph{IEEE Access}, vol.~7, pp. 116\,753--116\,773, Aug. 2019.

\bibitem{liang2019large}
Y.-C. Liang, R.~Long, Q.~Zhang, J.~Chen, H.~V. Cheng, and H.~Guo, ``Large
  intelligent surface/antennas {(LISA)}: Making reflective radios smart,''
  \emph{J. Commu. Info. Netw.}, vol.~4, no.~2, pp. 40--50, Jun. 2019.

\bibitem{arun2019rfocus}
V.~Arun and H.~Balakrishnan, ``{RFocus}: {Practical} beamforming for small
  devices,'' \emph{arXiv preprint arXiv:1905.05130}, 2019.

\bibitem{DOCOMO2018}
``{NTT DOCOMO} and {Metawave} announce successful demonstration of 28{GHz}-band
  {5G} using world's first meta-structure technology,'' Available:
  \url{https://www.businesswire.com/news/home/20181204005253/en/NTT-DOCOMO-Metawave-Announce-Successful-Demonstration-28GHz-Band}.

\bibitem{Lumotive2019}
``{TowerJazz} and {Lumotive} demonstrate solid-state beam steering for
  {LiDAR},'' Available: \url{https://techtime.news/2019/06/26/lidar/}.

\bibitem{PivotalCommware2019}
``Pivotal {Commware} achieves {gigabit} connectivity in live {5G} {mmWave} demo
  at mobile world congress {Los Angeles} 2019,'' Available:
  \url{https://pivotalcommware.com/2019/11/04/pivotal-commware-achieves-gigabit-connectivity-in-live-5g-mmwave-demo-at-mobile-world-congress-los-angeles-2019/}.

\bibitem{DOCOMO2020}
``{DOCOMO} conducts world's first successful trial of transparent dynamic
  metasurface,'' Available:
  \url{https://www.nttdocomo.co.jp/english/info/media_center/pr/2020/0117_00.html}.

\bibitem{greenerwave}
``A technological platform to simplify all {EM} infrastructures: {Replacing}
  hardware complexity by algorithms,'' Available:
  \url{http://greenerwave.com/our-technology/}.

\bibitem{visorsurf}
``A hardware platform for software-driven functional metasurfaces,'' Available:
  \url{http://www.visorsurf.eu}.

\bibitem{ariadne}
``Artificial intelligence aided {D}-band network for {5G} long term
  evolution,'' Available: \url{https://www.ict-ariadne.eu}.

\bibitem{PathFinder}
``Harnessing multipath propagation in wireless networks: {A} meta-surface
  transformation of wireless networks into smart reconfigurable radio
  environments,'' Available: \url{https://cordis.europa.eu/project/id/891030}.

\bibitem{he2019tunable}
Q.~He \emph{et~al.}, ``Tunable/reconfigurable metasurfaces: {Physics} and
  applications,'' \emph{Research}, vol. 2019, Article ID1849272, Jul. 2019.

\bibitem{hum2013reconfigurable}
S.~V. Hum and J.~Perruisseau-Carrier, ``Reconfigurable reflectarrays and array
  lenses for dynamic antenna beam control: {A} review,'' \emph{IEEE Trans.
  Antennas Propag.}, vol.~62, no.~1, pp. 183--198, Jan. 2013.

\bibitem{Liaskos2018}
C.~Liaskos, S.~Nie, A.~Tsioliaridou, A.~Pitsillides, S.~Ioannidis, and
  I.~Akyildiz, ``A new wireless communication paradigm through
  software-controlled metasurfaces,'' \emph{{IEEE} Commun. Mag.}, vol.~56,
  no.~9, pp. 162--169, Sep. 2018.

\bibitem{ntontin2019reconfigurable2}
K.~Ntontin \emph{et~al.}, ``Reconfigurable intelligent surfaces vs. relaying:
  {Differences}, similarities, and performance comparison,'' \emph{arXiv
  preprint arXiv:1908.08747}, 2019.

\bibitem{elmossallamy2020reconfigurable}
M.~A. ElMossallamy, H.~Zhang, L.~Song, K.~G. Seddik, Z.~Han, and G.~Y. Li,
  ``Reconfigurable intelligent surfaces for wireless communications:
  {Principles}, challenges, and opportunities,'' \emph{arXiv preprint
  arXiv:2005.00938}, 2020.

\bibitem{huang2019holographic}
C.~Huang, S.~Hu, G.~C. Alexandropoulos, A.~Zappone, C.~Yuen, R.~Zhang,
  M.~Di~Renzo, and M.~Debbah, ``Holographic {MIMO} surfaces for {6G} wireless
  networks: {Opportunities}, challenges, and trends,'' \emph{arXiv preprint
  arXiv:1911.12296}, 2019.

\bibitem{di2020smart_JSAC}
M.~Di~Renzo, A.~Zappone, M.~Debbah, M.-S. Alouini, C.~Yuen, J.~de~Rosny, and
  S.~Tretyakov, ``Smart radio environments empowered by reconfigurable
  intelligent surfaces: How it works, state of research, and road ahead,''
  \emph{arXiv preprint arXiv:2004.09352}, 2020.

\bibitem{yuan2020reconfigurable}
X.~Yuan, Y.-J. Zhang, Y.~Shi, W.~Yan, and H.~Liu,
  ``Reconfigurable-intelligent-surface empowered {6G} wireless communications:
  {Challenges} and opportunities,'' \emph{arXiv preprint arXiv:2001.00364},
  2020.

\bibitem{gong2019towards}
S.~Gong, X.~Lu, D.~T. Hoang, D.~Niyato, L.~Shu, D.~I. Kim, and Y.-C. Liang,
  ``Towards smart radio environment for wireless communications via intelligent
  reflecting surfaces: {A} comprehensive survey,'' \emph{arXiv preprint
  arXiv:1912.07794}, 2019.

\bibitem{bjornson2020reconfigurable}
E.~Bj{\"o}rnson, {\"O}.~{\"O}zdogan, and E.~G. Larsson, ``Reconfigurable
  intelligent surfaces: Three myths and two critical questions,'' \emph{arXiv
  preprint arXiv:2006.03377}, 2020.

\bibitem{baker2003mathematical}
B.~B. Baker and E.~T. Copson, \emph{The mathematical theory of Huygens'
  principle}.\hskip 1em plus 0.5em minus 0.4em\relax American Mathematical
  Soc., 2003, vol. 329.

\bibitem{garcia2019reconfigurable}
J.~B. Garcia, A.~Sibille, and M.~Kamoun, ``Reconfigurable intelligent surfaces:
  Bridging the gap between scattering and reflection,'' \emph{arXiv preprint
  arXiv:1912.05344}, 2019.

\bibitem{tang2019wireless}
W.~Tang, M.~Z. Chen, X.~Chen, J.~Y. Dai, Y.~Han, M.~Di~Renzo, Y.~Zeng, S.~Jin,
  Q.~Cheng, and T.~J. Cui, ``Wireless communications with reconfigurable
  intelligent surface: Path loss modeling and experimental measurement,''
  \emph{arXiv preprint arXiv:1911.05326}, 2019.

\bibitem{ozdogan2019intelligent}
{\"O}.~{\"O}zdogan, E.~Bj{\"o}rnson, and E.~G. Larsson, ``Intelligent
  reflecting surfaces: Physics, propagation, and pathloss modeling,''
  \emph{IEEE Wireless Commun. Lett.}, vol.~9, no.~5, pp. 581--585, May 2020.

\bibitem{ellingson2019path}
S.~W. Ellingson, ``Path loss in reconfigurable intelligent surface-enabled
  channels,'' \emph{arXiv preprint arXiv:1912.06759}, 2019.

\bibitem{di2020analytical}
M.~Di~Renzo, F.~H. Danufane, X.~Xi, J.~de~Rosny, and S.~Tretyakov, ``Analytical
  modeling of the path-loss for reconfigurable intelligent surfaces--anomalous
  mirror or scatterer?'' \emph{arXiv preprint arXiv:2001.10862}, 2020.

\bibitem{cui2014coding}
T.~J. Cui, M.~Q. Qi, X.~Wan, J.~Zhao, and Q.~Cheng, ``Coding metamaterials,
  digital metamaterials and programmable metamaterials,'' \emph{Light: Science
  \& Applications}, vol.~3, e218, Oct. 2014.

\bibitem{nayeri2018reflectarray}
P.~Nayeri, F.~Yang, and A.~Z. Elsherbeni, \emph{Reflectarray antennas:
  {T}heory, designs, and applications}.\hskip 1em plus 0.5em minus 0.4em\relax
  John Wiley \& Sons, 2018.

\bibitem{zhang2018space}
L.~Zhang \emph{et~al.}, ``Space-time-coding digital metasurfaces,'' \emph{Nat.
  Commun.}, vol.~9, no.~1, p. 4334, Oct. 2018.

\bibitem{yang2017design}
H.~Yang, X.~Chen, F.~Yang, S.~Xu, X.~Cao, M.~Li, and J.~Gao, ``Design of
  resistor-loaded reflectarray elements for both amplitude and phase control,''
  \emph{{IEEE} Antennas Wireless Propag. Lett.}, vol.~16, pp. 1159--1162, Nov.
  2017.

\bibitem{wu2018IRS_discrete}
Q.~Wu and R.~Zhang, ``Beamforming optimization for intelligent reflecting
  surface with discrete phase shifts,'' in \emph{Proc. IEEE ICASSP}, May 2019.

\bibitem{abeywickrama2019ICCintelligent}
S.~Abeywickrama, R.~Zhang, and C.~Yuen, ``Intelligent reflecting surface:
  Practical phase shift model and beamforming optimization,'' in \emph{Proc.
  IEEE ICC}, Jun. 2020.

\bibitem{abeywickrama2019intelligent}
S.~Abeywickrama, R.~Zhang, Q.~Wu, and C.~Yuen, ``Intelligent reflecting
  surface: Practical phase shift model and beamforming optimization,''
  \emph{IEEE Trans. Commun.}, Early Access.

\bibitem{zhu2013active}
B.~O. Zhu, J.~Zhao, and Y.~Feng, ``Active impedance metasurface with full 360
  reflection phase tuning,'' \emph{Sci. Rep.}, vol.~3, p. 3059, Oct. 2013.

\bibitem{kaina2014shaping}
N.~Kaina, M.~Dupr{\'e}, G.~Lerosey, and M.~Fink, ``Shaping complex microwave
  fields in reverberating media with binary tunable metasurfaces,'' \emph{Sci.
  Rep.}, vol.~4, p. 6693, Oct. 2014.

\bibitem{zhang2019mutual}
S.~Zhang, X.~Chen, and G.~F. Pedersen, ``Mutual coupling suppression with
  decoupling ground for massive {MIMO} antenna arrays,'' \emph{IEEE Trans. Veh.
  Technol.}, vol.~68, no.~8, pp. 7273--7282, Aug. 2019.

\bibitem{Hien2013}
H.~Q. Ngo, E.~Larsson, and T.~Marzetta, ``Energy and spectral efficiency of
  very large multiuser {MIMO} systems,'' \emph{{IEEE} Trans. Commun.}, vol.~61,
  no.~4, pp. 1436--1449, Apr. 2013.

\bibitem{hoydis2013massive}
J.~Hoydis, S.~Ten~Brink, and M.~Debbah, ``Massive {MIMO} in the {UL/DL} of
  cellular networks: How many antennas do we need?'' \emph{IEEE J. Sel. Areas
  Commun.}, vol.~31, no.~2, pp. 160--171, Feb. 2013.

\bibitem{so2007approximating}
A.~M.-C. So, J.~Zhang, and Y.~Ye, ``On approximating complex quadratic
  optimization problems via semidefinite programming relaxations,''
  \emph{Mathematical Programming}, vol. 110, no.~1, pp. 93--110, Jun. 2007.

\bibitem{JR:wu2019discreteIRS}
Q.~Wu and R.~Zhang, ``Beamforming optimization for wireless network aided by
  intelligent reflecting surface with discrete phase shifts,'' \emph{IEEE
  Trans. Commun.}, vol.~68, no.~3, pp. 1838--1851, Mar. 2020.

\bibitem{MIMO}
S.~Zhang and R.~Zhang, ``Capacity characterization for intelligent reflecting
  surface aided {MIMO} communication,'' \emph{IEEE J. Sel. Areas Commun.},
  {Early Access}.

\bibitem{OFDM_Protocol}
Y.~Yang, B.~Zheng, S.~Zhang, and R.~Zhang, ``{Intelligent reflecting surface
  meets OFDM: Protocol design and rate maximization},'' \emph{IEEE Trans.
  Commun.}, {Early Access}.

\bibitem{OFDM_BX}
B.~Zheng and R.~Zhang, ``{Intelligent reflecting surface-enhanced OFDM: Channel
  estimation and reflection optimization},'' \emph{IEEE Wireless Commun.
  Lett.}, vol.~9, no.~4, pp. 518--522, Apr. 2020.

\bibitem{NOMA_BX}
B.~Zheng, Q.~Wu, and R.~Zhang, ``{Intelligent reflecting surface-assisted
  multiple access with user pairing: NOMA or OMA?}'' \emph{IEEE Commun. Lett.},
  vol.~24, no.~4, pp. 753--757, Apr. 2020.

\bibitem{OFDMA}
Y.~Yang, S.~Zhang, and R.~Zhang, ``{IRS-enhanced OFDMA: Joint resource
  allocation and passive beamforming optimization},'' \emph{IEEE Wireless
  Commun. Lett.}, vol.~9, no.~6, pp. 760--764, Jun. 2020.

\bibitem{zhao2019intelligent}
M.-M. Zhao, Q.~Wu, M.-J. Zhao, and R.~Zhang, ``Intelligent reflecting surface
  enhanced wireless network: Two-timescale beamforming optimization,''
  \emph{arXiv preprint arXiv:1912.01818}, 2019.

\bibitem{zhao2020exploiting}
M.-M. Zhao, Q.~Wu, M.-J. Zhao, and R.~Zhang, ``Exploiting amplitude control in intelligent reflecting surface aided
  wireless communication with imperfect {CSI},'' \emph{arXiv preprint
  arXiv:2005.07002}, 2020.

\bibitem{wu2019jointSWIPT}
Q.~Wu and R.~Zhang, ``Joint active and passive beamforming optimization for
  intelligent reflecting surface assisted {SWIPT} under {QoS} constraints,''
  \emph{IEEE J. Sel. Areas Commun.}, Early Access.

\bibitem{di2019hybrid}
B.~Di, H.~Zhang, L.~Song, Y.~Li, Z.~Han, and H.~V. Poor, ``Hybrid beamforming
  for reconfigurable intelligent surface based multi-user communications:
  Achievable rates with limited discrete phase shifts,'' \emph{arXiv preprint
  arXiv:1910.14328}, 2019.

\bibitem{zhang2019reconfigurable}
H.~Zhang, B.~Di, L.~Song, and Z.~Han, ``Reconfigurable intelligent surfaces
  assisted communications with limited phase shifts: How many phase shifts are
  enough?'' \emph{arXiv preprint arXiv:1912.01477}, 2020.

\bibitem{xu2019discrete}
J.~Xu, W.~Xu, and A.~L. Swindlehurst, ``Discrete phase shift design for
  practical large intelligent surface communication,'' in \emph{Proc. IEEE
  PACRIM}, Aug. 2019.

\bibitem{feng2020deep}
K.~{Feng}, Q.~{Wang}, X.~{Li}, and C.~{Wen}, ``Deep reinforcement learning
  based intelligent reflecting surface optimization for {MISO} communication
  systems,'' \emph{IEEE Wireless Commun. Lett.}, vol.~9, no.~5, pp. 745--749,
  May 2020.

\bibitem{OFDM_GC}
Y.~Yang, S.~Zhang, and R.~Zhang, ``{IRS-enhanced OFDM: Power allocation and
  passive array optimization},'' in \emph{Proc. IEEE GLOBECOM}, Dec. 2019.

\bibitem{MIMO_ISIT}
S.~Zhang and R.~Zhang, ``{On the capacity of intelligent reflecting surface
  aided MIMO communication},'' in \emph{Proc. IEEE ISIT}, Jun. 2020.

\bibitem{huang2018largeRIS}
C.~Huang, A.~Zappone, G.~C. Alexandropoulos, M.~Debbah, and C.~Yuen,
  ``Reconfigurable intelligent surfaces for energy efficiency in wireless
  communication,'' \emph{{IEEE} Trans. Wireless Commun.}, vol.~18, no.~8, pp.
  4157--4170, Aug. 2019.

  \bibitem{yangliu2020}
Y.~Liu, J.~Zhao, M.~Li, and Q.~Wu, ``Intelligent reflecting surface aided {MISO} uplink communication network: Feasibility and SINR optimization,'' \emph{arXiv preprint arXiv:2007.01482}, 2020.
  

\bibitem{kammoun2020asymptotic}
Q.-U.-A. {Nadeem}, A.~{Kammoun}, A.~{Chaaban}, M.~{Debbah}, and M.~{Alouini},
  ``Asymptotic max-min {SINR} analysis of reconfigurable intelligent surface
  assisted {MISO} systems,'' \emph{IEEE Trans. Wireless Commun.}, Early Access.

\bibitem{guo2019weighted}
H.~Guo, Y.-C. Liang, J.~Chen, and E.~G. Larsson, ``Weighted sum-rate
  optimization for intelligent reflecting surface enhanced wireless networks,''
  \emph{arXiv preprint arXiv:1905.07920}, 2019.

\bibitem{MIMO_Pan}
C.~Pan, H.~Ren, K.~Wang, W.~Xu, M.~Elkashlan, A.~Nallanathan, and L.~Hanzo,
  ``Multicell {MIMO} communications relying on intelligent reflecting
  surface,'' {\emph{arXiv preprint arXiv:1907.10864}}, 2019.

\bibitem{hua2020intelligent}
M.~Hua, Q.~Wu, D.~W.~K. Ng, J.~Zhao, and L.~Yang, ``Intelligent reflecting
  surface-aided joint processing coordinated multipoint transmission,''
  \emph{arXiv preprint arXiv:2003.13909}, 2020.

\bibitem{yu2020optimal}
X.~Yu, D.~Xu, and R.~Schober, ``Optimal beamforming for {MISO} communications
  via intelligent reflecting surfaces,'' \emph{arXiv preprint
  arXiv:2001.11429}, 2020.

\bibitem{yang2020intelligent}
H.~Yang, Z.~Xiong, J.~Zhao, D.~Niyato, Q.~Wu, H.~V. Poor, and M.~Tornatore,
  ``Intelligent reflecting surface assisted anti-jamming communications: A fast
  reinforcement learning approach,'' \emph{arXiv preprint arXiv:2004.12539},
  2020.

\bibitem{zhang2019analysis}
Z.~Zhang, Y.~Cui, F.~Yang, and L.~Ding, ``Analysis and optimization of outage
  probability in multi-intelligent reflecting surface-assisted systems,''
  \emph{arXiv preprint arXiv:1909.02193}, 2019.

\bibitem{SpatialThroughput}
J.~Lyu and R.~Zhang, ``{Spatial throughput characterization for intelligent
  reflecting surface aided multiuser system},'' \emph{IEEE Wireless Commun.
  Lett.}, {Early Access}.

\bibitem{zheng2020intelligent}
B.~Zheng, C.~You, and R.~Zhang, ``Intelligent reflecting surface assisted
  multi-user {OFDMA}: Channel estimation and training design,'' \emph{arXiv
  preprint arXiv:2003.00648}, 2020.

\bibitem{alexandropoulos2020hardware}
G.~C. {Alexandropoulos} and E.~{Vlachos}, ``A hardware architecture for
  reconfigurable intelligent surfaces with minimal active elements for explicit
  channel estimation,'' in \emph{Proc. IEEE ICASSP}, May 2020.

\bibitem{Tahaa2019enabling}
A.~Taha, M.~Alrabeiah, and A.~Alkhateeb, ``Enabling large intelligent surfaces
  with compressive sensing and deep learning,'' \emph{arXiv preprint
  arXiv:1904.10136}, 2019.

\bibitem{Taha2019Deep}
A.~{Taha}, M.~{Alrabeiah}, and A.~{Alkhateeb}, ``Deep learning for large
  intelligent surfaces in millimeter wave and massive {MIMO} systems,'' in
  \emph{Proc. IEEE GLOBECOM}, Dec. 2019.

\bibitem{taha2020deep}
A.~Taha, Y.~Zhang, F.~B. Mismar, and A.~Alkhateeb, ``Deep reinforcement
  learning for intelligent reflecting surfaces: Towards standalone operation,''
  \emph{arXiv preprint arXiv:2002.11101}, 2020.

\bibitem{Han2020deep}
W.~Han, P.~Chen, and Z.~Cao, ``Deep neural network-based quantized signal
  reconstruction for {DOA} estimation,'' \emph{arXiv preprint
  arXiv:2005.01102}, 2020.

\bibitem{Mishra2019Channel}
D.~Mishra and H.~Johansson, ``Channel estimation and low-complexity beamforming
  design for passive intelligent surface assisted {MISO} wireless energy
  transfer,'' in \emph{Proc. IEEE ICASSP}, May 2019.

\bibitem{jensen2019optimal}
T.~L. Jensen and E.~De~Carvalho, ``An optimal channel estimation scheme for
  intelligent reflecting surfaces based on a minimum variance unbiased
  estimator,'' in \emph{Proc. IEEE ICASSP}, May 2020.

\bibitem{wang2019compressed}
P.~{Wang}, J.~{Fang}, H.~{Duan}, and H.~{Li}, ``Compressed channel estimation
  for intelligent reflecting surface-assisted millimeter wave systems,''
  \emph{IEEE Signal Process. Lett.}, vol.~27, pp. 905--909, May 2020.

\bibitem{he2019cascaded}
Z.-Q. He and X.~Yuan, ``Cascaded channel estimation for large intelligent
  metasurface assisted massive {MIMO},'' \emph{IEEE Wireless Commun. Lett.},
  vol.~9, no.~2, pp. 210--214, Feb. 2020.

\bibitem{de2020parafac}
G.~T. de~Ara{\'u}jo and A.~L. de~Almeida, ``{PARAFAC}-based channel estimation
  for intelligent reflective surface assisted {MIMO} system,'' \emph{arXiv
  preprint arXiv:2001.06554}, 2020.

\bibitem{mirza2019channel}
J.~Mirza and B.~Ali, ``Channel estimation method and phase shift design for
  reconfigurable intelligent surface assisted {MIMO} networks,'' \emph{arXiv
  preprint arXiv:1912.10671}, 2019.

\bibitem{he2020channel}
J.~He, M.~Leinonen, H.~Wymeersch, and M.~Juntti, ``Channel estimation for
  {RIS}-aided mmwave {MIMO} channels,'' \emph{arXiv preprint arXiv:2002.06453},
  2020.

\bibitem{chen2019channel}
J.~Chen, Y.-C. Liang, H.~V. Cheng, and W.~Yu, ``Channel estimation for
  reconfigurable intelligent surface aided multi-user {MIMO} systems,''
  \emph{arXiv preprint arXiv:1912.03619}, 2019.

\bibitem{hu2019two}
C.~Hu and L.~Dai, ``Two-timescale channel estimation for reconfigurable
  intelligent surface aided wireless communications,'' \emph{arXiv preprint
  arXiv:1912.07990}, 2019.

\bibitem{liu2019matrix}
H.~Liu, X.~Yuan, and Y.~Jun, ``Matrix-calibration-based cascaded channel
  estimation for reconfigurable intelligent surface assisted multiuser
  {MIMO},'' \emph{arXiv preprint arXiv:1912.09025}, 2019.

\bibitem{wei2020parallel}
L.~Wei, C.~Huang, G.~C. Alexandropoulos, and C.~Yuen, ``Parallel factor
  decomposition channel estimation in {RIS}-assisted multi-user {MISO}
  communication,'' \emph{arXiv preprint arXiv:2001.09413}, 2020.

\bibitem{wan2020broadband}
Z.~Wan, Z.~Gao, and M.-S. Alouini, ``Broadband channel estimation for
  intelligent reflecting surface aided mmwave massive {MIMO} systems,''
  \emph{arXiv preprint arXiv:2002.01629}, 2020.

\bibitem{khan2019deep}
S.~Khan and S.~Y. Shin, ``Deep-learning-aided detection for reconfigurable
  intelligent surfaces,'' \emph{arXiv preprint arXiv:1910.09136}, 2019.

\bibitem{elbir2020deep}
A.~M. Elbir, A.~Papazafeiropoulos, P.~Kourtessis, and S.~Chatzinotas, ``Deep
  channel learning for large intelligent surfaces aided {mm-Wave} massive
  {MIMO} systems,'' \emph{arXiv preprint arXiv:2001.11085}, 2020.

\bibitem{ning2019channel}
B.~Ning, Z.~Chen, W.~Chen, and Y.~Du, ``Channel estimation and transmission for
  intelligent reflecting surface assisted {THz} communications,'' \emph{arXiv
  preprint arXiv:1911.04719}, 2019.

\bibitem{ning2019channelMU}
B.~Ning, Z.~Chen, W.~Chen, Y.~Du, and J.~Fang, ``Channel estimation and hybrid
  beamforming for reconfigurable intelligent surfaces assisted {THz}
  communications,'' \emph{arXiv preprint arXiv:1912.11662}, 2019.

\bibitem{nadeem2019intelligent}
Q.-U.-A. Nadeem, A.~Kammoun, A.~Chaaban, M.~Debbah, and M.-S. Alouini,
  ``Intelligent reflecting surface assisted wireless communication: Modeling
  and channel estimation,'' \emph{arXiv preprint arXiv:1906.02360}, 2019.

\bibitem{wang2019channel}
Z.~Wang, L.~Liu, and S.~Cui, ``Channel estimation for intelligent reflecting
  surface assisted multiuser communications: Framework, algorithms, and
  analysis,'' \emph{IEEE Trans. Wireless Commun.}, {Early Access}.

\bibitem{you2019intelligent}
C.~You, B.~Zheng, and R.~Zhang, ``Intelligent reflecting surface with discrete
  phase shifts: Channel estimation and passive beamforming,'' in \emph{Proc.
  IEEE ICC}, Jun. 2020.

\bibitem{you2019progressive}
------, ``Channel estimation and passive beamforming for intelligent reflecting
  surface: Discrete phase shift and progressive refinement,'' \emph{IEEE J.
  Sel. Areas Commun.}, Early Access.

\bibitem{BS_Shen}
J.~Pan, Y.~T. Hou, L.~Cai, Y.~Shi, and S.~X. Shen, ``{Optimal base-station
  locations in two-tiered wireless sensor networks},'' \emph{IEEE Trans. Mobile
  Comput.}, vol.~4, no.~5, pp. 458--473, Sep. 2005.

\bibitem{BS_Youssef}
K.~Akkaya, M.~Younis, and W.~Youssef, ``{Positioning of base stations in
  wireless sensor networks},'' \emph{IEEE Commun. Mag.}, vol.~45, no.~4, pp.
  96--102, Apr. 2007.

\bibitem{Relay_Yu}
M.~Yu, J.~Li, and H.~Sadjadpour, ``{Amplify-forward and decode-forward: The
  impact of location and capacity contour},'' in \emph{Proc. IEEE MILCOM}, Oct.
  2005.

\bibitem{Relay_Tapolcai}
B.~Lin, P.~H. Ho, L.~Xie, S.~X. Shen, and J.~Tapolcai, ``{Optimal relay station
  placement in broadband wireless access networks},'' \emph{IEEE Trans. Mobile
  Comput.}, vol.~9, no.~2, pp. 259--269, Feb. 2010.

\bibitem{DoubleIRS}
Y.~Han, S.~Zhang, L.~Duan, and R.~Zhang, ``{Cooperative double-IRS aided
  communication: Beamforming design and power scaling},'' \emph{IEEE Wireless
  Commun. Lett.}, {Early Access}.

\bibitem{MAC}
S.~Zhang and R.~Zhang, ``Intelligent reflecting surface aided multiple access:
  Capacity region and deployment strategy,'' in \emph{Proc. IEEE SPAWC}, May
  2020.

\bibitem{Deployment_Journal}
------, ``Intelligent reflecting surface aided multi-user communication:
  Capacity region and deployment strategy,'' {to be submitted}.

\bibitem{Exploiting}
M.~A. Kishk and M.-S. Alouini, ``Exploiting randomly-located blockages for
  large-scale deployment of intelligent surfaces,'' {\emph{arXiv preprint
  arXiv:2001.10766}}, 2020.

\bibitem{Deployment_YW}
X.~Liu, Y.~Liu, Y.~Chen, and H.~V. Poor, ``{RIS} enhanced massive
  non-orthogonal multiple access networks: Deployment and passive beamforming
  design,'' {\emph{arXiv preprint arXiv: 2001.10363}}, 2020.

\bibitem{leung1978gaussian}
S.~Leung-Yan-Cheong and M.~Hellman, ``The {Gaussian} wire-tap channel,''
  \emph{IEEE Trans. Info. Theory}, vol.~24, no.~4, pp. 451--456, Jul. 1978.

\bibitem{cui2019secure}
M.~{Cui}, G.~{Zhang}, and R.~{Zhang}, ``Secure wireless communication via
  intelligent reflecting surface,'' \emph{IEEE Wireless Commun. Lett.}, vol.~8,
  no.~5, pp. 1410--1414, Oct. 2019.

\bibitem{yu2019enabling}
X.~Yu, D.~Xu, and R.~Schober, ``Enabling secure wireless communications via
  intelligent reflecting surfaces,'' in \emph{Proc. IEEE GLOBECOM}, Dec. 2019.

\bibitem{chen2019intelligent}
J.~Chen, Y.-C. Liang, Y.~Pei, and H.~Guo, ``Intelligent reflecting surface: {A}
  programmable wireless environment for physical layer security,'' \emph{IEEE
  Access}, vol.~7, pp. 82\,599--82\,612, Jul. 2019.

\bibitem{lu2020robust}
X.~Lu, W.~Yang, X.~Guan, Q.~Wu, and Y.~Cai, ``Robust and secure beamforming for
  intelligent reflecting surface aided {mmWave} {MISO} systems,'' \emph{arXiv
  preprint arXiv:2003.11195}, 2020.

\bibitem{guan2019intelligent}
X.~{Guan}, Q.~{Wu}, and R.~{Zhang}, ``Intelligent reflecting surface assisted
  secrecy communication: Is artificial noise helpful or not?'' \emph{IEEE
  Wireless Commun. Lett.}, vol.~9, no.~6, pp. 778--782, Jun. 2020.

\bibitem{xu2019resource}
D.~Xu, X.~Yu, Y.~Sun, D.~W.~K. Ng, and R.~Schober, ``Resource allocation for
  secure {IRS}-assisted multiuser {MISO} systems,'' in \emph{Proc. IEEE
  GLOBECOM}, Dec. 2019.

\bibitem{yu2019robust}
X.~Yu, D.~Xu, Y.~Sun, D.~W.~K. Ng, and R.~Schober, ``Robust and secure wireless
  communications via intelligent reflecting surfaces,'' \emph{arXiv preprint
  arXiv:1912.01497}, 2019.

\bibitem{yang2020deep}
H.~Yang, Z.~Xiong, J.~Zhao, D.~Niyato, L.~Xiao, and Q.~Wu, ``Deep reinforcement
  learning based intelligent reflecting surface for secure wireless
  communications,'' \emph{arXiv preprint arXiv:2002.12271}, 2020.

\bibitem{zeng2017communications}
Y.~Zeng, B.~Clerckx, and R.~Zhang, ``Communications and signals design for
  wireless power transmission,'' \emph{IEEE Trans. Commun.}, vol.~65, no.~5,
  pp. 2264--2290, May 2017.

\bibitem{wu2019weighted}
Q.~{Wu} and R.~{Zhang}, ``Weighted sum power maximization for intelligent
  reflecting surface aided {SWIPT},'' \emph{IEEE Wireless Commun. Lett.},
  vol.~9, no.~5, pp. 586--590, May 2020.

\bibitem{pan2019intelligent}
C.~Pan, H.~Ren, K.~Wang, M.~Elkashlan, A.~Nallanathan, J.~Wang, and L.~Hanzo,
  ``Intelligent reflecting surface enhanced {MIMO} broadcasting for
  simultaneous wireless information and power transfer,'' \emph{arXiv preprint
  arXiv:1908.04863}, 2019.

\bibitem{tang2019joint}
Y.~Tang, G.~Ma, H.~Xie, J.~Xu, and X.~Han, ``Joint transmit and reflective
  beamforming design for {IRS}-assisted multiuser {MISO} {SWIPT} systems,''
  \emph{arXiv preprint arXiv:1910.07156}, 2019.

\bibitem{lu2020enabling}
H.~Lu, Y.~Zeng, S.~Jin, and R.~Zhang, ``Enabling panoramic full-angle
  reflection via aerial intelligent reflecting surface,'' \emph{in Proc. IEEE
  ICC Workshop, arXiv preprint arXiv:2001.07339}, 2020.

\bibitem{li2020reconfigurable}
S.~{Li}, B.~{Duo}, X.~{Yuan}, Y.~{Liang}, and M.~{Di Renzo}, ``Reconfigurable
  intelligent surface assisted {UAV} communication: Joint trajectory design and
  passive beamforming,'' \emph{IEEE Wireless Commun. Lett.}, vol.~9, no.~5, pp.
  716--720, May 2020.

\bibitem{ma2020enhancing}
D.~Ma, M.~Ding, and M.~Hassan, ``Enhancing cellular communications for {UAVs}
  via intelligent reflective surface,'' in \emph{Proc. IEEE WCNC}, May 2020.

\bibitem{zeng2019accessing}
Y.~Zeng, Q.~Wu, and R.~Zhang, ``Accessing from the sky: A tutorial on {UAV}
  communications for {5G} and beyond,'' \emph{Proceedings of the IEEE}, vol.
  107, no.~12, pp. 2327--2375, Dec. 2019.

\bibitem{you20193d}
C.~You and R.~Zhang, ``{3D} trajectory optimization in {Rician} fading for
  {UAV}-enabled data harvesting,'' \emph{IEEE Trans. Wireless Commun.},
  vol.~18, no.~6, pp. 3192--3207, Jun. 2019.

\bibitem{you2020hybrid}
C.~You and R.~Zhang, ``Hybrid offline-online design for {UAV}-enabled data harvesting in
  probabilistic {LoS} channel,'' \emph{IEEE Trans. Wireless Commun.}, vol.~19,
  no.~6, pp. 3753--3768, Jun. 2020.

\bibitem{wang2019intelligent}
P.~Wang, J.~Fang, X.~Yuan, Z.~Chen, H.~Duan, and H.~Li, ``Intelligent
  reflecting surface-assisted millimeter wave communications: Joint active and
  passive precoding design,'' \emph{arXiv preprint arXiv:1908.10734}, 2019.

\bibitem{jamali2019intelligent}
V.~Jamali, A.~M. Tulino, G.~Fischer, R.~M{\"u}ller, and R.~Schober,
  ``Intelligent reflecting and transmitting surface aided millimeter wave
  massive {MIMO},'' \emph{arXiv preprint arXiv:1902.07670}, 2019.

\bibitem{you2020fast}
C.~You, B.~Zheng, and R.~Zhang, ``Fast beam training for {IRS}-assisted
  multiuser communications,'' \emph{IEEE Wireless Commun. Lett.}, Early Access.

\bibitem{mao2017mobile2}
Y.~Mao, C.~You, J.~Zhang, K.~Huang, and K.~B. Letaief, ``A survey on mobile
  edge computing: The communication perspective,'' \emph{IEEE Commun. Surveys
  Tuts.}, vol.~19, pp. 2322--2358, Aug. 2017.

\bibitem{cao2019intelligent}
Y.~Cao and T.~Lv, ``Intelligent reflecting surface enhanced resilient design
  for {MEC} offloading over millimeter wave links,'' \emph{arXiv preprint
  arXiv:1912.06361}, 2019.

\bibitem{liu2020intelligent}
Y.~Liu, J.~Zhao, Z.~Xiong, D.~Niyato, Y.~Chau, C.~Pan, and B.~Huang,
  ``Intelligent reflecting surface meets mobile edge computing: Enhancing
  wireless communications for computation offloading,'' \emph{arXiv preprint
  arXiv:2001.07449}, 2020.

\bibitem{bai2019latency}
T.~Bai, C.~Pan, Y.~Deng, M.~Elkashlan, and A.~Nallanathan, ``Latency
  minimization for intelligent reflecting surface aided mobile edge
  computing,'' \emph{arXiv preprint arXiv:1910.07990}, 2019.

\bibitem{guan2020joint}
X.~Guan, Q.~Wu, and R.~Zhang, ``Joint power control and passive beamforming in
  {IRS}-assisted spectrum sharing,'' \emph{IEEE Commun. Lett.}, Early Access.

\bibitem{xu2020resource}
D.~Xu, X.~Yu, and R.~Schober, ``Resource allocation for intelligent reflecting
  surface-assisted cognitive radio networks,'' \emph{arXiv preprint
  arXiv:2001.11729}, 2020.

\bibitem{wymeersch2019radio}
H.~Wymeersch, J.~He, B.~Denis, A.~Clemente, and M.~Juntti, ``Radio localization
  and mapping with reconfigurable intelligent surfaces,'' \emph{arXiv preprint
  arXiv:1912.09401}, 2019.

\bibitem{bourdoux20206g}
A.~Bourdoux \emph{et~al.}, ``{6G} white paper on localization and sensing,''
  \emph{arXiv preprint arXiv:2006.01779}, 2020.


\bibitem{chaccour2020risk}
C.~Chaccour, M.~N. Soorki, W.~Saad, M.~Bennis, and P.~Popovski, ``Risk-based  optimization of virtual reality over terahertz reconfigurable intelligent
  surfaces,'' \emph{arXiv preprint arXiv:2002.09052}, 2020.


\bibitem{basar2020reconfigurable}
E.~Basar, ``Reconfigurable intelligent surface-based index modulation: A new
  beyond {MIMO} paradigm for {6G},'' \emph{IEEE Trans. Commun.}, vol.~68,
  no.~5, pp. 3187--3196, May 2020.
 \end{thebibliography}
\end{document}